\documentclass[
10pt, 
a4paper, 
oneside, 
headinclude,footinclude, 
BCOR5mm, 
]{scrartcl}
\usepackage{tabularx,colortbl}
\usepackage{setspace}
\usepackage{caption}
\usepackage[
nochapters, 
beramono, 
eulermath,
pdfspacing, 
dottedtoc 
]{classicthesis} 

\usepackage{arsclassica} 

\usepackage[T1]{fontenc} 

\usepackage[utf8]{inputenc} 

\usepackage{graphicx} 
\graphicspath{{Figures/}} 

\usepackage{enumitem} 

\usepackage{lipsum} 

\usepackage{subfig} 

\usepackage{amsmath,amssymb,amsthm} 

\usepackage{varioref} 

\theoremstyle{definition} 

\theoremstyle{plain} 

\theoremstyle{remark} 

\hypersetup{
colorlinks=true, breaklinks=true, bookmarks=true,bookmarksnumbered,
urlcolor=webbrown, linkcolor=RoyalBlue, citecolor=webgreen, 
pdftitle={}, 
pdfauthor={\textcopyright}, 
pdfsubject={}, 
pdfkeywords={}, 
pdfcreator={pdfLaTeX}, 
pdfproducer={LaTeX with hyperref and ClassicThesis} 
} 
\usepackage{amsfonts}
\usepackage{amssymb}
\usepackage{amsmath}
 \newcommand{\ie}{\mbox{\textit{i.e.}, }}
\newcommand{\eg}{\mbox{\textit{e.g.}, }}
\newcommand{\La }{\mbox{$\mathcal{L}$} }       

\newcommand{\A}{\mathcal{A}}

\newcommand{\PA}{\mathcal{A}^+}

\newcommand{\SA}{\mathcal{A}}

\newcommand{\rest}[2]{{#1}_{\downarrow#2}}

\newcommand{\ai}{\varphi}

\newcommand{\Js}{J}

\newcommand{\Ct}{\Gamma}

\newcommand{\Dmc}{\mathbb{J}}

\newcommand{\Pmc}{\mathbb{V}(\mathcal{O})}

\newcommand{\Pf}{P}

\newcommand{\Qf}{Q}
 
\newcommand{\Alt}{\mathcal{O}}

\newcommand{\V}{V}

\newcommand{\R}{F}
\newcommand{\F}{F}

\newcommand \argmax[1] {\underset{{#1}}{\mbox{argmax}}}
\newcommand \argmin[1] {\underset{{#1}}{\mbox{argmin}}}

 \newcommand{\npf}[2]{N({#1},{#2})}
 
  \newcommand{\pp}[2]{p_{{#1}\succ{#2}}}

\newcommand{\RMWA}{{{\textsc{med}}}}
\newcommand{\RY}{{{\textsc{y}}}}

\newcommand{\RMSA}{{{\textsc{mc}}}}
\newcommand{\RMNAC}{{\textsc{full$_H$}}}
\newcommand{\RMCSA}{{{\textsc{mcc}}}}
\newcommand{\RRA}{{{\textsc{ra}}}}

\newcommand{\REVS}{\textsc{\R}_{rev}}

\newcommand{\MC}[1]{{max(#1, \subseteq)}}

\newcommand{\MCC}[1]{{max(#1, |.|)}}

\newcommand{\Atoms}{{\mathtt{Atoms}}}

\newcommand{\new}[1]{{\color{Mahogany}#1}}

\newtheorem{example}{Example}

\title{\normalfont\spacedallcaps{An Introductory Course to Judgment Aggregation}} 

\author{\spacedlowsmallcaps{Marija Slavkovik\textsuperscript{1}}} 

\date{} 

\begin{document}


\renewcommand{\sectionmark}[1]{\markright{\spacedlowsmallcaps{#1}}} 
\lehead{\mbox{\llap{\small\thepage\kern1em\color{halfgray} \vline}\color{halfgray}\hspace{0.5em}\rightmark\hfil}} 

\pagestyle{scrheadings} 


\maketitle 

\setcounter{tocdepth}{3} 

 {\em These lecture notes accompany the  course "An Introduction to Judgment Aggregation for Multi Agent Systems" given at the \href{http://easss2016.dmi.unict.it}{18th European Agent Systems Summer  School}. }

\section*{Abstract} 
 Reaching some form of consensus is often necessary for  autonomous agents  that  want to coordinate their actions or otherwise engage in  joint activities. One way to reach a consensus is by aggregating individual information, such as decisions, beliefs, preferences  and constraints. Judgment aggregation is a social choice method, which generalises voting, that studies the aggregation of individual judgments regarding the truth-value  of logically related propositions. As such, judgment aggregation is applicable for consensus reaching problems in multi agent systems. As other social choice theory, judgment aggregation research is abundant with impossibility results. However, the aim of this tutorial is to give an introduction to the methods of judgment aggregation, not the impossibility results.  In particular, the tutorial will introduce the basic frameworks that model judgment aggregation problems  and give an overview of the judgment aggregation functions so far developed as well as their social theoretic and computational complexity properties. The focus of the tutorial are  consensus reaching problems in multi agent systems that can be modelled as judgment aggregation problems. The desirable properties of a judgment aggregation method applied to these problems are not necessarily the same as properties desirable in legal or political contexts, which is considered to be the native domain of judgment aggregation. After this tutorial the participants are expected to be able to read and understand judgment aggregation literature and have a grasp on the state-of-the-art and open questions in judgment aggregation research of interest to multi agent systems.


{\let\thefootnote\relax\footnotetext{\textsuperscript{1} \textit{Department of Information Science and Media Studies,   University of Bergen, Norway}}}


\newpage 
\tableofcontents 
\newpage
\listoffigures 
\newpage
\listoftables 

\newpage 


\section{Introduction}  

\label{ch:introduction}  
Social choice theory \cite{Arrow:2002} is concerned with transforming individual decision into collective ones. As a research area it spans mathematics, economics, political science, philosophy, and as of recently, computer science. The interest for social science in computer science stems from the modern development of the computer science field itself.  One of the characteristics of modern computer science is an increased distributivity both of computation and information, but also the increase of intelligence and autonomy of computational entities. Past passive web pages are increasingly becoming intelligent trackers of behaviour and extractors of information from the visitors. Intelligent agents gather information from various sources and dynamically construct prices for services.  These are all situations in which intelligent autonomous agents would need to make collective decisions and merge information, which can be modelled as well known social choice theory problems. This gave rise to a new interdisciplinary field of research -- {\em computational social choice} \cite{handbook}.

The goal of this document is to acquaint  students and researchers that work in computer science, and multi agent systems in particular, with a lesser known area of social choice - judgment aggregation. Judgment aggregation is an abstract aggregation framework in which many complex collective decision problems can be modelled and it is based on a logic framework, making it somewhat more natural choice for a social choice method used in multi agent systems. We aim to acquaint the readers with the state of the art in judgment aggregation enabling them to independently follow the literature in judgment aggregation and perhaps pursue some of the numerous open problems in this field. 

\subsection{What are social choice problems}

In most democracies elections are held at regular intervals to select members of parliaments, presidents, prime ministers and various other representatives and leaders. This is the perhaps the  best known example of social choice problems,   the problems of voting and preference aggregation. The legitimacy of the elected representative is based on confidence in the fairness of the procedure that is used to aggregate the individual votes in the electorate. A fair and overall good aggregation  procedure  is one that selects winners in lines with what each of the agents in  the electorate  prefers. We give an overview of preference aggregation and voting in Section~\ref{sec:pref}. Much of social choice theory is concerned with  study of aggregation procedures - designing good procedures and analysing their properties. 

Social choice methods are not limited to  voting problems. Another well known social choice problem is {\em matching}: each from one group  agents needs to be paired off with someone from a second group of agents based on preferences. For example consider a list of candidates for residency positions in hospitals and a list of hospitals. Each hospital has a preference over which resident it would like to hire, while  each resident has a preference over for which hospital she would like to work. The problem of finding a match such that no two agents would preferred to be matched with each other rather than with the agents with which they are already matched is known as the {\em stable marriage problem}.

Well known social choice problems with established applicability in  multi agent systems  are the {\em resource allocation problem} \cite{Chevaleyre:2005}. Given a set of goods, that can be divisible, indivisible, or both, and a set of agents each with different preferences over  (combinations of) the goods, we need to find a way to divide the goods so that no agent would rather have someone else's allocation. This is the envy-free resource allocation problem.

Judgment aggregation is the problem of finding collective answers to a set of questions. The questions, or {\em issues} admit binary answer, "yes" and "no", or alternatively "true" or "false". The questions are such that an answer to some questions constraints the answers that can be given to others; we typically say that the issues are {\em logically interrelated}. The problem is to find a function that aggregates  sets of answers, called {\em judgment sets} into a logically consistent, but also representative,  collective set of answers.

\subsection{ Social choice terminology}
We  now introduce some basic terminology used when discussing social choice and social choice problems. 

On a general abstract level, a social choice problem is the problem of combining individual decisions into a representative collective decision. The individual decisions are combined by applying an aggregation function, or an aggregator, to them.   The term representative is intuitively equated with {\em majority supported} or {\em plurality supported} - the collective decision is surely representative if it coincides with the  majority of the individual decisions, or with the plurality of theme. A decision is supported by a majority if out of $n$ individuals more than $\frac{n}{2}$ support it. A decision is supported by a plurality if there are more individual supporting this decision then any other. 

However, representative does not only mean majority supported. Furthermore, a collective choice by some majoritarian aggregator, even if appealing but not always possible for many reasons.  Different ideas of what a good aggregator is or is not have been floated in the literature throughout the history.  Social choice theory aims to axiomatise these concepts and study them formally.

We can often find mentions of  {\em  characterisation results } in social choice theory. This term refers to works in which a  list of aggregator properties are considered together and shown that there exists precisely one aggregator that simultaneously satisfies them. Another frequently used term is {\em impossibility result}. This term refers to theorems  which show that no aggregator can simultaneously satisfy a particular list of aggregator properties. The most famous of the impossibility results is that of   Kenneth Arrow  \cite{Arrow:63}  regarding  preference aggregation. Arrow shows that a very small set of some rather simple properties cannot be simultaneously satisfied by an aggregator of preferences.

What impossibility results actually show is that an ideally representative, or fair, or democratic collective decision maybe does  not exist for every social choice problem. However, this does not mean that no aggregation is possible, rather it means that too many options for aggregating individual decisions are available. Since  each  aggregator may identify  a different collective decision,   which aggregator is best for a given aggregation problem is also within the scope of interest for social choice theory.

 \subsection{Short history of judgment aggregation}
We give the short history of the judgment aggregation as a research field. Short because the history of judgment aggregation is short, particularly when compared with that of other problems such as the aggregation of votes. The origins of aggregating  collections of binary answers can be found in  \cite{RuFi86,Wilson1975}. These works consider sets of allowed lists (of fixed length) of binary evaluations. An example of a list of binary evaluations of length three is  $(1,0,0)$. It is shown that aggregating a collection of  lists from the allowed set,  for example $\langle (1,0,0), (0,1,0), (1,1,1) \rangle$ does not necessarily yield a list of evaluations from the  allowed set. 

 It is widely considered \cite{ListPuppe2009} that interest in the field was sparked by the papers on how collective decisions are made by collegiate courts \cite{KS86,KS93}. A court needs to not only make a verdict but also justify that verdict with reasons, based on the low. How the reasons determine the verdict is set out by the legal doctrine, or in plain terms "the law".  Korhauser and Sager \cite{KS86,KS93} use examples from legal cases to show that  the direction in which the verdict is reached can influence which verdict is reached.  We give an example of the problem they consider. 

 The verdict the each judge  needs to make is 
 \begin{description}
\item `Do you (the judge) rule for the plaintiff?'. 
 \end{description} 
 Each judge must rule in favour of the plaintiff when there was both a contract and a breach of contract, otherwise the judge must rule for the defendant (and against the plaintiff).  So effectively the judges also nee to make a judgment regarding two issues: 
 \begin{description}
\item `Are you convinced  there is a contract between the plaintiff and the defendant?' 
\item  `Are you convinced  the defendant breached the contract?'. 
\end{description}     
  A problem arrises on how to reach a verdict when:
  \begin{description}
\item the first judge believes there was a contract but no breach of contract and thus rules for the defendant, 
\item the second judge believes there was no contract, but should there had been one it would have been breached, 
 thus rules for the defendant, 
 \item the third judge rules for the plaintiff since she is convinced there was both a contract and a breach of contract.
 \end{description}
  If the judges first look at the reasons for the verdict, they find that 2 out of 3 judges believe there was a contract and 2 out of 3 judges believe there was a breach of contract. The collective decision on the reasons implies that  the court should rule in favour of the plaintiff. However, 2  out of 3 judges individually have ruled for the defendant, therefore the court should rule in favour of the defendant. 
 Much of the literature on this topic in legal theory is concerned with identifying when the the judges should base their verdict on individual opinions regarding the reasons and when directly on individual verdicts \cite{Nash2003}.
 
 Judgment aggregation took its present form  when the aggregation problem was modelled using propositional logic to  represent the issues on which individual answers are given. The   purpose of using logic representation is to explicitly represent  the logic relations between the issues of interest and be able to  check the consistency of the combinations of answers. 
 
 It was shown that aggregating judgments is a different  problem than generalises aggregating preferences \cite{CLiPe02}. Judgment aggregation is also subject to impossibility results, the kind that Arrow proved for preference aggregation problems, \cite{Dietrich07}. The first decade of judgment aggregation yielded many impossibility results, comprehensive overview of which can be found in \eg~ \cite{ListPuppe2009,GrossiP:2014}. The properties used in attaining impossibility results are heavily inspired by those used for the same purpose in preference aggregation.  

While much attention was devoted to showing that the perfect judgment aggregator cannot exist,  very little actual  aggregators have been developed until very recently \cite{LangPSTV15} and even less non-preference aggregation-inspired properties of aggregators have been studied. One of the reason for this state of affairs can be found in the abstractness of the framework - it is not obvious what are the natural judgment aggregation examples. Outside of collegiate courts, people seem to avoid dealing with complex collective decisions and reasoning about them. In comparison voting is very cognitively simple and intuitive. Instead of stating their opinions and then aggregating them people seem to prefer to talk things over until a consensus is reached one way or another, not necessarily by fair and democratic means \cite{march}. Intelligent artificial agents however are not yet capable of taking collective decision shortcuts. 

Although we devote an entire section to the judgment aggregation for multi agent systems, we would like to briefly mention here the place that judgment aggregation can have as a tool for collective decision making for agents. 
When a collective decision is mentioned, the intuitive mental image that arises is one of people on a round table attempting to reach an agreement. Agents that cooperate would need to make collective decisions about what to believe, which goal to pursue, which plan to jointly choose, how to divide tasks and revenues between them etc. 
Although making agreements is one application for judgment aggregators it is not the only one. 

Artificial agents would need to combine information from various sources to form their beliefs about the world or other agents. This is a form of learning about the world. For example an agent should recommend "Lord of the Rings" as a movie to watch to a user  if the user is older than thirteen, has watched  at least two other fantasy movies and has not expressed dislike of the "Lord of the Rings" books. The information about the age, movie preferences and book preferences of the user may come from various different sources on the web and may contradict each other. 

A strong shortcoming of judgment aggregation is that we have at present sufficient results \cite{LangECAI14,EndrissGP12,EndrissDeHaanAAMAS2015} to be sure that the problem of aggregating judgments  is  computationally expensive. This should not be taken as a discouragement. Many logic reasoning problems are computationally expensive, nonetheless  optimisation methods have been found. Designing efficient judgment aggregators is still an open territory in judgment aggregation research. 

\subsection{Beyond these lecture notes}
The goal of these lecture notes is to give a comprehensive introduction to judgment aggregation and our focus is on specific judgment aggregators and their properties. The list of aggregators and properties is very small and almost none of the aggregators have been characterised.  Although the field of judgment aggregation is small, it was still not possible to cover all of it within this document. 

The complexity theoretic analysis of various aspects of the judgment aggregators is very much in the scope of interest of computational social choice. We have results regarding almost all judgment aggregators at present and we can direct the reader towards \eg~\cite{LangECAI14,EndrissGP12,EndrissDeHaanAAMAS2015}, which also contain references to further work in this area. After covering this course and with some background in complexity theory a reader should be able to independently follow the literature.

One area of research involving judgment aggregation and falling within the scope of multi agent systems is the use of judgment aggregation in abstract argumentation \cite{CaminadaP11}. Works in this area include, but are not limited to \cite{AwadBTR14,BoothAR14,Booth14,RahwanT10}. To follow these works the reader is expected to have some background in abstract argumentation theory.
 
 The assumed background for these lecture notes is basic understanding of classical propositional logic. We start with introducing the two most popular frameworks for judgment aggregation. No agreement on notation exists in the literature, with each group of researchers devising their own, but the concepts used remain the same. We make an attempt to give numerous examples and intuitive insights  which may at time detract from mathematical rigour.   
 \newpage
 


\section{Aggregation frameworks} 

\label{ch:framework}  

We begin by introducing the framework, or rather frameworks,  for judgment aggregation. Several judgment aggregation frameworks have been proposed in the literature.  The two arguably most frequently used are the propositional logic framework and the binary framework. It is common in the literature to use the term {\em  judgment aggregation} when studying aggregation problems in the  propositional logic framework and {\em binary aggregation} when  studying aggregation problems in the binary framework.  We predominantly use the logic framework within the scope of this document, but introduce both to help highlight the differences and equip the reader to follow with ease all scientific articles within the judgment aggregation discipline.  The choice between the two frameworks is a matter of convenience and personal preference for a particular research problem and personal preference. It was shown in \cite{EndrissEtAlKR2016} that the two frameworks are equally expressive and only slightly differ in terms of succinctness of representation. In this section we introduce both the propositional logic framework and the binary framework and briefly discuss the relations between them. For completeness, we also discuss and give references to other frameworks for judgment aggregation.


\subsection{Propositional logic framework}
The propositional logic framework for judgment aggregation is used extensively in judgment aggregation works in economics and political science such as \cite{CLiPe02,Dietrich07,DuddyP:2012,Dietrich:2013,Mongin2008}, as well as in artificial intelligence such as \cite{GrossiP:2014,LangPSTV15} and \cite[Chapter 17]{handbook}. This framework utilises  a set of well formed formulas of propositional logic $\La_p$. The propositional logic framework is built around the representation of an issue  as a pair\footnote{Frequently we refer just to the non negated formula $\ai$ as issue.} of formulas $\{\ai,\neg \ai\} \subset \La_p$. Now, making a  judgment on an issue equates to making  a choice between the two formulas $\ai$ and $\neg \ai $.   A judgment aggregation problem is a tuple of $\langle \A, \Ct, \Pf \rangle$, where $\A \subset \La_p$ is an {\em agenda}\footnote{In the literature  the  letters $\Phi$ and $X$ are also is used for the agenda. We reserve $\Phi$ for agendas in the binary framework and use $\A$   to distinguish the agendas in the logic framework.} of issues, $\Ct \subset \La_p$ is a {\em set of constraints}, and $\Pf$ is a {\em profile of judgments}  that is a list (an ordered multi-set) of {\em judgment sets}. We define the rest of the concepts.

 \subsubsection{Judgment aggregation problem}
\paragraph{Agenda} An   agenda  $\A = \{ \ai_1, \neg \ai_1, \ldots, \ai_m, \neg \ai_m\}$ is a finite set of issues $\{ \ai_i, \neg \ai_i\} \subset \La_p$. The issues in the agenda may be literals, but also more complex propositional formulas. Given an agenda $\A$, the {\em pre-agenda} $\PA$ associated with it\footnote{For the pre-agenda the notation $[\A]$ is sometimes used in the literature.} is the set of the non-negated formulas from $\A$, namely $\PA=\{\ai_1, \ldots, \ai_m\}$. It is assumed that the formulas in the pre-agenda are satisfiable, namely they are neither tautologies, nor contradictions. An agenda subset $\A_s \subseteq \A$ is a subset of issues in $\A$ such that necessarily if $\ai \in \A_s$, then $\neg \ai \in \A_s$. 
 
 The agenda represents all the issues on which binary collective decisions need to be made. Within the propositional logic framework,  the issues can be formulas, not only atoms, the logic relations between such issues can be captured already in the agenda, as  illustrated by the following example, slightly modified  from the example in \cite{ListPuppe2009}.
 
 \begin{example}\label{ex:co2} Let the atom  $p$ denote  the proposition "Current \textsc{CO}$_2$ emissions lead to global warming". Let the formula $(p \wedge r)\rightarrow q$ denote the proposition "If current \textsc{CO}$_2$ emissions lead to global warming and global warming will be disastrous for humanity, then we should reduce \textsc{CO}$_2$ emissions". Lastly let the atom $q$ denote the proposition "We should reduce \textsc{CO}$_2$ emissions". The agenda formed of these three issues is \linebreak $\A = \{p, \neg p, (p \wedge r)\rightarrow q, \neg \big( (p \wedge r)\rightarrow q\big), q, \neg q\}$.  The pre-agenda for this agenda is $\PA = \{p, (p \wedge r)\rightarrow q, q\}$.  
   The atom $r$, which clearly denotes the proposition  "Global warming will be disastrous for humanity" is not explicitly considered an issue in the agenda,  since the statement it represents  is  a common assumption and requires no judgment. Nonetheless it is included in the formula representing the second agenda issue to make the assumption explicit in determining a judgment for this issue. 
 \end{example} 

In addition to implicitly expressing logic relations among the agenda issues by using non-atomic formulas, the judgment aggregation problem can have explicitly mandated agenda issue relations, or constraints. 

\paragraph{Constraints} Given a set of formulas $S \subset \La_p$, let $\Atoms(S)$ be the set of all propositional variables that occur in a formula in $S$. The  set of constraints for an agenda $\A$ is a given nonempty set  $\Ct \subset \La_p \cup \{ \top\} $ such that $\Atoms(\A) \cap \Atoms(\Ct) \neq \emptyset$ and if $\ai \in \A$, then $\ai \not \in \Ct$ and $\neg \ai \not \in \Ct$. It is assumed that the formulas in $\Ct$ are satisfiable and that $\Ct$ is a consistent set of formulas. 

The condition  $\Atoms(\A) \cap \Atoms(\Ct) \neq \emptyset$ ensures that the constraints are relevant to the agenda, while the condition  if $\ai \in \A$, then $\ai \not \in \Ct$ and $\neg \ai \not \in \Ct$ ensures that judgments can be cast for issues in the agenda. If  either  $\ai   \in \Ct$ or  $\neg \ai   \in \Ct$, the issue $\{ \ai, \neg \ai\}$ is already {\em resolved}.

The constraints ensure that certain combination of judgments cannot be made concurrently for issues in the constrained agenda. The case when no constraints are specified is represented with $\Ct = \{ \top\}$.  The difference between non-atomic formulas in the agenda and constraints is illustrated in Example~\ref{ex:fjs}.

\begin{example}\label{ex:fjs} Consider two pre-agendas $\A^+_1 = \{ p, p\rightarrow q, q\}$ with constraints $\Ct_1 = \{ \top\}$ and $\A^+_2 = \{ p, q\}$ with constraints $\Ct_2 = \{ p\rightarrow q\}$. When giving judgments on the first agenda, the agents can choose whether the relation $p \rightarrow q$ holds between $p$ and $q$ -- the relation between issues is an issue itself.  When giving judgments on the second agenda however, $p \rightarrow q$ has to hold. Thus a combination of judgments $p$ and $\neg q$ is allowed for judgments on the first agenda, but not on the second. 
\end{example}

 In the propositional logic framework, the decision-makers, or agents, express their judgments on a given agenda, respecting relevant given constraints for that agenda. The input from each individual agent is a judgment set.



\paragraph{Judgment set.} A judgment set is an agenda subset $\Js \subset \A$. A judgment set is {\em complete}  for an agenda $\A$ if it contains a judgment for every issue in $\A$, or in other words,  for every $\ai \in \PA$, either $\ai \in \Js$ or $\neg \ai \in \Js$. A judgment set is {\em consistent} if it is a consistent set of classical logic formulas and it is consistent with the constraints $\Ct$, namely $\Js \cup \Ct \nvDash \bot$, where $\models$ is the classical logic consequence operator.  A judgment set is {\em rational} if and only if it is consistent and complete, of course all with respect to given $\A$ and $\Ct$.

In general,   a subset of the agenda $S \subset \A$ is {\em inconsistent} if $S  \cup \Ct \models \bot$. A set $S$ is a {\em consistent subset of the agenda} if and only of it is not an inconsistent subset of the agenda.   The set $S$ is a {\em minimally  inconsistent subset} of the agenda if it is an  inconsistent subset of the agenda and for all $S' \subset S$, $S'$ is a consistent subset of the agenda.  Namely $S$ is minimally  inconsistent if it can be made consistent by adding any one element from the agenda. 

In a judgment aggregation problem, the agents are asked to give their judgment sets for an agenda $\A$ and associated constraints $\Ct$. Typically, the agents are allowed to only contribute complete and consistent judgment sets. 
For a given agenda $\A$ and associated constraints $\Ct$ we can construct the set of all complete and consistent judgment sets. Within the scope of this document, we will call this set  {\em the codomain}. 

\paragraph{Codomain.} Given an agenda $\A$ and associated constraints $\Ct$, the codomain $\Dmc(\A,\Ct)$ is the set  of all complete and consistent  judgment sets  that can be constructed for  $\A$ and $\Ct$ . 

We will sometimes need to refer to the set of all consistent, but incomplete, judgments for an agenda $\A$ and associated constraints $\Ct$. We denote this set with $\mathcal{\Js}(\A,\Ct)$. Clearly $\Dmc(\A,\Ct) \subset \mathcal{\Js}(\A,\Ct)$. As a convention, we adopt that the empty set is in  $\mathcal{\Js}(\A,\Ct)$. 

Although it would not be very succinct in general to do so, a judgment aggregation problem can also be represented as 
a pair of codomain and profile, namely $\langle \Dmc(\A,\Ct), \Pf \rangle$.

\begin{example} Consider again the pre-agenda $\PA = \{ p,  (p \wedge r)\rightarrow q, q\}$ and two possible constraints $\Ct_1 = \{ \top\}$ and $\Ct_2=\{r\}$. We obtain the following codomains. 

\begin{minipage}{0.4\textwidth}
\[
\Dmc(\A,\Ct_1)=\left\{
\begin{array}{l}
\{p,  (p \wedge r)\rightarrow q, q\},\\ 
\{p,  (p \wedge r)\rightarrow q, \neg q\},\\ 
\{p,  \neg \big( (p \wedge r)\rightarrow q\big),  \neg q\},\\ 
\{\neg p,  (p \wedge r)\rightarrow q, q\},\\ 
\{\neg p,  \neg \big( (p \wedge r)\rightarrow q\big), q\}
\end{array}\right\}
\]
\end{minipage}

\begin{minipage}{0.4\textwidth}
\[
\Dmc(\A,\Ct_2)=\left\{
\begin{array}{l}
\{p,  (p \wedge r)\rightarrow q, q\},\\
\{p,  \neg \big( (p \wedge r)\rightarrow q\big), \neg q\},\\ 
\{\neg p,  (p \wedge r)\rightarrow q, q\},\\ 
\{\neg p,  (p \wedge r)\rightarrow q, \neg q\},\\  
\end{array}\right\}
\]
\end{minipage}
\end{example}

The last element of the judgment aggregation problem is the profile of judgments. The profile is nothing but a collection 
of judgment sets, one representing each decision-maker or agent. 

\paragraph{Profile.} Consider an agenda  $\A$ and associated constraints $\Ct$.  Given a set of $n$ agents, each represented with a  judgment set $\Js_i \in \Dmc(\A,\Ct)$, a profile $\Pf$ is a list $\Pf \in \Dmc(\A, \Ct)^n$, $\Pf= (\Js_1, \ldots, \Js_n)$. We slightly abuse notation and use $\Js_i \in \Pf$ to denote that $\Js_i$ is agent $i$'s judgment set in $\Pf$. 

Within the scope of this document, we use subscripts to denote judgment sets that are associated with an agent, thus  $\Js_i$ is a judgment set that represents agent $i$. We use superscript to denote judgment sets in the codomain. 

The next example is an instance of the "doctrinal paradox" judgment aggregation problem. 

\begin{example}\label{ex:doctrinal} The "doctrinal paradox" judgment aggregation problem is the problem of deciding whether a defendant is liable or not (guilty or innocent) with respect to a case of breached contract. The following issues are considered: $p$ representing "a contract exist between the plaintive and the defendant", $q$ representing  "the defendant had breached   the contract (assuming one existed)" , and $d$ representing "the defendant is liable for a breach of contract with the plaintive". The legal doctrine stipulates the following constraint $(p\wedge q) \leftrightarrow d$ that represents that the defendant is liable for a breach of contract if and only if a contract existed and that contract was breached. Thus the pre-agenda is $\PA = \{ p,q,d\}$ and the set of constraints is $\Ct = \{ (p\wedge q) \leftrightarrow d\}$. Now we can determine the codomain \linebreak $\Dmc(\A,\Ct) = \{ \{ p, q, d\}, \{ p, \neg q, \neg d\},\{ \neg p, q, \neg d\},\{ \neg p, \neg q, \neg d\}\}$. The profile of the "doctrinal paradox" judgment aggregation problem is \linebreak $\Pf = (\{ p, q, d\}, \{ p, \neg q, \neg d\},\{ \neg p, q, \neg d\})$.
\end{example}

Various operations on profiles can be defined.  We give the definitions of ones that we will use most often within this document. 

\subsubsection{Operations on profiles} \label{sec:op}

\begin{description}
\item The number $N(\ai, \Pf)$, is the number of times a judgment $\ai \in \A$ is endorsed in a profile $\Pf \in \Dmc(\A,\Ct)^n$. Formally $N(\ai, \Pf)= \#\{ i \mid \ai \in \Js_i, \Js_i \in \Pf\}$. Clearly $N(\ai, \Pf) \geq 0$ and $N(\ai, \Pf) \leq n$. If $N(\ai, \Pf) = n$ for some $\ai \in \A$, then we say that there is a {\em unanimity on $\ai$ in $\Pf$}.  If $N(\ai, \Pf) = n$ for every $\ai \in \A$, then we say that $\Pf$ is a {\em unanimous profile}. For a unanimous profile $\Pf$ there exists a  judgment set $\Js \in \Dmc(\A,\Ct)$ such that for every $i$,  $1 \leq i \leq n$, it holds $\Js_i = \Js$. This judgment set $\Js$ is called the {\em unanimity judgment}.

\item The partial profile $\Pf^{\downarrow \A_S}$ is defined for a profile $\Pf \in \Dmc(\A,\Ct)^n$, $\Pf = (\Js_1, \ldots, \Js_n)$ and an agenda subset $\A_S \subseteq \A$ as follows 
\[\Pf^{\downarrow \A_S} = (\Js_1 \cap \A_S, \ldots , \Js_n \cap \A_S). \]
A partial profile $\Pf^{\downarrow \A_S}$ is obtained from a profile $\Pf$ when in each judgment set $\Js_i$  in $\Pf$, the judgments that are not elements of $\A_S$ are removed from $\Js_i$. We give an example. 
\begin{example} Consider again "doctrinal paradox" problem given in Example~\ref{ex:doctrinal}. Let $\A_S=\{ p,\neg p,  d, \neg d\}$ and let  $\Pf$ be  the judgment profile in this example. For the  partial profile $\Pf^{\downarrow \A_S}$ we have 
  $\Pf^{\downarrow \A_S} =  (\{ p,  d\}, \{ p, \neg d\},\{ \neg p,  \neg d\})$.
  
  Visually, the partial profile $\Pf^{\downarrow \A_S}$. See Figure~\ref{fig:profvert}.
  
  \begin{figure}[h!]
    \begin{center}
        \includegraphics[scale=0.35]{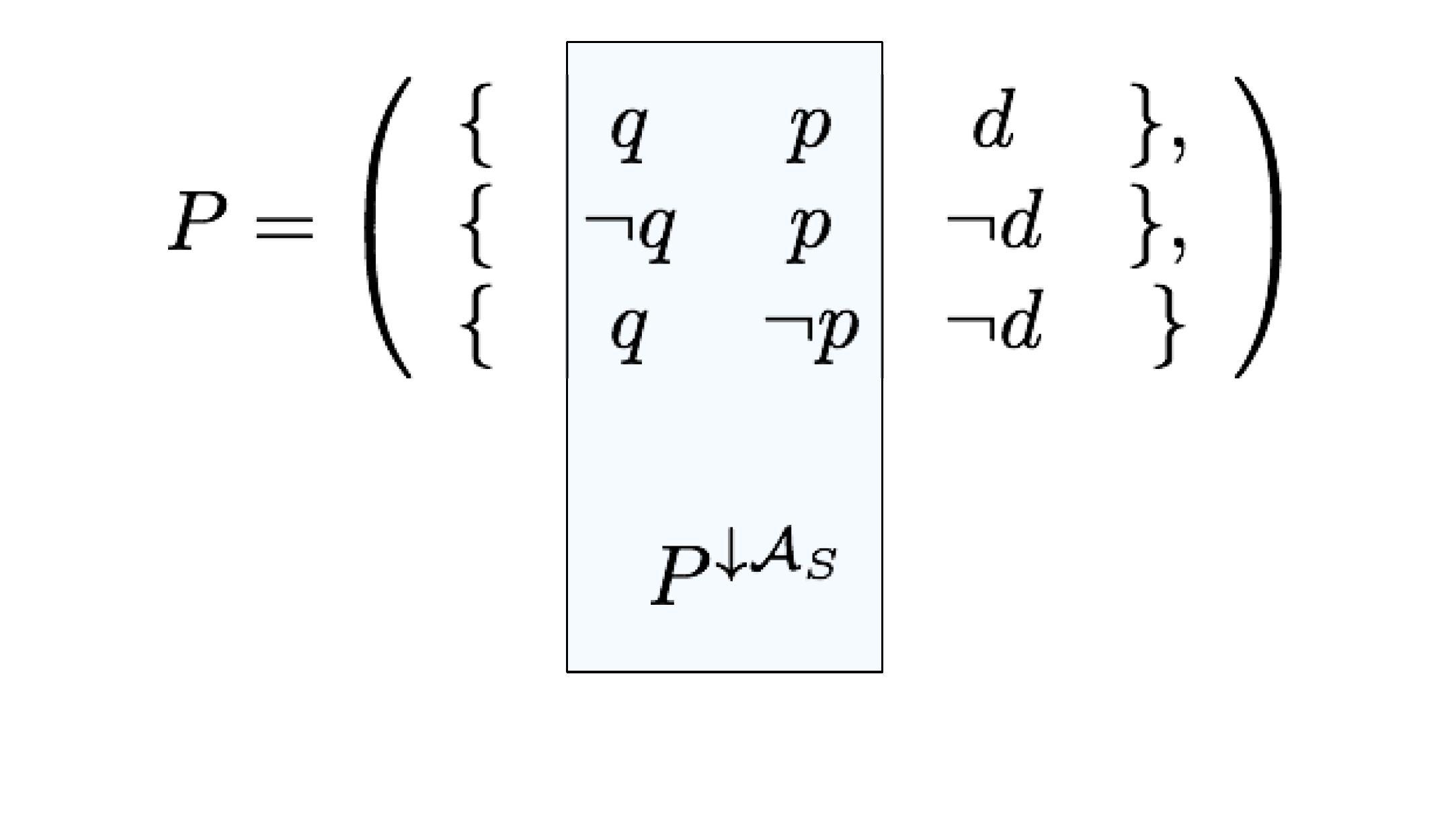}   
\caption{Visualising partial profiles.}\label{fig:profvert} 
\end{center}
\end{figure}
  
\end{example} 

\item The sum of profiles  $\Pf_1$ and $\Pf_2$ where $\Pf_1 \in \Dmc(\A,\Ct)^{n_1}$, $\Pf_1=(\Js_1, \ldots, \Js_{n_1})$  and $ \Pf_2 \in \Dmc(\A,\Ct)^{n_2}$, $\Pf_2=(\Js'_1, \ldots, \Js'_{n_2})$  is a profile $\Pf = \Pf_1 + \Pf_2$, such that   $\Pf \in \Dmc(\A,\Ct)^n$, where $n = n_1 + n_2$ and $\Pf= (\Js_1, \ldots, \Js_{n_1}, \Js'_1, \ldots, \Js'_{n_2})$.

\item The operators $\sqsubseteq$,  and $\sqcap$ we define for profiles in the following way. 
\begin{itemize}
\item For two profiles $\Pf_1 \in \Dmc(\A,\Ct)^{n_1}$ and $\Pf_2 \in \Dmc(\A,\Ct)^{n_2}$, where $n_1 \leq n_2$, it holds that $ \Pf_1 \sqsubseteq \Pf_2$ if and only if every agent that is in $\Pf_1$ is also in $\Pf_2$.
\item Given two profiles  $\Pf_1 \in \Dmc(\A,\Ct)^{n_1}$ and $\Pf_2 \in \Dmc(\A,\Ct)^{n_2}$, for profile $\Pf \in \Dmc(\A,\Ct)^{n}$, where $n \leq \min(n_1, n_2)$, it holds $\Pf = \Pf_1 \sqcap \Pf_2$ if and only if $\Pf$ contains all the agents, and only these agents, that are both in $\Pf_1$ and in $\Pf_2$. 
\end{itemize}
\end{description}

\begin{example}Consider the pre-agenda $\A^+=\{ p, p\wedge q, r\vee s\}$ and \linebreak $\Ct = \{ q \rightarrow r\}$. We have that
\[
 \Dmc(\A, \Ct) = \left\{
 \begin{array}{l}
 \Js^a=\{ \neg p, p\wedge q, \neg(r\vee s)\},\\
 \Js^b=  \{ \neg p, p\wedge q,  r\vee s \},\\
   \Js^c=  \{ p, \neg (p\wedge q),  \neg(r\vee s )\},\\
  \Js^d=      \{ p, \neg (p\wedge q),   r\vee s  \},\\
     \Js^e=      \{ p, p\wedge q,  r\vee s \}
 \end{array}
 \right\}
 \]
 
 Let $\Pf_1 = (\Js^a, \Js^b, \Js^a, \Js^e)$ and $\Pf_2=(\Js^b,\Js^a)$. We have that $\Pf_1 + \Pf_2 = (\Js^a, \Js^b, \Js^a, \Js^e,\Js^b,\Js^a)$ and it holds that $\Pf_2 \sqsubseteq \Pf_1$. 
\end{example}

We conclude this section by introducing various properties of the agenda that occur in the literature, including here the definition of special types of agendas.

\subsubsection{Agenda properties}\label{sec:agprop}
We start with the {\em premises - conclusions agenda}, which is the type of agenda we find in the "doctrinal paradox" example.  The judgment aggregation problem can sometimes be given as a tuple $\langle \A_p, \A_c, \Ct, \Pf\rangle$, where $\A_p$ is an agenda of premises and $\A_c$ is an agenda of conclusions. Necessary  $\A_p \cap \A_c = \emptyset$ and the agenda is actually the union of the $\A_p$ and $\A_c$. The judgments on the issues that are premises serve as explanations for the judgments on the issues of the conclusion.  In the "doctrinal paradox" example, $\A_p = \{p, \neg p, q, \neg q\}$ and $\A_c=\{d, \neg d\}$. Some aggregation problems naturally give rise to  a {\em premises - conclusions agenda} and as we shall see later on, some judgment aggregation functions are applicable only to these types of judgment aggregation problems. 

Given an agenda, we cannot test, we cannot determine whether it is a { premises - conclusions agenda}. A  premises - conclusions agenda has to be specified as such in the specification of the judgment aggregation problem.  The next agenda properties we define can be identified for any agenda. 

It has to be emphasised that most of these properties have been defined only for the aggregation problems in which the agenda is not associated  with constraints, \ie~$\Ct=\{ \top\}$. We extend the definitions here to include any $\Ct$. This extension is straightforward and intuitive when the consistency of $S \subset \A$ is defined as $S \cup \Ct \nvDash \bot$.

\begin{description}
\item An agenda is {\em closed under propositional variables} \cite{Mongin2008} when \linebreak $\Atoms(\A) \subseteq \A$. For example, the agenda in Example~\ref{ex:doctrinal} is closed under propositional variables, while the agenda Example~\ref{ex:co2} is not closed under propositional variables, since the variable $r$ does not occur as an issue in the agenda. The agenda we give in Example~\ref{ex:scw} later is also not closed under propositional variables.

\item An agenda satisfies {\em non-simplicity} \cite{GrossiP:2014} if and only if there exists a subset $S \subseteq \A$ of at least three elements, \ie~$|S| \geq 3$, that is a {\em minimally inconsistent subset} of $\A$. An agenda is {\em simple} otherwise.

Observe that the reason why $|S| \geq 3$ is that any agenda has a minimally inconsistent subset of size two because  every issue is a  minimally inconsistent subset of size two. If an agenda does not satisfy non-simplicity, we say it is {\em simple}.  In \cite{EndrissGP12} the term {\em median property} is used. The median property is  synonymous with agenda simplicity \ie~an agenda is simple if and only if it satisfies the median property. 

\item An agenda satisfies {\em the k-median property} \cite{EndrissGP12} if and only if  every
inconsistent subset of $\A$  has itself an inconsistent subset of size at most k, where $k\geq 2$. Clearly for $k = 2$, the agenda is simple, so the 2-median property is the median property, while for $k =1 $, is not possible since any set of one formula  is consistent.

\item To define when an   agenda is {\em path-connected} \cite{Dietrich2013}, also called  {\em totally-blocked} in \cite{NehringP:2002}, we first define {\em conditional entailment between two judgments in an agenda}. 

Let $\ai, \psi \in \A$. We say that $\ai$  conditionally entails $\psi$, written $\ai \vdash^{\ast} \psi$ if $\{\ai\} \cup S \models \{ \psi\}$ for some $S \subset \A$ that is a consistent agenda subset which is consistent with $\ai$ and consistent with $\neg \psi$. Furthermore, we write $\ai \vdash\vdash^{\ast} \psi$ if there is a sequence of judgments $\{\ai_i,\ldots \ai_k\} \in \A$ such that $\ai = \ai_i \vdash^{\ast} \cdots \vdash^{\ast} \ai_k=\psi$. Thus $\vdash\vdash^{\ast}$ is the transitive closure of $\vdash^{\ast}$. 

An agenda is {\em path-connected} \cite{Dietrich2013} if for any two judgments $\ai,\psi \in \A$ it holds $\ai \vdash\vdash^{\ast} \psi$. Intuitively, an agenda is path-connected when the judgment on  each issue is a logic consequence of the judgment of some other issues in the agenda.    

\begin{example}\label{ex:scw} An example of an agenda that is not path-connected is the agenda $\A = \{  p\wedge r, \neg (p \wedge r), p \wedge s, \neg (p\wedge s), q, \neg q, p\wedge q, \neg (p \wedge q), t, \neg t\}$  introduced in  \cite{LangPSTV15}. In this agenda $t$, and respectively $\neg t$, is not conditionally independent on any agenda subset that does not include this judgment. 
\end{example}
 
\item We now define two {\em separability properties} of the agenda. A partition $\{ \A_1, \A_2\}$ of agenda $\A$ is an {\em independent partition of $\A$} \cite{AAAI} if for every two judgment sets $\Js^1 \in \Dmc(\A_1, \Ct)$ and $\Js^2 \in \Dmc(\A_2, \Ct)$, $\Js^1 \cup \Js^2$ is a consistent and complete judgment set from   $\Dmc(\A_1, \Ct)$. If $\Ct = \top$,  a partition $\{ \A_1, \A_2\}$ of agenda $\A$ is  a {\em syntactical  independent partition of $\A$} \cite{AAAI} if and only if  $\Atoms(\A_1) \cap \Atoms(\A_2) = \emptyset$. Clearly,  a syntactical independent partition of $\A$ is also a   independent partition of $\A$.
\end{description}

\subsection{Binary framework}
The binary framework is   frequently used in judgment aggregation works in artificial intelligence such as \cite{Grandi2013,ecai16,EndrissG14}, but also in economics \cite{Dokow2010a}.
The core difference between the binary and the logic aggregation frameworks  is  how a judgment is represented. Intuitively in both framework a judgment is an answer to a binary question, which we call  "issue". In the logic framework a judgment is a binary answer to a single issue represented as a logic formula or a negated logic formula. In the binary framework a judgment is an answer to {\em all} the issues in the agenda, which  here are only allowed to be propositional variables,  represented as a truth-value assignment function. 

A judgment aggregation problem in the binary framework is a tuple $\langle \Phi, \textrm{IC}, \mathbf{B}\rangle$, where $\Phi$ is an agenda of issues, $\textrm{IC}$ is a set of associated constraints for $\Phi$, which are also called {\em integrity constraints} and $\mathbf{B}$ is a profile of judgments.  We define each of the components. 

\paragraph{Agenda. } An agenda $\Phi = \{ p_1, \ldots, p_m\}$ is a set of propositional variables, \ie~$\Phi \subset \La_v$, where $\La_v$ is a set of propositional variables. 
 
\paragraph{(Integrity) constraints.} Let $\La_w$ be a set of well formed propositional logic formulas of propositional variables from $\La_v$ and the logic connectives $\neg$, $\wedge$, $\vee$, $\rightarrow$,  and $\leftrightarrow$. The integrity constraints  for an agenda $\Phi=\{ p_1, \ldots, p_m\}$ is a set of formulas $\textrm{IC} \subset \La_w$ such that $\Atoms(\textrm{IC} ) \cap \Phi \neq \emptyset$. It is assumed that $\Phi$ is satisfiable. The set $\textrm{Mod}(\textrm{IC}) \in \{0,1\}^m$ is a set of models for $\textrm{IC}$ if and only if every one of its elements is a truthful assignment for $\textrm{IC}$ for the classical logic sense. 

\begin{example}\label{ex:bscw} Let $\Phi=\{p_1,p_2,p_3,p_4,p_5\}$ and $\textrm{IC} = \{  ((p_1 \vee p_2) \wedge p_3) \rightarrow p_4, p_4 \rightarrow p_3\}$. The set of all models for $\textrm{IC}$ is 
\[
\textrm{Mod}(\textrm{IC})=\left\{
\begin{array}{l}
(1,1,1,1,1), (1,1,1,1,0),(1,1,0,0,1),(1,1,0,0,0),\\ 
(1,0,1,1,1), (1,0,1,1,0),(1,0,0,0,1),(1,0,0,0,0),\\ 
(0,1,1,1,1), (0,1,1,1,0),(0,1,0,0,1),(0,1,0,0,0),\\ 
(0,0,1,1,1), (0,0,1,1,0),(0,0,1,0,1),(0,0,1,0,0),\\ 
(0,0,0,0,1), (0,0,0,0,0)
\end{array}\right\}
\]
\end{example}

The difference between the logic and binary frameworks emerges in the representation of judgments. 

\paragraph{Judgment.} Given an agenda $\Phi=\{ p_1, \ldots, p_m\}$, a judgment for $\Phi$, also called a  a ballot \cite{EndrissG14}, is a
vector $B \in \{0,1\}^m$  that has an assignment of   either $0$ or $1$  for each agenda issue, $0$ indicating a disagreement with or a negative answer to the issue and $1$  indicating an agreement or a positive answer to the issue.  A judgment $B$ is {\em rational} for an agenda $\Phi$ and associated constrains  $\textrm{IC}$, when $B \in \textrm{Mod}(\textrm{IC})$, namely when $j$ is a truthful assignment for $\textrm{IC}$. We use $B(p)$ to denote the truth-value assigned to $p\in\Phi$ in  $B$. 

Note that in the binary framework, a judgment is a vector of length $m = |\Phi|$ populated with values $0$ and $1$. It is by definition complete in the sense that it has one truth-value assignment for each agenda issue. In the logical aggregation, a judgment is a set of cardinality  $m = |\A^+|$.  The set $ \textrm{Mod}(\textrm{IC})$ in the binary framework thus plays the same role as the codomain set in the logic framework. In general, regardless of framework, judgments that are complete and consistent are called {\em rational judgments}.

As in the logic framework, here too a judgment profile is a list of rational judgments, one representing each from a list of agents. 

\paragraph{Profile.} Consider an agenda  $\Phi$ and associated constraints $\textrm{IC}$.  A profile $\mathbf{B}$ is a list of $n$ agents, each represented with a rational judgment    $B_i \in  \textrm{Mod}(\textrm{IC})$. Thus $\mathbf{B} \in \textrm{Mod}(\textrm{IC})^n$.  We slightly abuse notation and use $B_i \in \mathbf{B}$ to denote that $B_i$ is agent $i$'s judgment set in $\mathbf{B}$. 

In the binary framework, again, we can also define the judgment aggregation problem as a pair $\langle \textrm{Mod}(\textrm{IC}), \mathbf{B} \rangle$, but this is in general not a succinct representation when the set $\textrm{Mod}(\textrm{IC})$ has many elements. 

\paragraph{Agenda properties.} The   agenda properties can be considered in the binary framework, with some rewriting of the definitions to include the integrity constraints, which we already did. Of course, premise-conclusion agendas can be specified here as well.   

\subsection{Relations between frameworks}
The binary and the logic framework are equally expressive, as shown by Proposition 1 in \cite{EndrissEtAlKR2016}. This relationship between the two frameworks is directly observable as it is relatively straightforward to transform judgment aggregation problems from one framework to the other. 

As argued in \cite{EndrissEtAlKR2016},  the agenda and integrity constraints in the binary aggregation framework can be seen as a special case of agenda and constraints in the logic aggregation framework.  Thus every binary judgment aggregation problem is a logic judgment  aggregation problem.  As an illustration consider the "doctrinal paradox" Example~\ref{ex:doctrinal}.  

The transformation from a problem in the logic to a problem in the binary framework is not as direct, but it is easy to see how it can be done. Given a logic framework agenda $\A$, a binary framework agenda $\Phi$ can be obtained by assigning a new propositional variable to represent each element in $\A^+$. The same rewriting can be directly applied to the elements of $\Ct$. The  integrity constraints would now contain the rewritten $\Ct$ and also a set of constraints describing the relation between the agenda items in $\Phi$ making them equivalent to the relations between issues in $\A$. 

This problem transformation processed is precisely how the agenda and integrity constraints in Example~\ref{ex:bscw} were obtained from the agenda and constraints $\Ct$ (which recall were $\Ct = \{\top\}$) in Example~\ref{ex:scw}. More precisely $p_1 \equiv ( p\wedge  r)$,  $p_2 \equiv ( p\wedge s)$,  $p_3 \equiv q$,  $p_4 \equiv ( p\wedge  q)$ and $p_5 \equiv t$. Observe that there exists an isomorphism between   $\textrm{Mod}(\textrm{IC})$ and $\Dmc(\A,\Ct)$. For every $\Js \in \Dmc(\A,\Ct)$ there is  a corresponding $B \in \textrm{Mod}(\textrm{IC})$ (and vice versa) such that for every $\ai \in \A^+$ and corresponding  $p \in \Phi$ that represent it,   $\ai \in \Js$ if and only if $B(p)=1$ or   $\neg\ai \in \Js$ if and only if $B(p)=0$. An example of such pair is a judgment set $\Js= \{ \neg(p\wedge r), \neg (p\wedge s), q, \neg (p\wedge q), t\}$ and judgment $B=(0,0,1,0,1)$. As shown by Proposition 9 in \cite{EndrissEtAlKR2016}, the algorithm for transforming any  $\A$ and $\Ct$ into corresponding $\Phi$ and $\textrm{IC}$, in the worst case, runs in a non-deterministic polynomial time.

By Theorem 6 in \cite{EndrissEtAlKR2016} it is shown that the logic aggregation framework is strictly more succinct than the binary aggregation frameworks, under the common assumption that the polynomial hierarchy does not collapse ($P\neq NP$). This means that representing a judgment aggregation problem in the logic framework requires less computational space than representing the same problem in the binary framework.   However, in \cite{EndrissEtAlKR2016} it is also shown the difficulty of finding result from the aggregation of the profile,  for the known judgment aggregation functions, in both frameworks remains the same. We  discuss  aggregation functions at length in Chapter~\ref{ch:mathtest}. Next we make a brief overview of, and giver references to, other judgment aggregation frameworks that appear in the literature.

Lastly we must mention the framework used by Nehring, Pivato, and Puppe \cite{NehringPP14,NehringPivato16} who work with a binary framework, but have $1$ and $-1$ as values assigned to issues instead of $1$ and $0$. Most of the other definitions they use in their model correspond to the definitions in the binary aggregation framework. 

\subsection{Other frameworks}
Outside of artificial intelligence, exploring other-than-classical logic frameworks had been primarily motivated by the  search for \linebreak (im)possibility results, \ie ~sets of properties that no judgment aggregation function can  satisfy at the same time, rather than by the search of elegant modelling of various judgment aggregation problems. 
 
In \cite{Franz2007} a general logic framework for judgment aggregation was studied. For this general logic, which   is (afore most) monotonic, non para-consistent and compact, the (im)possibility results already shown in the literature, such as those shown in \cite{Dietrich07}, persist. Most relevant perhaps from the view-point of multi agent systems is that the general logic of \cite{Franz2007} includes all two-valued monotonic logics such as the description logics and the modal logics, as well as some of the multi-valued logics. Not subsumed by the general logic   in \cite{Franz2007} is the multi-valued logic framework considered \cite{PaHe2006} in (also for work regarding im)possibility results) built over the 
multi-valued Post logic \cite{post21} and  three-valued {\L}ukasiewicz logic  \cite{urquhart} studied in  \cite{Dokow2010}.
  
 Some judgment aggregation work has been done with frameworks built upon three-valued logics.  In \cite{SCAI11} a ternary-logic framework was proposed for design of aggregation functions,  further specified for the    Kleeny logic and the {\L}ukasiewicz logic  \cite{urquhart}. 
 
 Multi-valued logic frameworks are of interest when the agents are allowed to express other judgments regarding an agenda issue except agreement and disagreement. To allow the agents to {\em abstain } on an issue, we need to use a three-valued logic to model related judgment aggregation problems. The semantics of the used three-valued logics will capture the precise meaning of an abstention on an issue. Choosing the right three-valued logic semantics only appears to be trivial. We will illustrate this observation with an example from experiments described in \cite{AAMAS-2-12}. 
 
 Consider a robot that needs to determine whether she has heard a sound. The robot has a reasoning rule which says that proposition $B \ai$, denoting a "I believe there was a sound", is evaluated to true if a collection of input samplings from the robot's microphone averages above a certain decibel threshold. The same rule assigns a value "false" to  $B \ai$ when the microphone input averages bellow the threshold.  It is unrealistic to require that the robot would always be able to evaluate $B \ai$ either to true or to false. One reason that the robot may fail to do so is an average  decibel value of microphone sound samples that falls exactly on the threshold. Another reason is that the robot's microphone is damaged or otherwise out of commission. In both these cases the robot will need to abstain regarding the issue $B \ai$, however the usefulness of these abstentions is not the same. In the first case, the robot's abstention in a judgment profile contributes valuable information that needs to be taken into consideration when aggregating the profile. In the second case, the abstention as information should be disregarded when the profile is aggregated. 

In the case of multi-valued judgments, the frameworks representing  judgments as value functions offer an intuitive way to model judgment aggregation problems, while this is not the case with frameworks that represents    judgments as  formulas.

\newpage



\section{Judgment Aggregators}  
\label{ch:mathtest} 


The primary aim of judgment aggregation research is the design and analysis of  aggregation functions, or {\em aggregators}. A {\em solution} to a judgment aggregation problem is a set of judgments, also called {\em collective judgment set} that are {\em most-representative} of the profile given in the problem, but also a complete and consistent judgment set.   A judgment aggregation function maps a profile of judgments to such a collective judgment set.  Depending on how the concept of "most-representative" is defined, various judgment aggregation functions can be defined. In voting theory, voting methods have been defined and discussed since the 18th century, and many specific methods are available. In judgment aggregation, comparatively few aggregators have been defined, almost all within the past decade.   We present almost all known aggregators  in this section, in historical order. 

Recall:
\begin{itemize}
 \item $N(\ai, \Pf)$ is the number of agents in $\Pf$ that have selected $\ai \in \A$ in their judgment set, \ie~the number of agents in the profile that {\em support} $\ai$, 
 \item $\mathcal{\Js}(\A,\Ct)$ is the set of all consistent, but possibly incomplete, judgment sets for $\A$ and $\Ct$ including the $\emptyset$, while 
 \item $\Dmc(\A,\Ct)\subset \mathcal{\Js}(\A,\Ct)$ is the set of all complete and consistent  judgment sets for $\A$ and $\Ct$.
 \item Given a subagenda $\A_S$, a partial profile for it of profile $\Pf$ is $\Pf^{\downarrow \A_S} = (\Js_1 \cap \A_S, \ldots , \Js_n \cap \A_S)$
  \end{itemize}

\subsection{Types of aggregators}

Let $\langle \A, \Ct, \Pf \rangle$ be any judgment aggregation problem, where $\Pf$ is a profile of a finite number of $n$ agents with $\Dmc(\A,\Ct)$ the codomain set. Given a set $S \subseteq \Dmc(\A,\Ct)$, we use  $\mathcal{P}^{\ast} (S)$ to denote the power set of $S$ excluding the empty set.  Recall that a judgment set is rational if and only if it is complete and consistent for $\A$ and $\Ct$. 

A judgment aggregation function $\F: S\rightarrow \mathcal{P}^{\ast}(\Dmc(\A,\Ct)) $, or aggregator, is a function that maps a set of rational judgment sets to a judgment  profile from $S \subseteq \Dmc(\A,\Ct)^n $. If $\F(\Pf)$ is a singleton for every $\Pf \in S$, then $\F$ is called {\em a resolute aggregator}, otherwise, we say that $\F$ is an irresolute aggregator. When $\F$ is defined for $S = \Dmc(\A,\Ct)^n $, then we say that $\F$ satisfies {\em universal domain}. If $\F$ is defined only for some $S \subset \Dmc(\A,\Ct)^n $, then we say that $\F$ is a partial aggregator. 

When judgment aggregation is used as a collective decision-making \linebreak method, it is clearly best to use resolute aggregators that satisfy universal domain. However, as shown by numerous impossibility results\footnote{For a thorough and comprehensive overview of impossibility results we point the reader to \cite{GrossiP:2014}.}, such as the ones in \cite{Dietrich07}, have shown that either universal domain or resoluteness needs to be "sacrificed". Since it most domains, one cannot reasonably guarantee that a profile will always be of a certain kind, universal domain is deemed more important than resoluteness. Although most properties have been defined for resolute aggregators, due to legacy reasons we discuss in more detail in Chapter~\ref{ch:name}, most specific judgment aggregation functions defined are irresolute. 

The first aggregators defined in the literature were resolute, but did not satisfy universal domain. In the next section we present this first class of aggregators. 

Before we proceed, it is be useful to have the tools to compare aggregators. We define what it means for one aggregator to refine another and for two aggregators to be different. 

An aggregator $\F_1$ {\em refines} an aggregator $\F_2$, denoted $\F_1 \subseteq \F_2$ if and only if, for every $\Pf \in \Dmc(\A, \Ct)^n$ it holds that $\F_1(\Pf) \subseteq \F_2(\Pf)$. 

Two aggregators  $\F_1$ and $\F_2$ are the same, denoted $\F_1 = \F_2$, if and only if $\F_1 \subseteq \F_2$ and $\F_2 \subseteq \F_1$. 

Two aggregators $\F_1$ and $\F_2$ are different, denoted $\F_1 \neq \F_2$ if and only if neither $\F_1$ refines $\F_2$, nor $\F_2$ refines $\F_1$. Or in other words, $\F_1 \neq \F_2$ if and only if there exists a profile $\Pf \in \Dmc(\A, \Ct)^n$ and a judgment sets $\Js,\Js' \in  \Dmc(\A, \Ct)$ such that $\Js \in \F_1(\Pf) $, $\Js \not\in \F_2(\Pf) $, $\Js' \in \F_2(\Pf) $, $\Js' \not\in \F_1(\Pf) $.

Intuitively, an aggregator $\F_1$ refines $\F_2$ when the collective judgments produced by $\F_1$ are always a selection from the collective judgments produced by $\F_2$. For two aggregators to be considered different, there should be at least one profile on which they give different collective judgments. 

In the  rest of this section $\A$ is an agenda of issues in the logic framework, $\A^+ = \{\ai_1, \ldots, \ai_m\}$ the respective pre-agenda, $\Ct$ a set of associated constraints and $\Pf \in \Dmc(\A, \Ct)^n$, $\Pf = \{\Js_1, \ldots, \Js_n\}$ is a profile of judgments. 
 
\subsection{Partial-aggregators}
We begin by defining the quintessential partial aggregator the {\em issue-by-issue majority} function $m$ which is defined as the set of judgments supported by a strict majority in $\Pf$ or formally: 
\begin{equation}  
m(\Pf) = \{\ai \mid  \ai \in \A, N(\ai, \Pf)>\frac{n}{2}\}
\end{equation}

We call $m(\Pf)$ the {\em majoritarian set}. It is easy to observe that $m(\Pf)$ would not always be a complete set of judgments, particularly when the number of agents in the profile is even. A bit less obvious is that  $m(\Pf)$ is not always a consistent judgment set, although all the agents in the profile have rational judgment sets. We illustrate this "paradox" with two examples given in Tables~\ref{tab:doctrinal}~and~\ref{tab:scw}. The first row of the tables lists the elements of the pre-agenda. The second row of the tables presents the  associated constraints.
Each subsequent row represents an agent $\Js_i$, with "$+$" under the corresponding judgment $\ai$ if  and only if $\ai \in \Js_i$ and "$-$" when $\neg \ai \in \Js_i$.   The last row of the tables contains the respective majoritarian set. 

In  Table~\ref{tab:doctrinal}, the first and second rows of the table give  the pre-agenda and associated constraints of the "doctrinal paradox" example given in Example~\ref{ex:doctrinal}, while in Table~{tab:scw}, the first two rows  contain the pre-agenda and associated constraints from  Example~\ref{ex:scw}.

\begin{table}[h!]
\centering
\begin{tabular}{r|ccc}
Agents       &\{ $\;\;\;p $,& $\;\;\;  q$, &  $\;\;\; d\}$\\\hline
 \multicolumn{4}{c}{      $\Ct = \{d \leftrightarrow (p\wedge q)\}$}\\\hline
  $\Js_1$          &  $\;\;\;\;$+           &  $\;\;\;$+             &  $\;\;\;$+      \\
  $\Js_2$        & $\;\;\;\;$+           & $\;\;\;$-             & $\;\;\;$ -        \\
 $\Js_3$        & $\;\;\;\;$-           & $\;\;\;$+            &   $\;\;\;$-       \\\hline
m(\Pf) & $\;\;\;$ + & $\;\;\;\;$+ & $\;\;\;\;$- 
\end{tabular}\caption{Judgment aggregation problem for Example~\ref{ex:doctrinal}.}\label{tab:doctrinal}
\end{table}

\begin{table}[h!]
\centering
\begin{tabular}{r|ccccc}
 Agents       &\{ $\;\;\; p\wedge r $,& $\;\;\;  p \wedge s$, &  $\;\;\; q$, &   $\;\;p\wedge q$, &   $\;\;\;t$ \} \\\hline
  \multicolumn{6}{c}{      $\Ct = \{\top\}$}\\\hline
  $\Js_1 -\Js_6$          &  $\;\;\;\;$+           &  $\;\;\;$+             &  $\;\;\;$+   &  $\;$ +           &  +    \\
  $\Js_7 -\Js_{10}$        & $\;\;\;\;$+           & $\;\;\;$+             & $\;\;\;$ -   &  $\;$ -           &  +    \\
 $\Js_{11} -\Js_{17}$        & $\;\;\;\;$-           & $\;\;\;$-             &   $\;\;\;$+   &  $\;$ -           &  -    \\\hline
m(\Pf) & $\;\;\;$ + & $\;\;\;\;$+ & $\;\;\;\;$+ & $\;\;\;$- &  +
\end{tabular}
\caption{Judgment aggregation problem for Example~\ref{ex:scw}.}\label{tab:scw}
\end{table}

The issue-by-issue majority function $m()$ is an example of a partial aggregator -- it is only defined for some profiles, which we will call {\em majority-consistent profiles}. The function $m()$ we will also call {\em the majority aggregator} Formally $\Pf \in \Dmc(\A,\Ct)$ is majority-consistent if and only if $m(\Pf)$ is a consistent judgments set, which may or may not be complete for $\A$. 

To deal with consistent but incomplete judgment set, which define the  function $\textrm{ext}: \mathcal{\Js}(\A,\Ct) \rightarrow \mathcal{P}^{\ast}(\Dmc(\A,\Ct))$ that returns all possible extensions of  an incomplete but consistent judgment set into complete and consistent judgment sets. 

\begin{equation}
\textrm{ext}(\Js) = \{\Js' \mid \Js' \in \Dmc(\A,\Ct), \Js \subset \Js'\}
\end{equation}

\subsubsection{Premise-based procedure}
The first aggregators that were considered, already in the works of Kornhauser and Sager \cite{KS86,KS93} were defined for the special type of premise-conclusions agenda, building upon the  majority aggregator. These were the {\em premise-based  procedure} $\textrm{PBP}$ and the {\em conclusion-based procedure}  $\textrm{CBP}$. We define them both here. 

Let $\A = \A_p \cup \A_c$ be an agenda partitioned in sub-agendas of premises $\A_p$ and sub-agenda of conclusions. The premise based procedure first applies issue-by-issue majority to aggregate the judgments on the issues in  $\A_p$ and then uses this partial majoritarian set of premises together with the constraints to deduce the judgments on the issues from $\A_c$, or alternatively we extend the incomplete judgment set of premises using the $\textrm{ext}$ function.  

The conclusion based procedure always produces incomplete collective judgment sets, as it applies the issue-by-issue majority only on the profile of conclusions. Whereas this is perhaps sufficient in  law, it is unacceptable for  solving general 
judgment aggregation problems, which is why the conclusion based procedure has been unexplored. An exception is \cite{ADT09} where a method for "completing" the conclusion-based procedure was proposed, which we will come back to in Section~\ref{sec:misc}.

We illustrate the premise-based procedure and the conclusion-based procedure on the "doctrinal paradox" example. 

\begin{example}
 Let $\A_p = \{p, \neg p, q, \neg q\}$, $\A_c = \{d, \neg d\}$ and  $\Ct = \{d \leftrightarrow (p\wedge q)\}$. Consider the profile for this agenda and constraints. Using the premise-based procedure In the "doctrinal paradox" example, for the $\Pf$ given  in Table~\ref{tab:doctrinal}, we first aggregate the judgments on the premises and obtain $m(\Pf^{\downarrow \A_p}) =\{ p, q\}$. We can now use the classical consequence operator $\models$ and obtain that $\textrm{PBP}(\Pf) = \{p, q, d\}$. 
 
Applying the conclusion based procedure on this profile $\Pf$ means applying issue-by-issue majority on the profile of conclusions, thus $\textrm{CBP}(\Pf)= m(\Pf^{\downarrow \A_p}) =\{ \neg d\}$. 
\end{example}

Formally, the premise-based procedure can be defined as   follows:
\begin{equation}
\textrm{PBP}(\Pf) = \textrm{ext}(m(\Pf^{\downarrow \A_p})).
\end{equation}

The premise-based procedure is a very simple and intuitive way to aggregate judgments. However, it is not without shortcomings. For one, it is limited to agendas that can be partitioned into premises and conclusions, and this is not  feasible for every judgment aggregation problem. Furthermore, 
and it is not difficult to observe this,  the premise-based procedure does not always return singleton sets. If, for example, the agenda is not totally-blocked and there exists an issue in the conclusions whose judgment cannot be deduced by judgments on the premises, the $\textrm{PBP}$ will not be resolute. The premise-based procedure is one of the best studied procedures in the literature. An  detailed analysis of the properties of this procedure can be found in \cite{premise10}. As to the question of deciding whether an agenda is such that the premise-based procedure is resolute, \cite{Endrissaamas2010} show that this problem is computationally not easy. 

\subsubsection{Uniform quota rules}
Last in this collection of aggregators we present the {\em quota rule}. The quota rule is a generalisation of the issue-by-issue majority. Instead of requiring that a judgment should have the support of majority in order to be included in the collective judgment set, the quota rule changes the threshold, \ie {the quota} from $\frac{n}{2}$ to a value $0< k \leq n$. If the same threshold is used for each issue in the agenda, then the quote rule is called {\em the uniform quota rule}. Formally, the uniform quota rule $\textrm{UQ}_k$ is defined as

\begin{equation}
\textrm{UQ}_k(\Pf) =  \{\ai \mid  \ai \in \A, N(\ai, \Pf)>k\}, \mbox{where $0< k \leq n$ is given}.
\end{equation}

The uniform quota rules are one of the best explored judgment aggregation rules in the literature, starting with \cite{DietrichList07} that analyse their social theoretic properties. In \cite{EndrissGP12} their computational theoretic properties were explore, more  precisely, how difficult it is to determine whether the agenda is such that the uniform quota rule returns a single rational judgment set. The highlight of these results, with respect to this document, is identifying that  $\textrm{UQ}_k$ "behaves well" when the agenda satisfies the the k-median property (Theorem 12 in \cite{EndrissGP12}). Some of the other  works that study the uniform quota rules and quota rules in general are \cite{Bozbay2014571,Baumeister2015,BotanEtAlAAMAS2016}.

We should lastly mention the special uniform quota arrgregator, called {\em issue-by-issue} unanimity, defined for $k =n$ where $n$ is the number of agents in the aggregated profile. This function is simply defined, for $\Pf \in \Dmc(\A,\Ct)^n$ as

\begin{equation}\label{eq:unanimity}
u(\Pf) = \{ \ai \mid \ai \in \A,  N(\ai, \Pf) =n\}.
\end{equation}

\subsection{Majority-preserving aggregators}

The next collection of functions we consider are the what we call  {\em majority-preserving aggregators}. The basic idea behind these aggregators is that the majoritarian set is the "ideal" collective judgment set for a profile and when this set is not consistent we should look for ways of minimally  "adjusting" the profile so that it becomes majority-consistent. Based on how these "adjustments"  are operationalised, different aggregators are obtained. 
A through analysis of the majority-preserving aggregators can be found in \cite{LangPSTV15}. We start with defining majority-preservation as a property of aggregators and then present a list  of such operators, using the names used in \cite{LangPSTV15}. The reader should be aware that since different groups of researchers arrived at the same aggregators virtually simultaneously, often different names are used for the same aggregator. 

Intuitively an aggregator is majority-preserving if it always returns the extensions of $m(\Pf)$ whenever this majoritarian set is consistent. We need the extensions of the majoritarian set since this set may not be complete. We give a formal definition

An aggregator $\F$ is majority-preserving if and only if, for every $\Pf$ such that $m(\Pf) \in \mathcal{\Js}(\A,\Ct)$   $\F(\Pf) = \textrm{ext}(m(\Pf))$.

 \subsubsection{Maximum Condorcet rule  ($\RMSA$)}
We start with  the {\em maximum Condorcet rule}  ($\RMSA$) since many of the subsequent aggregators we introduce are a refinement of this rule. The $\RMSA$ aggregator operationalises the profile "adjustment" by trying to use as many of the majority supported judgments in the majoritarian set as possible. Thus the   $\RMSA$ aggregator basically returns  the maximally consistent subset of $m(\Pf)$, with respect to set inclusion.  

Given a set of formulas $S \subset \A$, the maximally consistent subset of $S$ with respect to set inclusion, is a set $S' \subseteq S$ such that there exists no other consistent agenda subset $S''$ that is a superset of $S'$, \ie~ $S' \subset S''  \subseteq S$.  The set $\MC{S}$ is the set of of all maximally consistent subsets of $S$ with respect to set inclusion. 

Consider for example the profile given on Table~\ref{tab:scw}. For this profile $\Pf$ we have that $m(\Pf) =\{\{ p\wedge r, p\wedge s, q, \neg (p \wedge q), t  \}$.  This majoritarian set is not a consistent agenda subset. The set $\MC{m(\Pf)}$ contains the following sets: $\{ p\wedge r, p\wedge s, q, t  \}$, $\{p\wedge r, p\wedge s, \neg (p \wedge q), t \}$ and $\{ q, \neg (p \wedge q), t \}$. 

An intuitive way to think about the  $\MC{m(\Pf)}$ sets is to look at the profile as a table, see Table~\ref{tab:scwmc} for example. To obtain a  sets  $\MC{m(\Pf)}$  we are looking for a minimal number of columns in the table to "ignore" (other than the first column of course) so that $m(\Pf)$ becomes consistent. But, we want to be "fair", so for each "consistent" combination of columns there should be a set in which that column is not "ignored". So by ignoring the fifth column (just cover it with a strip of paper as an exercise), we obtain the  first consistent subset of $m(\Pf)$. By ignoring the fourth column, we obtain the second consistent subset of $m(\Pf)$. However, we have now either "ignored" either the fourth or the fifth column, although they support judgments that are consistent together, $\{ q, \neg (p \wedge q)\}$ is a consistent agenda subset. If we want to "give these columns a chance" we have to ignore the second and third column. This is how the last set in $\MC{m(\Pf)}$ is obtained. 

\begin{table}[h!]
\centering
\begin{tabular}{r|ccccc}
 Agents       &\{ $\;\;\; p\wedge r $,& $\;\;\;  p \wedge s$, &  $\;\;\; q$, &   $\;\;p\wedge q$, &   $\;\;\;t$ \} \\\hline
  \multicolumn{6}{c}{      $\Ct = \{\top\}$}\\\hline
  $\Js_1 -\Js_6$                 &  $\;\;\;\;$+           &  $\;\;\;$+             &  $\;\;\;$+   &  $\;$ +           &  +    \\
  $\Js_7 -\Js_{10}$            & $\;\;\;\;$+           & $\;\;\;$+             & $\;\;\;$ -   &  $\;$ -           &  +    \\
 $\Js_{11} -\Js_{17}$        & $\;\;\;\;$-           & $\;\;\;$-             &   $\;\;\;$+   &  $\;$ -           &  -    \\\hline
m(\Pf)                               & $\;\;\;$ + & $\;\;\;\;$+ & $\;\;\;\;$+ & $\;\;\;$- &  +\\ \hline
$\MC{m(\Pf)}$               & $\;\;\;$ + & $\;\;\;\;$+ & $\;\;\;\;$+ & $\;\;\;$  &  +\\
                                           & $\;\;\;$ + & $\;\;\;\;$+ & $\;\;\;\;$  & $\;\;\;$- &  +\\
                                           & $\;\;\;$   & $\;\;\;\;$  & $\;\;\;\;$+ & $\;\;\;$- &  +\\
\end{tabular}
\caption{Maximally consistent subsets of $m(\Pf)$, agenda from  Example~\ref{ex:scw}.}\label{tab:scwmc}
\end{table}

The $\RMSA$ aggregator is obtained by simply extending the sets in   $\MC{m(\Pf)}$ to complete sets using the $\textrm{ext}$ function: 
\begin{equation}\label{def:rmsa}
  \RMSA(\Pf) = \{\textrm{ext}(\Js) \mid \Js \in \MC{m(\Pf)}\}.
\end{equation}

\subsubsection{Maxcard Condorcet rule  $(\RMCSA)$}
Note that the set  $\{ p, \neg (p \wedge q), t \}$ in $\MC{m(\Pf)}$ uses less majority supported judgments than the other sets in  $\MC{m(\Pf)}$. The next aggregator, the {\em maxcard Condorcet rule} $(\RMCSA)$,  is obtained by refining   $\RMSA(\Pf)$, taking only those elements from $\MC{m(\Pf)}$ to extend that are of maximal cardinality. 

Given a set of formulas $S \subset \A$, the maximally consistent subset of $S$ with respect to set cardinality, is a set $S'$ that is a  consistent a consistent agenda subset and for which  there exists no other consistent agenda subset set $S'' \subseteq S$ such that $|S''| > |S'|$. The set $\MCC(S)$ is the set of all maximally consistent subsets of $S$ with respect to set cardinality. Now we can define the $\RMCSA$ aggregator. 
 
 \begin{equation}\label{def:rmcsa}
  \RMCSA(\Pf) = \{\textrm{ext}(\Js) \mid \Js \in \MCC{m(\Pf)}\}.
\end{equation}

\begin{example} For the profile given in Table~\ref{tab:scw},  agenda and constraints given in Example~\ref{ex:scw} we obtain: 

\[
  \RMSA(\Pf) =\left\{
\begin{array}{l}
 \{ p\wedge r, p\wedge s, q, p\wedge q, t  \},\\
 \{p\wedge r, p\wedge s, \neg q, \neg (p \wedge q), t \},\\
 \{ \neg(p\wedge r), \neg (p\wedge s), q, \neg (p \wedge q), t \}
\end{array}\right\}
\]

 \[
  \RMCSA(\Pf) =\left\{
\begin{array}{l}
 \{ p\wedge r, p\wedge s, q, p\wedge q, t  \},\\
 \{p\wedge r, p\wedge s, \neg q, \neg (p \wedge q), t \} \}
\end{array}\right\}
\]
\end{example}

 In \cite{LangPSTV15} it was shown that $\RMCSA \subseteq \RMSA$. 
The output of the aggregator $\RMSA$ is called {\em Condorcet admissible set}  by Nehring et al. \cite{PuppeNehring2011}, while   $\RMCSA$ is called {\em Slater} rule  \cite{PuppeNehring2011}. It is also a special case of the {\sc Endpoint} aggregator defined in  \cite{MillerOsherson08}.

The next aggregator we present is also   refines  $\RMSA$, however  not  in such a obvious way as $\RMCSA$.  This aggregator is called {\em ranked agenda rule} ($\RRA$). 

\subsubsection{Ranked agenda rule  ($\RRA$)}

Instead of just looking at the elements of the majoritarian set, we can also consider how much support each judgment obtained in the profile. Intuitively, the $\RRA$ constructs the collective judgment set by adding in it judgments one by one, starting with the judgment that has the "strongest support" in the profile, \ie~for which $N(\ai, \Pf)$ is highest. A judgment is only added if it is consistent (with respect to the constraints as well) with the judgments already added. Let us run this procedure with the profile given in Table~\ref{tab:scw}. 

We start with an empty collective judgment set $\Js = \emptyset$. For the profile in Table~\ref{tab:scw}, the judgment with the strongest support in the profile is  $q$ which appears in 13 of the 17 judgment sets. So now $\Js =\{ q\}$. The judgment $\neg (p \wedge q)$ has the next strongest support with 11 of the 17 judgment sets containing it. Since  $\neg (p \wedge q)$ is consistent with $q$ we now have $\Js =  \{\neg (p \wedge q) ,q\}$. The judgments $p\wedge r$,  $p\wedge s$ and $t$ each have the support of 10 judgment sets. However, only $t$ can be added to $\Js$, since adding either 
$p\wedge r$,  or $p\wedge s$ would cause $\Js$ to become inconsistent. Now $\Js =  \{\neg (p \wedge q) ,q, t\}$. The next judgments to be considered are $\neg  (p\wedge r)$,  or $\neg (p\wedge s)$ which each have the support of 7 judgment sets. Adding these two to $\Js$ completes the collective judgment set and now  $\RRA(\Pf) = \{ \{\neg (p \wedge q) ,q, t, \neg(p\wedge r), \neg(p\wedge s)\}\}$. 

However, $\RRA$ is not a resolute operator. Sometimes, a choice has to be made on which of two judgments with the same support should be added to the collective judgment set. Each time  this happens the collective judgment set "forks" into as many judgment sets as options, one judgment set for each option.  To illustrate this case we give an example from \cite{LangPSTV15}.

\begin{example}\label{ex:ra} Consider the pre-agenda $\A^+ =\{p\wedge q, p, q, p\wedge r, q\wedge r, s \}$ and constraints $\Ct=\{\top\}$. Let $\Pf$ be a profile for this agenda be as given in Table~\ref{tab:raleximax}. 

\begin{table}[h!]
\centering
\begin{tabular}{r|cccccc}
Voters & $p \wedge q$ & $p$ & $q$ & $p \wedge r$ & $q \wedge r $& $s$\\ \hline
  \multicolumn{7}{c}{ $\Ct = \{\top\}$}\\\hline
$\Js_1 -\Js_5 $        & - & + & -  & +& - & +\\
$\Js_6 - \Js_{10}$& - & - & +   & - & + & -\\
$\Js_{11} - \Js_{14}$ & + & + & + & +& + & +\\
$\Js_{15}$ & + & + & + & - & - & -\\ \hline
$m(\Pf)$              & -  & + &+ & + & + & +
\end{tabular}\caption{A profile from \cite{LangPSTV15} showing that $\RRA$ is not resolute.}\label{tab:raleximax}
\end{table}

Table~\ref{tab:stats} gives the support that each judgment has in this profile in descending order (left to right, top to bottom). 
\begin{table}[h!]
\centering
\begin{tabular}{r|cccccc}
Judgment  $\ai$& $\neg (p \wedge q)$ & $p$ & $q$ & $p \wedge r$ & $q \wedge r $& $s$ \\
$N(\ai,\Pf)$ & 10  & 10 & 10  & 9& 9 & 9\\\hline
Judgment $\ai$  &  $\neg (p \wedge r)$ & $\neg (q \wedge r)$ & $\neg s$ & $p \wedge q$& $\neg p$ & $\neg q$\\ 
$N(\ai,\Pf)$ & 6 & 6 & 6  & 5 & 5 & 5\\
\end{tabular}\caption{The support of each judgment in the profile of Table~\ref{tab:raleximax}.}\label{tab:stats}
\end{table}

Following the procedure, we should add $\neg(p\wedge q)$, $p$ and $q$ to the collective judgment, but this is not possible since this set of three judgments is not consistent. We must create three collective judgment sets, one for  each judgment: $\Js^1 = \{ \neg(p\wedge q)\}$, $\Js^2 = \{ p\}$ and $\Js^3 = \{ q\}$. We proceed "filling" each of these three judgment sets in parallel. Figure~\ref{fig:parallel} illustrates each step of this process for each starting judgment set. 

\begin{figure}
\begin{minipage}{\textwidth}
    \begin{center}
        \includegraphics[width=\textwidth]{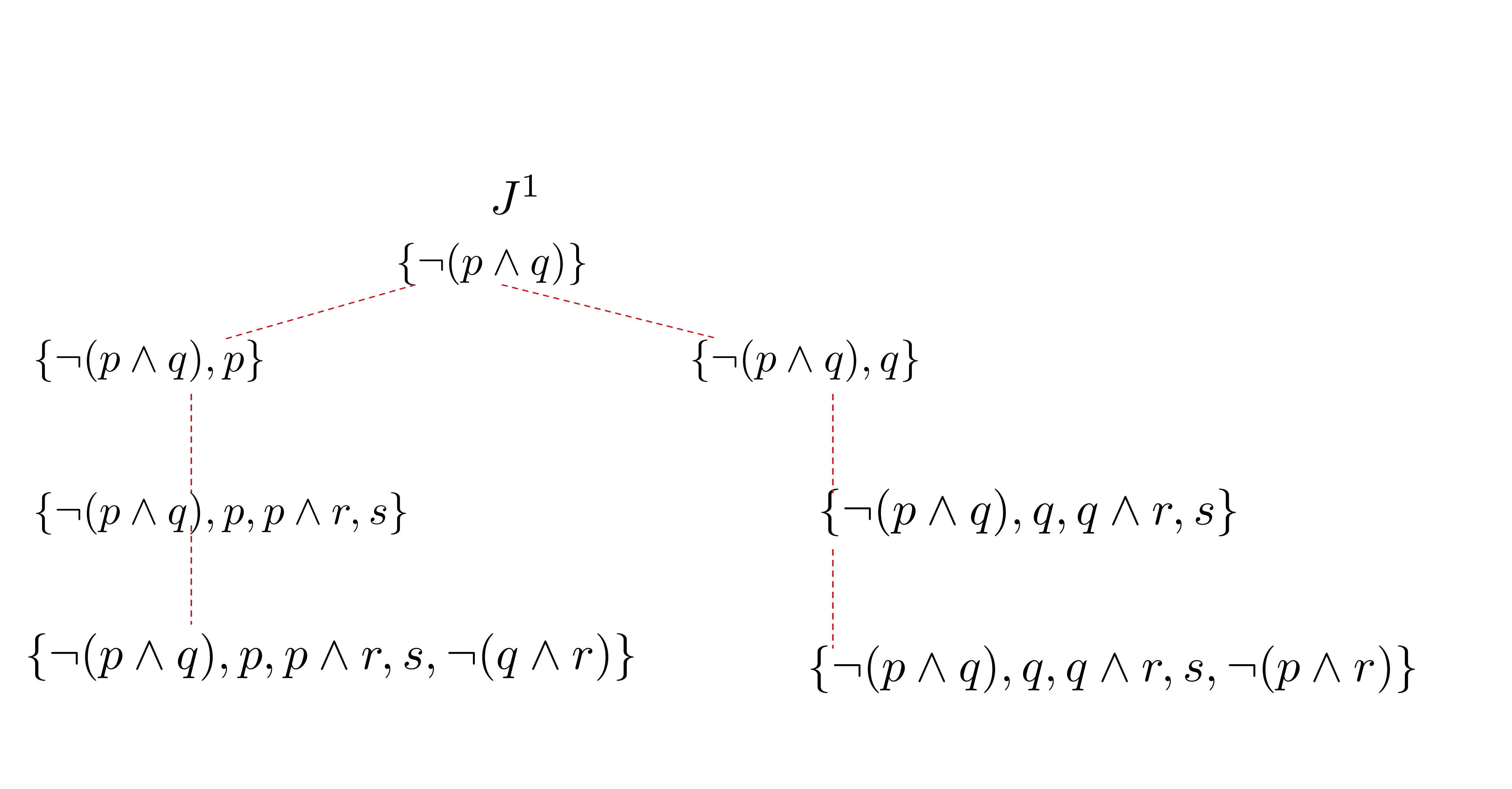} 
    \end{center}
\end{minipage}
\begin{minipage}{\textwidth}
    \begin{center}
        \includegraphics[width=\textwidth]{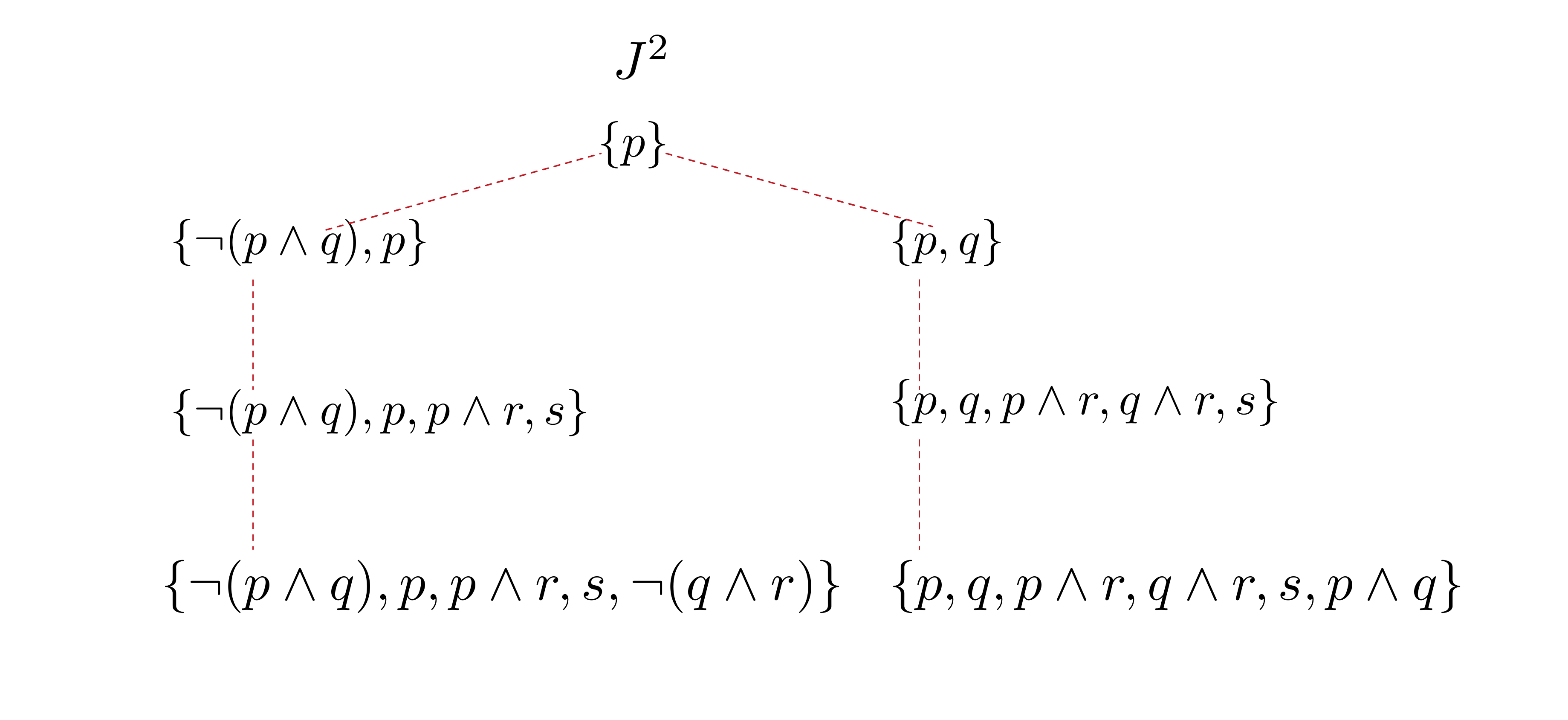} 
    \end{center}
\end{minipage}
\begin{minipage}{\textwidth}
    \begin{center}
        \includegraphics[width=\textwidth]{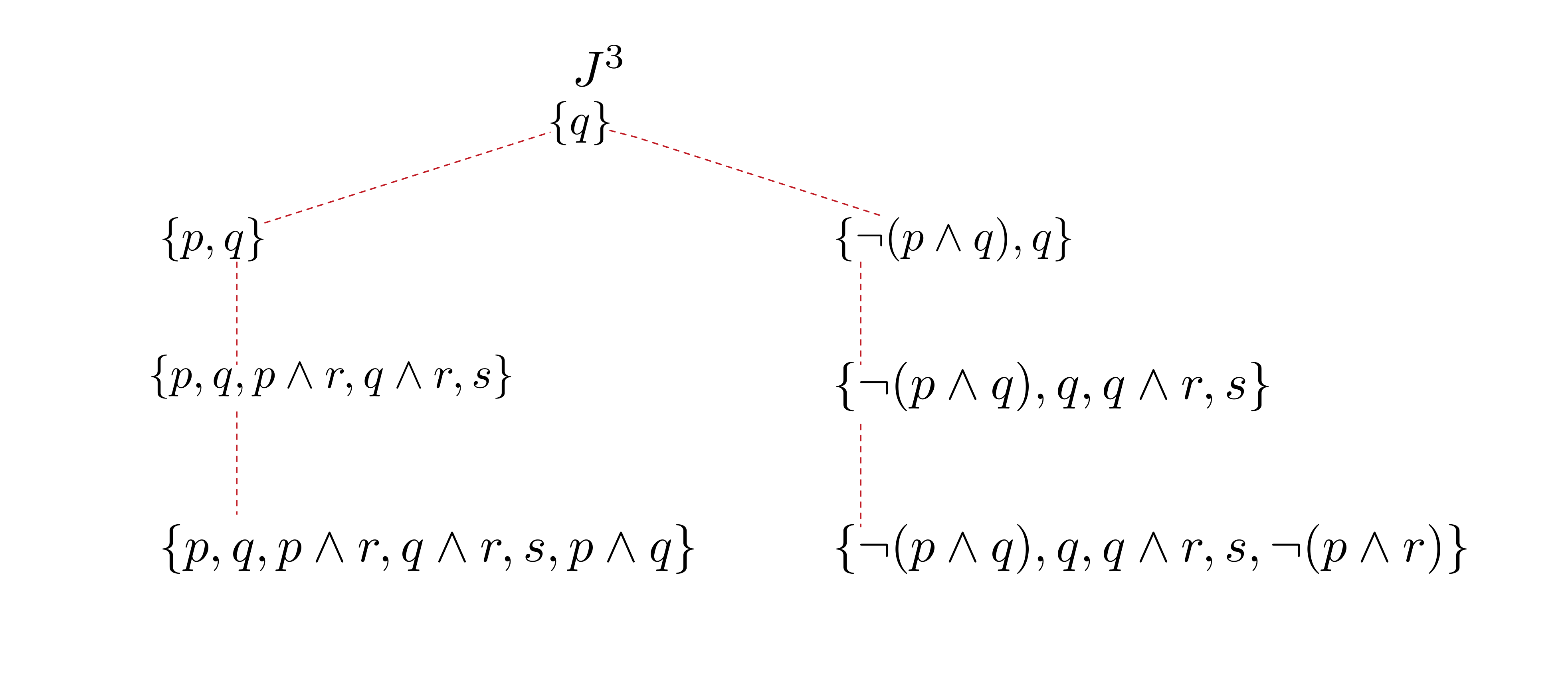} 
    \end{center}
\end{minipage}
\caption{Creating the collective judgment sets by $\RRA(\Pf)$ for $\Pf$ of Table~\ref{tab:raleximax}.}\label{fig:parallel}
\end{figure}

We now have 

$\RRA(\Pf) =
 \left\{ 
 \begin{array}{l}
 \{ p \wedge q, p, q, p \wedge r, q \wedge r, s\}, \\
\{ \neg(p \wedge q), p, \neg q, p \wedge r, \neg (q \wedge r), s\}, \\
 \{ \neg (p \wedge q), \neg p, q, \neg (p \wedge r), q \wedge r, s\}\}
 \end{array}\right\}
 $ 
\end{example} 

Formally the $\RRA$ rule is defined in \cite{LangPSTV15} in the following way.

Let $\A = \{\psi_1, \ldots, \psi_{2m}\}$. For  every 
 profile $\Pf \in \Dmc(\A,\Ct)^n$, $\RRA$ consists of those judgment sets
$\Js \in \Dmc(\A,\Ct) $ for which there exists a permutation $(\ai_1, \ai_2,  \ldots, \varphi_{2m})$ of the propositions in
$\A$ such that $\npf{\Pf}{\ai_1} \geq\npf{\Pf}{\ai_2} \geq \cdots\geq \npf{\Pf}{\ai_{2m}} $ and $\Js$ is obtained by the
following  algorithmic procedure:
\begin{center}
\begin{tabular}{l}
$S := \emptyset$\\
$\mbox{\bf for } k = 1, \ldots,2m\mbox{ do}$\\
~~~~$\mbox{\bf if } S \cup \{ \ai_k\} \mbox{ is consistent}$ $ \mbox{\bf then } S \leftarrow S \cup \{ \ai_k\}$ \\
~~~~$\mbox{\bf end if }$\\
{\bf end for}\\
$\Js := S$
\end{tabular}
\end{center}

The formal proof that the $\RRA$ aggregator is a refinement of the $\RMSA$ aggregator is given in \cite{LangPSTV15}. To gain an intuitive insight into why this is so, consider again the little trick with "ignoring" columns of the profile table we did before.Here again we can see the operation of the $\RRA$ aggregator as   "ignoring" issue columns, but now we "play favourites": we start by ignoring all the columns, and then we "let them in" one by one,  preferring the most "popular" judgments and "letting in" the  others that are not in conflict with the ones that are already in. The $\RRA$ aggregator behaves sort of like a bouncer at a night club. 

The $\RRA$ aggregator has its own refinement. The aggregator we consider next, the {\sc leximax} rule defined in \cite{NehringPivato2011,EKM13} refines the $\RRA$ aggregator, the formal  proof of this relationship is also given in \cite{LangPSTV15}.

\subsubsection{The {\sc leximax} rule }
The {\sc leximax} rule again considers the support a judgment has in the profile and  is very similar to the $\RRA$ aggregator, to the point that they are sometimes mistakenly considered to be the same. To use the same analogy as before, the  {\sc leximax} bouncer is more picky than the $\RRA$ bouncer: if a set of judgments have the same support in a profile, but is inconsistent,  it lets in first those judgments that have a more popular posey, or in logic-speak  consistent with more majority supported judgments. 

Let us look at again the judgment aggregation problem from Example~\ref{ex:ra}. Observe that if we start constructing a collective judgment set by adding either $p$ or $q$ first, we can next add all the majority-supported judgments except $\neg (p \wedge q)$. If instead, we start constructing a collective judgment set with $\neg (p \wedge q)$, we see that two of the majority-supported judgments cannot be added. The   {\sc leximax} rule further  selects from the output of the $\RRA$ aggregator, those collective judgment sets that contain a maximal number of majority-supported judgments. 
This $\textrm{\sc leximax}(\Pf) =\{  \{ p \wedge q, p, q, p \wedge r, q \wedge r, s\}\} $ for the profile $\Pf$  from Example~\ref{ex:ra}. 

Since we use the same framework as the one used in \cite{LangPSTV15},  we give almost verbatim the formal definition  of  {\sc leximax} from there. 

Given a profile $\Pf \in \Dmc(\A,\Ct)^n$   and a  judgment set $\Js \in \Dmc(\A,\Ct)$, define $S_k(\Pf) = \{ \ai \in \A \mid N(\Pf, \ai) = k, \frac{n}{2} \leq k \leq n\}$ and $s_k(\Js,\Pf) = |S_k(\Pf) \cap \Js |$.  Given two rational judgment sets $\Js, \Js'$, let $\Js >^{\textrm{\sc leximax}}_\Pf \Js'$ if and only if there is a $k \in \{\frac{n}{2}, \ldots, n\}$ such that
$s_k(\Js, \Pf) > s_k(\Js',\Pf)$ and 
for all $i > k$,  $s_i(\Js,\Pf) = s_i(\Js',\Pf)$.
$\textrm{\sc leximax}(\Pf)$ is the set of all undominated rational judgment sets in with respect to $>^{\textrm{\sc leximax}}_\Pf $.

\subsubsection{The median rule ($\RMWA$)}
We consider the last aggregator that refines $\RMSA$ know in the literature at present. This is aggregator was defined under many names and in many different ways by various authors. Here, we use for it the name {\em median rule} ($\RMWA$).   This aggregator is  a special type of the  {\sc Prototype} aggregator defined in \cite{MillerOsherson08}, {\em median rule} \cite{PuppeNehring2011}, {\em maximum weighted agenda rule} \cite{TARK11}, {\em simple scoring rule} \cite{Dietrich:2013} and {\em distance-based procedure} \cite{EndrissGP12}. Variants of this rule have been defined by Konieczny and Pino-P\' erez \cite{KPP02} and Pigozzi \cite{Pigozzi2006}. 

As the $\RRA$ aggregator, the $\RMWA$ aggregator also considers the number of agents that have $\ai$ in their judgment set when choosing which  judgments to include in  the collective judgment set. Instead of sequentially considering the agenda judgments one by one in order of strength of support, the $\RMWA$ aggregator  assigns a value to each judgment set   $\Js \in \Dmc(\A,\Ct)$, based on the number $N(\ai,\Pf)$ of each $\ai \in \Js$.  The collective judgment sets according to the  $\RMWA$ aggregator are those with assigned highest value. We illustrate this process with an example. 

\begin{example} Consider the judgment profile given in Table~\ref{tab:scw},  for the agenda and constraints given in Example~\ref{ex:scw}. Table~\ref{tab:statsscw} gives the $N(\ai,\Pf)$ values for each $\ai \in \A$; the first two rows belong to the majority-supported judgments. 

\begin{table}[h!]
\centering
\begin{tabular}{r|ccccc}
Judgment  $\ai$& $p\wedge r$ & $p\wedge s$ & $q$ & $\neg (p \wedge q)$ & $t $  \\
$N(\ai,\Pf)$ & 10  & 10 & 13  &11& 10  \\\hline
Judgment $\ai$  &  $\neg (p \wedge r)$ & $\neg (p \wedge s)$ & $\neg q$ & $p \wedge q$& $\neg t$  \\ 
$N(\ai,\Pf)$ & 7 & 7 & 4  & 6 & 7  \\
\end{tabular}\caption{The support of each judgment in the profile of Table~\ref{tab:scw}.}\label{tab:statsscw}
\end{table}

Table~\ref{tab:scwmed} gives the  values for each $\Js \in \Dmc(\A,\Ct)$; the first  row belonging to a  $\Js \in \Dmc(\A,\Ct)$ and the second to its value. 
 
 \begin{table}[h!]
\centering
\begin{tabular}{r|l}
 $\Js$& Value \\ \hline
$\{ p\wedge r, p\wedge s, q, p\wedge q, t\}$ &  49  \\
$\{ p\wedge r, p\wedge s, q, p\wedge q, \neg t\}$&  46  \\

$\{ p\wedge r, p\wedge s, \neg q, \neg (p\wedge q), t\}$&  45  \\
$\{ p\wedge r, p\wedge s, \neg q, \neg (p\wedge q), \neg t\}$&  42  \\

$\{ \neg (p\wedge r), \neg (p\wedge s), q, \neg (p\wedge q), t\}$&  48  \\
$\{ \neg (p\wedge r), \neg (p\wedge s), q, \neg (p\wedge q), \neg t\}$&  46  \\

$\{\neg (p\wedge r), \neg (p\wedge s), q, p\wedge q, t\}$&  43  \\
$\{ \neg (p\wedge r), \neg (p\wedge s), q, p\wedge q, \neg t\}$&  40  \\

$\{ p\wedge r, \neg (p\wedge s), q, p\wedge q, t\}$&  46  \\
$\{ p\wedge r, \neg (p\wedge s), q, p\wedge q, \neg t\}$&  43  \\

$\{\neg (p\wedge r),  p\wedge s, q, p\wedge q, t\}$&  46  \\
$\{ \neg (p\wedge r),  p\wedge s, q, p\wedge q, \neg t\}$&  43  \\

$\{ \neg( p\wedge r), p\wedge s, \neg q, \neg (p\wedge q), t\}$&  42  \\
$\{ \neg (p\wedge r), p\wedge s, \neg q, \neg (p\wedge q), \neg t\}$&  39  \\

$\{ p\wedge r, \neg (p\wedge s), \neg q, \neg (p\wedge q), t\}$&  42  \\
$\{ p\wedge r, \neg (p\wedge s), \neg q, \neg (p\wedge q), \neg t\}$&  39  \\

$\{ \neg( p\wedge r), \neg (p\wedge s), \neg q, \neg (p\wedge q), t\}$&  39  \\
$\{ \neg (p\wedge r), \neg (p\wedge s), \neg q, \neg (p\wedge q), \neg t\}$&  36  \\

\end{tabular}\caption{The value of each judgment in the codomain for the profile of Table~\ref{tab:scw}.}\label{tab:scwmed}
\end{table}

The judgment set with the highest value assigned is $\{ p\wedge r, p\wedge s, q, p\wedge q, t\}$, thus $\RMWA(\Pf) = \{  \{ p\wedge r, p\wedge s, q, p\wedge q, t\}\}$. 
\end{example}

Despite this example, the $\RMWA$ aggregator is not resolute. It is not difficult to imagine that there will be profiles for which two judgments in $ \Dmc(\A,\Ct)$ will be assigned the same, and maximal, values. 

We now give the formal definition of the $\RMWA$ aggregator, as stated in \cite{LangPSTV15}.

\begin{equation}\label{eq:mwa}
\RMWA(\Pf) = \argmax{\Js \in \Dmc(\A,\Ct)}  \sum_{\ai \in \Js} N(\ai,\Pf).  \end{equation}
 
 Of all the aggregators that refine $\RMSA$, the $\RMWA$ does so perhaps in the most unintuitive of ways. For a formal proof that t $\RMWA \subseteq \RMSA$, the reader is advised to consult \cite{TARK11} or \cite{LangPSTV15}. To intuitively grasp why   $\RMWA \subseteq \RMSA$ is the case, observe first that the majoritarian set will always have the highest value assigned with respect to a profile, even if the majoritarian set is not rational. The rational judgment sets with maximal assigned values have to be those that include as many as possible of the majority-supported judgments, \ie~ the set with the maximal value must be among the extensions of the maximally consistent subsets of the majoritarian set, with respect to set inclusion. 
 
 The $\RMWA$ aggregator is the last of the aggregators in the literature, that we are aware of, that refines  $\RMSA$.
 
\subsubsection{The Young rule ($\RY$)}
 Not all majority-preserving aggregators are a refinement of $\RMSA$. Recall the analogy of "minimally changing" the profile by choosing to "ignore" some issue columns in the table representation of a profile.  What if we "minimally change" the profile by ignoring rows in the table representation of the profile instead? This is how we obtain the {\em Young} rule ($\RY$), or aggregator.  The $\RY$ aggregator was defined in \cite{TARK11} and further studied in \cite{LangPSTV15}
 
 Intuitively, the Young rule looks for a smallest number of {\em agents} to remove from the profile so that the new, smaller  profile is majority-consistent. We illustrate this process with an example from  \cite{LangPSTV15}. 
 
 \begin{example}\label{ex:y} Consider the judgment profile given in Table~\ref{tab:scw},  for the agenda and constraints given in Example~\ref{ex:scw}. This profile is majority-inconsistent. Removing just one, any one, agent still gives a majority-inconsistent profile. Removing any combination of two agents also yields a majority-inconsistent profile. The smallest number of agents to remove to "gain" majority-\linebreak consistency  for this profile is three. There are many ways to remove three agents from this profile to get a majority consistent profile. We enlist all possible combinations and resulting profiles in Tables~\ref{tab:y1},~\ref{tab:y2},~\ref{tab:y3}~and~\ref{tab:y4}. 
 
 \begin{table}[h!]
\centering
\begin{tabular}{r|ccccc}
 Agents       &\{ $\;\;\; p\wedge r $,& $\;\;\;  p \wedge s$, &  $\;\;\; q$, &   $\;\;p\wedge q$, &   $\;\;\;t$ \} \\\hline
  \multicolumn{6}{c}{      $\Ct = \{\top\}$}\\\hline
  $\Js_1 -\Js_3$                 &  $\;\;\;\;$+           &  $\;\;\;$+             &  $\;\;\;$+   &  $\;$ +           &  +    \\
  $\Js_7 -\Js_{10}$            & $\;\;\;\;$+           & $\;\;\;$+             & $\;\;\;$ -   &  $\;$ -           &  +    \\
 $\Js_{11} -\Js_{17}$        & $\;\;\;\;$-           & $\;\;\;$-             &   $\;\;\;$+   &  $\;$ -           &  -    \\\hline
m(\Pf')                               & $\;\;\;$   & $\;\;\;\;$  & $\;\;\;\;$+ & $\;\;\;$- &   \\  
\end{tabular}
\caption{Profile resulting from removing any three of $\Js_1-\Js_6$ in the $\Pf$ from  Example~\ref{ex:scw}.}\label{tab:y1}
\end{table}

 \begin{table}[h!]
\centering
\begin{tabular}{r|ccccc}
 Agents       &\{ $\;\;\; p\wedge r $,& $\;\;\;  p \wedge s$, &  $\;\;\; q$, &   $\;\;p\wedge q$, &   $\;\;\;t$ \} \\\hline
  \multicolumn{6}{c}{      $\Ct = \{\top\}$}\\\hline
  $\Js_1 -\Js_4$                 &  $\;\;\;\;$+           &  $\;\;\;$+             &  $\;\;\;$+   &  $\;$ +           &  +    \\
  $\Js_7 -\Js_{9}$            & $\;\;\;\;$+           & $\;\;\;$+             & $\;\;\;$ -   &  $\;$ -           &  +    \\
 $\Js_{11} -\Js_{17}$        & $\;\;\;\;$-           & $\;\;\;$-             &   $\;\;\;$+   &  $\;$ -           &  -    \\\hline
m(\Pf'')                               & $\;\;\;$   & $\;\;\;\;$  & $\;\;\;\;$+ & $\;\;\;$- &   \\  
\end{tabular}
\caption{Profile resulting from removing any two agents  from the $\Js_1-\Js_6$ group  and any one agent from the $\Js_7-\Js_{10}$ group in the $\Pf$ from  Example~\ref{ex:scw}.}\label{tab:y2}
\end{table}
 
  \begin{table}[h!]
\centering
\begin{tabular}{r|ccccc}
 Agents       &\{ $\;\;\; p\wedge r $,& $\;\;\;  p \wedge s$, &  $\;\;\; q$, &   $\;\;p\wedge q$, &   $\;\;\;t$ \} \\\hline
  \multicolumn{6}{c}{      $\Ct = \{\top\}$}\\\hline
  $\Js_1 -\Js_5$                 &  $\;\;\;\;$+           &  $\;\;\;$+             &  $\;\;\;$+   &  $\;$ +           &  +    \\
  $\Js_7 -\Js_{8}$            & $\;\;\;\;$+           & $\;\;\;$+             & $\;\;\;$ -   &  $\;$ -           &  +    \\
 $\Js_{11} -\Js_{17}$        & $\;\;\;\;$-           & $\;\;\;$-             &   $\;\;\;$+   &  $\;$ -           &  -    \\\hline
m(\Pf''')                               & $\;\;\;$   & $\;\;\;\;$  & $\;\;\;\;$+ & $\;\;\;$- &   \\  
\end{tabular}
\caption{Profile resulting from removing any one  agents   from the $\Js_1-\Js_6$ group  and any two agents from the $\Js_7-\Js_{10}$ group in the $\Pf$ from  Example~\ref{ex:scw}.}\label{tab:y3}
\end{table}

  \begin{table}[h!]
\centering
\begin{tabular}{r|ccccc}
 Agents       &\{ $\;\;\; p\wedge r $,& $\;\;\;  p \wedge s$, &  $\;\;\; q$, &   $\;\;p\wedge q$, &   $\;\;\;t$ \} \\\hline
  \multicolumn{6}{c}{      $\Ct = \{\top\}$}\\\hline
  $\Js_1 -\Js_6$                 &  $\;\;\;\;$+           &  $\;\;\;$+             &  $\;\;\;$+   &  $\;$ +           &  +    \\
  $\Js_7$            & $\;\;\;\;$+           & $\;\;\;$+             & $\;\;\;$ -   &  $\;$ -           &  +    \\
 $\Js_{11} -\Js_{17}$        & $\;\;\;\;$-           & $\;\;\;$-             &   $\;\;\;$+   &  $\;$ -           &  -    \\\hline
m(\Pf'''')                               & $\;\;\;$   & $\;\;\;\;$  & $\;\;\;\;$+ & $\;\;\;$- &   \\  
\end{tabular}
\caption{Profile resulting from removing any three  agents   from the $\Js_{7}-\Js_{10}$ group in the $\Pf$ from  Example~\ref{ex:scw}.}\label{tab:y4}
\end{table}
The  extensions of $m(\Pf')$, $m(\Pf'')$,$m(\Pf''')$ and $m(\Pf''')$, which happen to be the same set for this profile, are all consistent but not complete. To obtain the collective judgment sets assigned by $\RY$, we extend them using the $\textrm{ext}$ function. 
Thus 

\[\RY(\Pf) = \textrm{ext}(\{ q, \neg (p \wedge q) \})  =\left\{
\begin{array}{rrrrr}
 \{\neg (p\wedge r), & \neg (p\wedge s), & q,  &   \neg (p\wedge q), & t\},       \\
 \{\neg (p\wedge r), & \neg (p\wedge s), & q,  &   \neg (p\wedge q), & \neg t\}
\end{array}
\right\}.
\]
 \end{example}
 
 In the case of Example~\ref{ex:y} all four different reduced profiles have the same majoritarian set, but in this need not be the case for every judgment aggregation problem. The majoritarian sets we obtain from the reduced profiles in Example~\ref{ex:y} are consistent agenda subsets but not necessarily complete. Thus before outputting them as collective judgment sets, the Young aggregator needs to complete them using the $\textrm{ext}$ function.  Following is the formal definition of the young rule  as it appears in \cite{LangPSTV15}.

\begin{equation}
\RY(\Pf) =  \{ ext(m(\Qf) ) \mid   \Qf \in \underset{m(\Qf) \in \mathcal{\Js}(\A,\Ct)} {\argmax{\Qf \sqsubseteq \Pf,}}|\Qf|
\}.\end{equation}

The same profile that we used as a running example, the  profile given in Table~\ref{tab:scw},  for the agenda and constraints given in Example~\ref{ex:scw}, shows that $\RY \neq \RMSA$. 

The aggregators we presented so far were majority-preserving  in an obvious way. They were  individual aggregators designed by looking for minimal parts of the profile to "ignore". Next we look into classes of aggregators. Some of them are majority-preserving, but most are not. 

\subsection{Value aggregators}
When we  explained the $\RMWA$ aggregator we established that this aggregator "operates" by assigning {\em values} to each rational judgment set in $\Dmc(\A,\Ct)$ and choosing the highest valued sets for collective. The classes of aggregators we consider now are built by expanding  this idea of assigning values. The {\em scoring} judgment aggregators are built around the idea that values are assigned to individual judgments in a rational judgment set. The {\em distance-based} aggregators directly extend the core aggregation principle of the $\RMWA$ aggregator -- values are assigned to rational judgment sets. Lastly, the {\em rationalising aggregators} "operate" by assigning values to entire profiles that have the same number of agents as the aggregated $\Pf$.  We start with presenting the distance-based class. 

\subsubsection{Distance-based aggregators}
The class of distance-based aggregators includes the $\RMWA$ aggregator and this is its best known member. The value that we assign to each $\Js \in \Dmc(\A,\Ct)$ under $\RMWA$ is the sum of the  $N(\ai,\Pf)$ numbers of each $\ai \in \Js$. Another way to look at this  value is as a measure  of {\em similarity } between  $\Js$  and the aggregated profile $\Pf$. The distance-based aggregators are defined using measures of  {\em dissimilarity } between a  $\Js \in \Dmc(\A,\Ct)$  and an aggregated $\Pf    \in \Dmc(\A,\Ct)^n$. The concept of dissimilarity between a rational judgment set and a profile of rational judgments uses two building concept: a measure of dissimilarity between two rational judgment sets and a method for aggregating each dissimilarity value between one {\em candidate} judgment set  $\Js \in \Dmc(\A,\Ct)$  and each of the $\Js_i \in \Pf$. 

Distance functions are used to measure the dissimilarity between two rational judgment sets. Given a set $S$, a distance function $d: S\times S \rightarrow \mathbb{R}$, also called {\em a metric} \cite{Deza2009},  is a function that satisfies the following properties, for each $x,y,z \in S$:
\begin{enumerate}
\item $d(x,y) \geq 0$ (nonnegativity)
\item $d(x,y) = 0$ if and only if $x=y$ (identity of the indiscernible)
\item $d(x,y) = d(y, x)$ (symmetry)
\item $d(x,z) = d(x,y) + d(y,z)$ (triangle inequality)
\end{enumerate}

Functions that satisfy properties $(1) - (3)$ but not $(4)$ are called {\em pseudo-distances}, while those that satisfy properties $(1),(2), (4)$ but not $(3)$ are called {\em quasi-distances}. For comparing rational judgments, pseudo-distances suffice, as no-one has so far, to the best of our knowledge,  argued for the need of the triangle inequality in the context of judgment aggregation. 

There are many (pseudo)-distances defined in the literature in general, but only a handful of these are used in judgment aggregation. The question of what makes a distance function a good measure of dissimilarity between rational judgment sets remains far from closed.  An in-depth  discussion of this question,  as well as a list of distance functions used in  judgment aggregation can be found in \cite{JELIA2014}.

After we compare a   $\Js \in \Dmc(\A,\Ct)$ with each of the judgment sets $\Js_i$ in the profile, what we get is a list of $n$ values  $(d(\Js, \Js_1), d(\Js,\Js_2), \ldots, d(\Js,\Js_n))$. We now need a method to aggregate this list of values into a single value to assign the $\Js$. One class of functions that is typically used for such aggregations are the functions called {\em norms}.  A norm on $\mathbb{R}^n$ is  a mapping $\eta: \mathbb{R}^n \rightarrow \mathbb{R}$ that has the following properties,  for all $a\in \mathbb{R}^+$ and every $\mathbf{x}, \mathbf{y} \in  \mathbb{R}^n $ :
\begin{description}
\item[i.] $\eta(a \mathbf{x}) = |a| \eta (\mathbf{x})$ (absolute homogeneity),
\item[ii.] $\eta(\mathbf{x} + \mathbf{y}) \leq \eta(\mathbf{x}) + \eta(\mathbf{y})$ (subadditivity),
\item[iii.]  $\eta(\mathbf{x})=0$ if and only if  $\mathbf{x} = (0, \ldots, 0)$.
\end{description}

The best known norm functions, as well as the most used ones in judgment aggregation,  are the $\Sigma$ and the $\max$. 
The (i) and (iii) properties of norms are necessary to guarantee some very essential good properties of the resulting aggregator.  Past this insight, the more general question of which are the best value aggregation functions to be used in the context of judgment aggregation is altogether unexplored. 

We can now formally define the class of distance-based judgment aggregators as follows. 
\begin{equation}\F^{d,\eta}(\Pf) = \argmin{\Js \in \Dmc(\A,\Ct)}\;   \eta (d(\Js_1, \Js), \ldots, d(\Js,\Js_n))
\end{equation}

Let us define the three most frequently discussed and occurring distances that are featured in the distance-based aggregators. All distances are defined for any two $\Js, \Js' \in \Dmc(\A,\Ct)$, for any judgment aggregation problem $\langle\A,\Ct,\Pf\rangle$.

\paragraph{Drastic distance.} The drastic distance $d_D$ is a very crude  tool for measuring dissimilarity. It is defined as follows
\begin{equation}
d_D(\Js,\Js')=\left\{
\begin{array}{l}
1 \textrm{ if and only if} \Js \neq \Js',\\
0 \textrm{ if and only if} \Js = \Js'.
\end{array}\right.
\end{equation}
Intuitively, two judgment sets are either different or this same, this is the only thing that the drastic distance can determine. 
The next distance is more fine-grained. 
 
 \paragraph{Hamming distance.} The Hamming distance $d_H$ identifies the number of issues on which two rational judgment sets give different judgments. For example the judgment set $\{ p\wedge r, \neg (p\wedge s), \neg q, \neg (p\wedge q), t\}$ and the judgment set $\{ \neg (p\wedge r), p\wedge s, \neg q, \neg (p\wedge q), \neg t\}$ differ on three issues: $\{ p\wedge r, \neg (p\wedge r)\}$, $\{p\wedge s, \neg (p\wedge s)\}$, and $\{t,\neg t\}$. Thus, the Hamming distance between these two judgment sets is 3. The formal definition of $d_H$, for two {\em complete} rational judgment sets, is as follows. 
 
 \begin{equation}
 d_H(\Js,\Js') = \Js\setminus \Js' = \Js' \setminus\Js.
 \end{equation}
 
 It is, of course, highly arguable whether the Hamming distance is really the best choice for measuring dissimilarity. A difference in judgments on any  issue from  the agenda is taken to have an equal "weight" in determining the overall dissimilarity between two judgment sets. But not all issues are the same, some are logically related to many other issues in the agenda, while others are not related to any other agenda issues. As argued in \cite{JELIA2014},  the Hamming distance does not take into consideration these possible relevant issues, which was why other distances and pesudo-distances were defined  in \cite{JELIA2014} and somewhat earlier in \cite{DuddyP:2012}. The distance axiomatised in \cite{DuddyP:2012} is the one we present next. 
 
\paragraph{Geodesic distance} Whereas the drastic and the Hamming distances are very easy to define and understand, the geodesic distance $d_G$ is a bot more convoluted. The geodesic distance avoids some of the shortcomings of the Hamming distance, and, as shown in \cite{DuddyP:2012}, it is the only distance function that in addition to the standard properties $(1)-(4)$ also satisfies properties $(5)$ and $(6)$ defined bellow.  
Before we introduce these two new properties, we have to  introduce the concept of {\em betweenness} for rational judgment sets. 

Consider three distinct $\Js,\Js',\Js'' \in \Dmc(\A,\Ct)$,\ie~$\Js \neq \Js'$, $\Js \new \Js''$ and $\Js' \neq \Js''$. Intuitively, the judgment set $\Js'$ is said to be between judgment sets $\Js$ and $\Js''$ when $\Js'$ contains all judgments  on which $\Js$ and $\Js''$ coincide. Formally, $\Js'$ is said to be between judgment sets $\Js$ and $\Js''$ if and only of
 $(\Js \cap \Js'') \subset \Js'$. The judgment set between two judgment sets is in a way a compromise between them. 
We can now define  properties $(5)$ and $(6)$  of a distance function $d$ for any  three distinct $\Js,\Js',\Js'' \in \Dmc(\A,\Ct)$, for any $\A$ and $\Ct$. 
\begin{enumerate}\setcounter{enumi}{4}
\item If $\Js'$ is between $\Js$ and $\Js''$, then $d(\Js,\Js'') = d(\Js,\Js') + d(\Js',\Js'')$. 
\item If $\Js$ and $\Js''$ are such that there exists no $\Js' \in \Dmc(\A,\Ct)$ that is between them, then $d(\Js,\Js'') =1$.
\end{enumerate}

To intuitively explain property $(6)$ consider the following three rational  judgment sets for $\A$ and $\Ct$ of Example~\ref{ex:scw}:
\[
\begin{array}{l}
\Js = \{ p\wedge r, p\wedge s, q, p\wedge q,t\}\\
\Js' = \{ p\wedge r, \neg(p\wedge s), q, p\wedge q,t\}\\
\Js'' = \{ p\wedge r, p\wedge s, \neg q, \neg (p\wedge q),t\}\\
\end{array}
\]

The judgment sets $\Js$ and $\Js'$ differ on only one issue and therefore by definition there cannot exist a judgment set between them. Thus for a distance $d$ that satisfies property $(6)$, it must be the case that $d(\Js,\Js') = 1$. We also have  $d_H(\Js,\Js') = 1$, but the Hamming distance does not satisfy property $(6)$. Consider judgment sets $\Js$ and $\Js''$ instead. These two differ on two issues and therefore   $d_H(\Js,\Js'') =2$. However,  if we list out  all the judgment set in $\Dmc(\A,\Ct)$, we will not find a judgment set among them that is between $\Js$ and $\Js''$. For a distance $d$ that satisfies property $(6)$, it  thus must be the case that $d(\Js,\Js'') = 1$ and for the geodesic distance this is the case. 

The geodesic distance is defined as the path distance on an {\em agenda graph} $G_{\A,\Ct} = \langle V, E\rangle$ where the set of vertices $V$ is  precisely the set $\Dmc(\A,\Ct)$. The set of edges $E \subseteq \Dmc(\A,\Ct) \times \Dmc(\A,\Ct)$ is defined as follows. There exists an edge between vertices $\Js$ and $\Js''$ if and only if there exists no judgment set $\Js\in \Dmc(\A,\Ct)$ that is between $\Js$ and  $\Js''$. The geodesic distance between two rational $\Js$ and $\Js'$, is the minimal path between them in $G$.

An example is on order, but fist a note.  Note that the only difference between a  $G_{\A,\Ct}$ and  $G_{\Phi,\textrm{IC}}$ for corresponding agenda and (integrity) constraints, is that the nodes in the first graph have sets of formulas and in the second lists of $0$s and $1s$ of length the same as the cardinality of the sets. Thus the  graphs  $G_{\Phi,\textrm{IC}}$  for judgment aggregation problems in the binary framework are visually much neater than the equivalent $G_{\A,\Ct}$ for the same judgment aggregation problem represented in the logic framework.  Therefore exceptionally we give  the example  of agenda graph  for the binary framework.

\begin{example} Consider the agenda and integrity constraints from Example~\ref{ex:bscw}. Figure~\ref{fig:running} depicts the agenda graph for these agenda and integrity constraints. Recall that this judgment aggregation problem  
was a rewriting in the binary framework of  the agenda and   constraints in the logic framework given in  Example~\ref{ex:scw}. Observe that   every vertex $x_1\;x_2\;x_3\;x_4\;x_5$ corresponds to exactly one judgment set $\Js$ and:
\begin{itemize}
 \item[] $x_1=1$ if and only if $p\wedge r \in \Js$,  $x_1=0$ if and only if $\neg (p\wedge r )\in \Js$
  \item[] $x_2=1$ if and only if $p\wedge s\in \Js$,  $x_2=0$ if and only if $\neg (p\wedge s )\in \Js$
   \item[] $x_3=1$ if and only if $q\in \Js$,  $x_3=0$ if and only if $\neg q\in \Js$
    \item[] $x_4=1$ if and only if $p\wedge q\in \Js$,  $x_4=0$ if and only if $\neg (p\wedge q )\in \Js$
     \item[] $x_5=1$ if and only if $t \in \Js$,  $x_5=0$ if and only if $\neg t\in \Js$.
 \end{itemize}
\begin{figure}[h!]
    \begin{center}
        \includegraphics[width=1.2\textwidth]{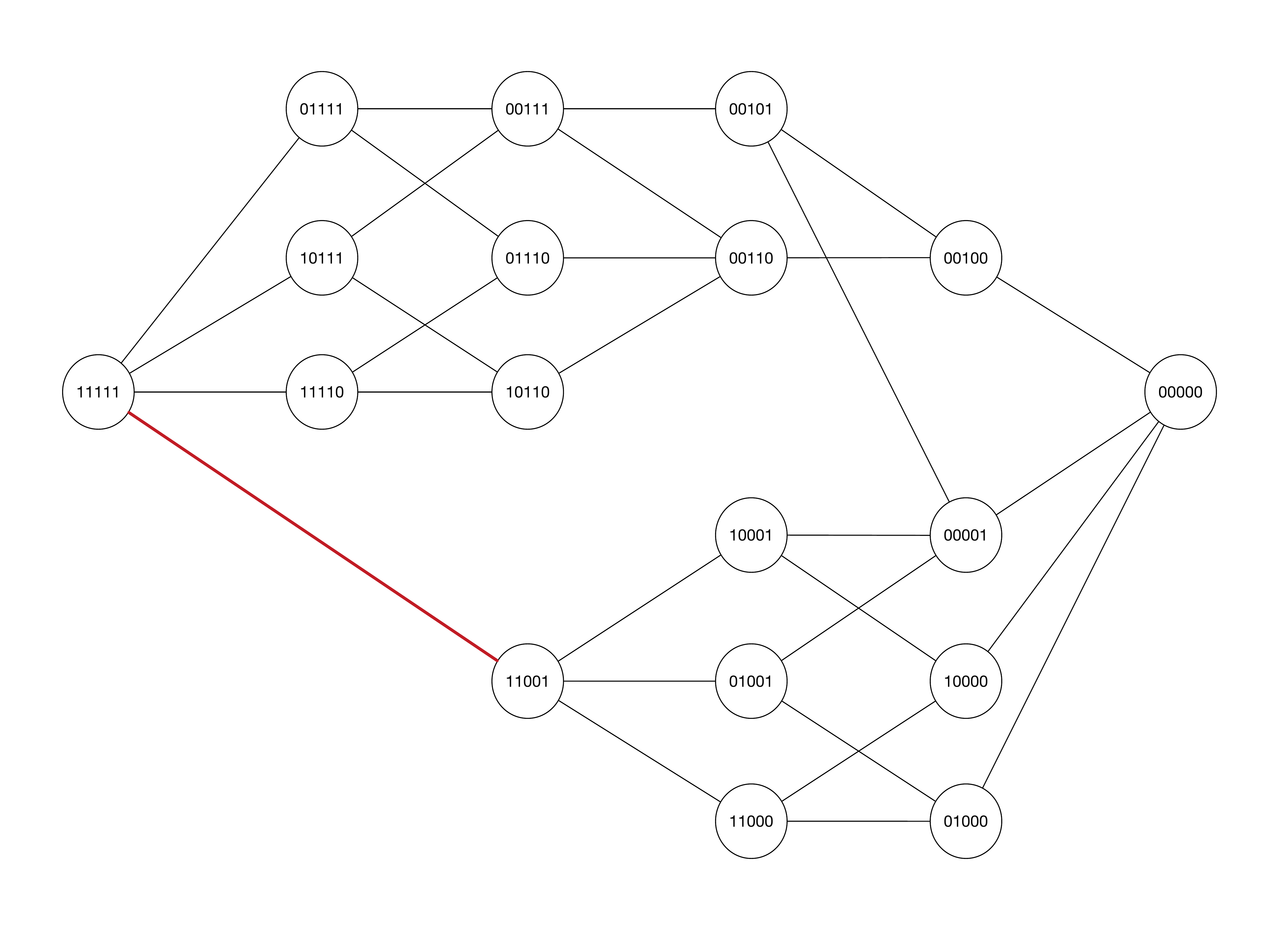} 
    \end{center}
\caption{Agenda graph for the judgment aggregation problem of Examples~\ref{ex:bscw}~and~\ref{ex:scw}. }\label{fig:running}
\end{figure}
In the graph on Figure~\ref{fig:running}, all the edges in black are between judgments (sets) that are at a Hamming distance one from each other,  while the edge in red is between judgments (sets) that are not at a Hamming distance one form each other. 
\end{example}
  
  We can say that in a way, the geodesic distance refines the Hamming distance. The following properties between the two distances are very easy to observe for any $\Js, \Js' \in \Dmc(\A,\Ct)$: if $d_H(\Js, \Js') = 1$, then $d_G(\Js,\Js') =1$ and, more generally, $d_H(\Js, \Js')  \geq d_G(\Js,\Js') $.

\paragraph{Specific distance-based aggregators}
We now give the specific \linebreak distance-based aggregators that have made an appearance in the literature. The sub-class of aggregators $\F^{d,\Sigma}$ was defined in \cite{MillerOsherson08} under the name \textsc{Prototype}. We already mentioned that the $\RMWA$ aggregator is distance-based. More precisely $\RMWA$ is the aggregator  $\F^{d_H,\Sigma}$. The proof that  $\F^{d_H,\Sigma} $ and $\RMWA$ is straightforward and we leave it as an exercise to the reader\footnote{Hint: observe that the Hamming distance can also be defined as \linebreak $d_H(\Js, \Js') = \underset{\ai \in \Js, \ai'\in\Js'}{\sum} d_D(\{\ai\},\{ \ai'\})$}. 
 
 The other two specific distance-base aggregators are $\F^{d_G,\Sigma}$, introduced in \cite{DuddyP:2012} and analysed in \cite{LangPSTV15}, as well as $\F^{d_H,\max}$, also analysed in \cite{LangPSTV15}. The aggregator $\F^{d_D,\Sigma}$ selects the judgment sets in the profile that are shared by the highest number of agents\footnote{Also called a {\em plurality-supported judgment set}.}, when such a judgment set exists and the entire profile otherwise. This aggregator is considered for historical reasons, it is related to a well known voting rule, but it is otherwise uninteresting. 

\subsubsection{Scoring aggregators}
The next class of aggregators we consider are the {\em scoring aggregators}. This class was introduced in \cite{Dietrich:2013}. This class is defined around the idea that each agent can set a value for every judgment in her judgment set. Intuitively, this value can be seen as an answer to the question:  "if I reverse my judgment on this issue, how much would it cost me?". The class of scoring aggregators is defined around a {\em scoring function}.

A scoring function  $s: \A \times \Dmc(\A,\Ct) \rightarrow \mathbb{R}$ assigns a value to a pair of judgment and a rational judgment set. Intuitively, the scoring function tells us how much is each judgment in a particular judgment set "worth". 

 Several scoring functions are introduced in \cite{Dietrich:2013}. We include here two that are more interesting: the {\em simple scoring function} and the {\em reversal scoring function} (defined for any $\A$ and $\Ct$).
 
 The simple scoring function $s_I$ is defined as follows.
 
 \begin{equation}
 s_I(\ai, \Js) = \left\{
 \begin{array}{l}
 1 \textrm{ if and only if} \ai \in \Js,\\
0 \textrm{ if and only if} \neg \ai \in \Js,
 \end{array}
 \right.
 \end{equation}
 
The reversal scoring function $s_{\textrm{rev}}$ is much more convoluted than the simple one. Intuitively, the reversal scoring function counts what is the smallest number of judgments that have to be "flipped" in $\Js$ for consistency to not be violated when $\ai $ in $\Js$ is replaced with $\neg \ai$. Formally this is the definition:

\begin{equation}
s_{\textrm{rev}}(\ai, \Js) = \argmin{\underset{\neg \ai \in\Js^{\ast}}{\Js^{\ast} \in \Dmc(\A,\Ct)}}\;\; d_H(\Js, \Js^{\ast}).
\end{equation} 

\begin{example}\label{ex:scores} Consider the judgment aggregation problem of the "doctrinal paradox", Example~\ref{ex:doctrinal}. Tables~\ref{tab:doctrinalscores1}~and~\ref{tab:doctrinalscores1} give the scores for each judgment in every rational judgment set both for the simple and the reversal scoring function. Recall that we had $\A^+=\{ p,q,d\}$ and $\Ct=\{ (p\wedge q) \leftrightarrow d\}$. 

\begin{table}[h!]
\centering
\begin{tabular}{r|cccccc}
$s_{I}(\ai,\Js_i)$& $p$ & $\neg p$ & $q$ & $\neg q$ & $d$ & $\neg d$ \\\hline
$\{ p, q, d\}$      & $1$ & $0$ &$1$& $0$ & $1$ & $0$  \\
$\{ p, \neg q, \neg d\}$      & $1$ & $0$ &$0$& $1$ & $0$ & $1$  \\
$\{ \neg p, q, \neg d\}$      & $0$ & $1$ &$1$& $0$ & $0$ & $1$  \\
$\{ \neg p, \neg  q, \neg d\}$      & $0$ & $1$ &$0$& $1$ & $0$ & $1$  \\
\end{tabular}
\caption{Simple scores for judgments in the  "doctrinal paradox", Example~\ref{ex:doctrinal}. }\label{tab:doctrinalscores1}
\end{table}
 
\begin{table}[h!]
\centering
\begin{tabular}{r|cccccc}
$s_{\textrm{rev}}(\ai,\Js_i)$& $p$ & $\neg p$ & $q$ & $\neg q$ & $d$ & $\neg d$ \\\hline
$\{ p, q, d\}$      & $2$ & $0$ &$2$& $0$ & $2$ & $0$  \\
$\{ p, \neg q, \neg d\}$      & $1$ & $0$ &$0$& $2$ & $0$ & $2$  \\
$\{ \neg p, q, \neg d\}$      & $0$ & $2$ &$1$& $0$ & $0$ & $2$  \\
$\{ \neg p, \neg  q, \neg d\}$      & $0$ & $1$ &$0$& $1$ & $0$ & $2$  \\
\end{tabular}
\caption{Reversal  scores for judgments in the  "doctrinal paradox", Example~\ref{ex:doctrinal}. }\label{tab:doctrinalscores2}
\end{table}
\end{example}

A scoring aggregator assigns the most similar judgment sets with respect to $\textrm{sim}_s$ to the aggregated profile. 

Formally 

\begin{equation}
\F_{s} = \argmax{\Js \in \Dmc(\A,\Ct)} \sum_{\Js_i \in \Pf} \sum_{\ai \in \Js_i \cap \Js} s(\ai,\Js_i)
\end{equation}

Having defined a scoring function $s$, we can compare how similar a judgment set $\Js \in \Dmc(\A,\Ct)$ is with a particular judgment set $\Js_i$ in a profile $\Pf$ by simply summing up the values of all the judgments that $\Js_i$ shares with $\Js$. Formally, we define the similarity measure $\textrm{sim}_s: \Dmc(\A,\Ct) \times \Dmc(\A,\Ct) \rightarrow \mathbb{R}$ as

\begin{equation}
\textrm{sim}_s(\Js_i,\Js) = \sum_{\ai \in \Js_i \cap \Js} s(\ai,\Js_i).
\end{equation}

Observe that if we count the values assigned to judgments that $\Js$ does not share with  $\Js_i$, we define a dissimilarity measure. However, it is not necessary that this dissimilarity measure is a distance function. 
Dietrich in  \cite{Dietrich:2013} does not set any requirements on the scoring functions, not even that they should be positive. However, it is not difficult to observe that a similarity measure will have the following properties, for every $\Js,\Js_i \in \Dmc(\A,\Ct)$, for any $\A$, $\Ct$:
\begin{itemize}
\item $\textrm{sim}_s(\Js_i,\Js)  \geq 0$,
\item $\textrm{sim}_s(\Js_i,\Js) = 0$  if and only if $\Js_i =\Js$.
\end{itemize} 
While $\textrm{sim}_s$ satisfies nonnegativity and identity of the indiscernible, it may not satisfy even symmetry, let alone triangle inequality. We have that \linebreak $\textrm{sim}_I(\Js,\Js')  =\textrm{sim}_I(\Js',\Js)$, but for $\textrm{sim}_{\textrm{rev}}$ this is not the case.  Consider Example~\ref{ex:scores}. We have that $\textrm{sim}_{\textrm{rev}}(\{ p, q, d\},\{ p, \neg q, \neg d\})  = 2$, while \linebreak   $\textrm{sim}_{\textrm{rev}}(\{ p, \neg q, \neg d\},\{ p, q, d\})  = 1$.

For $\textrm{sim}_I$ even triangular inequality holds. Observe that $\textrm{sim}_I(\Js,\Js') = m - d_H(\Js,\Js') $, where $m$ is the cardinality of the pre-agenda. Consequently, the aggregators $\F_{I}$, $\RMWA$ and $\F^{d_H,\Sigma} $  are the same aggregator: $\F_{I} = \RMWA = \F^{d_H,\Sigma} $.

\subsubsection{\RMNAC~and the rationalising aggregators class}

The last class of aggregators we consider, we call here {\em rationalising aggregators}.  The rationalising aggregators combine the two ideas for constructing aggregators: changing the profile in a minimal way to make it  majority-consistent and using (dis)similarity measures to compare judgment sets. Intuitively, the rationalising aggregators "assume" that the aggregated profile has "errors" and look for the most similar "ideal" profile to aggregate instead by using issue-by-issue majority.

Given an agenda $\A$ and associated constraints $\Ct$, let $\mathcal{M}(\A,\Ct,n) \subseteq \Dmc(\A,\Ct)^n$ be the set of all majority-consistent profiles for $\A$ and $\Ct$. More precisely $\mathcal{M}(\A,\Ct,n) = \{ \Pf \mid \Pf \in  \Dmc(\A,\Ct)^n \mbox{ and } m(\Pf) \in \Dmc(\A,\Ct)\}$.  To ease notation, we will write $\mathcal{M}$ when $\A, \Ct$ and $n$ are clear from the contest.  
For Miller and Osherson \cite{MillerOsherson08} this set $\mathcal{M}$ contained the "ideal" profiles. 
To compare two profiles for (dis)similarity they used a distance function $D$ defined over sets of profiles with the same number of agents. Namely,  $D$ is a distance function of type $D: S \times S \rightarrow \mathbb{R}$ where $S \subseteq \Dmc(\A,\Ct)^n$. 

The way  Miller and Osherson construct their $D$ functions in  \cite{MillerOsherson08} is by considering a distance function $d$ between judgment sets. For $\Pf , \Pf'   \in  \Dmc(\A,\Ct)^n$, $\Pf = (\Js_1, \ldots, \Js_n)$,   $\Pf' = (\Js'_1, \ldots, \Js'_n)$:

\begin{equation}
D(\Pf,\Pf')  = \sum_{i=1}^n d(\Js_i, \Js'_i)
\end{equation}

However, any combination of a norm $\eta$ and a distance function $d$ between judgment sets yields a distance function between profiles of judgment sets. So in general we can define

 \begin{equation}
D_{\eta,d}(\Pf,\Pf')  = \eta( d(\Js_1, \Js'_1), \ldots, d(\Js_n, \Js'_n)).
\end{equation}

Let  $\mathcal{C} \subseteq \Dmc(\A,\Ct)^n$ be the set of all "ideal"  profiles to which a profile $\Pf \in \Dmc(\A,\Ct)^n$  is compared. Let $c: \mathcal{C} \rightarrow  \Dmc(\A,\Ct)$ be a resolute aggregator that assigns a profile from $\mathcal{C}$ to a rational judgment set. 
We can now give a formal definition of the class of rationalising aggregators. 

\begin{equation}
\F^{d,\eta}_{\mathcal{C},c}(\Pf) = \{ c(\Pf') \mid \argmin{\Pf' \in \mathcal{C}}\;  D_{\eta,d}(\Pf,\Pf')\}.
\end{equation}

 We already saw that one such "ideal" collection of profiles is \linebreak $\mathcal{M}(\A,\Ct,n)$. The class of rationalising aggregators defined in \cite{MillerOsherson08},   $\textsc{full}_d$ is a special case of our class $\F^{d,\eta}_{\mathcal{M},m}$, where $m$ is the issue-by-issue majority. Of this class, the best known is the specific rule  $\RMNAC =  \F^{d_H,\Sigma}_{\mathcal{M},m}.$
 
  Another set of  intuitively "ideal" profiles is the set of profiles in which all the agents have the same judgment set, we can call this the {\em unanimity} profile collection\footnote{For each uniform quota aggregator we can define an "ideal" collection of profiles.}.  We can define this collection $\mathcal{U}(\A,\Ct,n) \subset \mathcal{M}(\A,\Ct,n) \subset \Dmc(\A,\Ct)^n$ as 
  
\begin{equation}
\mathcal{U}(\A,\Ct,n)  = \{\Pf \mid \Pf \in \Dmc(\A,\Ct)^n \textrm{ and } u(\Pf) \in  \Dmc(\A,\Ct)\}. \end{equation}
  
 Interestingly,  we obtain that $\F^{d,\eta}_{\mathcal{U},u} = \F^{d,\eta}$, namely that the distance-based aggregators are a sub-class of the rationalising aggregators. This relation between aggregators is not difficult to prove and we leave it as an exercise for the reader. 
 
Lastly, we need to mention that t here is considerable work on rationalising aggregators within the scope of preference aggregation. We would like to point the interested reader to \cite{Elkind2015}.

 \subsection{Other aggregators}
Apart from the aggregators we have so far introduced and classified, we need to mention some of the other aggregators that have appeared in the literature, but that do fit in the classifications we have introduced. We give here the intuition behind these aggregator. For the formal definitions, the reader should consult the respective cited publications. 

The first one in this collection is the {\em extended conclusion-based procedure} introduced in \cite{ADT09}. Recall that the conclusion-based procedure applied only to   agendas that can be partitioned into agenda of premises and agenda of conclusions. The conclusion-based procedure aggregates only the judgments on the conclusions, using the issue-by-issue majority,  thus always producing incomplete judgment sets.  In  \cite{ADT09} the conclusion-based procedure is extended with a distance-based aggregator.  First the judgments on the conclusion issues are aggregated and the  disjunction of the obtained collective judgment sets is added to $\Ct$. Next the whole profile is aggregated using the new constraints.   
This procedure is best to explain using the simple example of the "doctrinal paradox" judgment aggregation problem. 

\begin{example} Recall the "doctrinal paradox" judgment aggregation problem:   $\A^{+}_p = \{ p, q,\}$, $\A^+_c=\{ d\}$, $\Ct=\{ (p\wedge q) \leftrightarrow d\}$ and $\Pf = (\{ p,q,d\}, \{ \neg p,q,\neg d\}, $ \linebreak $\{ p,\neg q,\neg d\})$. Let us use the $\F^{d_H,\Sigma}$ aggregator on the profile of premises. We obtain $\F^{d_H,\Sigma} (\Pf^{\downarrow \A_c}) = \{ \neg d\}$. We now extend the set of constraints $\Ct' = \Ct \cup \{ \bigvee \F^{d_H,\Sigma} (\Pf^{\downarrow \A_c})\}$, which here simply yields $\Ct' = \{p\wedge q) \leftrightarrow d, d\} $. 

We now aggregate $\Pf$ with $\Ct'$ using again  $\F^{d_H,\Sigma}$. There are three rational judgment sets in $\Dmc(\A, \Ct')$, of which $ \{ \neg p,q,\neg d\}$ and $\{ p,\neg q,\neg d\}$ are closest to $\Pf$ so these are   the collective judgment sets selected by the extended conclusion based procedure.  
\end{example}

Note that in the extended conclusion based procedure, in the second step there might be judgment sets in the profile that are not rational for the new constraints, as indeed is the case with $\{ p,q,d\}$ in the above example. The distance used to define the aggregator should be able to handle comparing rational and irrational judgment sets. This is not a problem for the drastic and the Hamming distances, but the geodesic distance cannot be used.

The second aggregator we consider is the aggregator that chooses the  {\em most representative voter} proposed in \cite{EndrissG14}. This aggregator is distance-based aggregator that instead of choosing a collective judgment set from the entire set $\Dmc(\A,\Ct)$, it limits its choice from among the sets in the aggregated profile $\Pf$. We illustrate this approach with an example. 

\begin{example}Consider the  judgment aggregation problem of Example~\ref{ex:scw}, with the profile from Table~\ref{tab:scw}.  Let us use the $\RMWA$ aggregator.  There are three distinct rational judgment sets in the profile and the values for each are given in Table~\ref{tab:scwmrv}. The most representative voter procedure, when the  $\RMWA$ aggregator is used selects the set $\{ p\wedge r, p\wedge s, q, p\wedge q, t\}$ as a collective judgment set.

 \begin{table}[h!]
\centering
\begin{tabular}{r|l}
 $\Js$& Value \\ \hline
$\{ p\wedge r, p\wedge s, q, p\wedge q, t\}$ &  49  \\
$\{ p\wedge r, p\wedge s, \neg q, \neg (p\wedge q), t\}$&  45  \\
$\{ \neg (p\wedge r), \neg (p\wedge s), q, \neg (p\wedge q), \neg t\}$&  46   
\end{tabular}\caption{The value of each judgment in the  the profile of Table~\ref{tab:scw} for the same profile.}\label{tab:scwmrv}
\end{table}
\end{example}
 
 The last aggregator we consider here is a class of aggregators introduced in \cite{CostantiniEtAlAAAI2016} and called {\em binomial rules}. This class of aggregators are defined for the binary framework, but of course can be applied for judgment aggregation problems in the logic framework as well. The binomial rules assign 
scores, or values to patterns of accepted/rejected issues so instead of comparing whole judgment sets for similarity, they compare how similar are certain subsets (corresponding to particular subset of agenda issues) in the judgment sets, allowing for certain subsets to be more relevant than others.   The most representative voter and the medial rule are shown to be special cases of the binomial rules class.

\subsection{A note on tie-breaking and good domains}\label{sec:misc}

As we pointed out at the begining of this section, all of the specific judgment aggregation aggregators proposed in the literature are either irresolute, or partial, \ie~do not satisfy the universal domain property. 
When a single rational  judgment set is needed, there needs to be a way to break the tie between the collective judgment sets. Tie-breaking is not an uncommon practice in social choice, however it is significantly under-explored in judgment aggregation. 

The simplest approach to tie-breaking is to assign a default ordering over the set $\Dmc(\A,\Ct)$. Although effective this method is not very practical given that the set can have exponentially many members, with respect to the size of the agenda. Another simple approach is to assign a default ordering over the judgments in the agenda $\A$. However, although more practical, this method  is not as effective since it may not be able to completely break the ties. In the next section we will discuss properties of aggregators. How ties are broken can have an effect on the resulting properties of the combination aggregator, tie-breaking method, as observed in \cite{AAAI}. What is a good way to break ties in judgment aggregation remains an open research problem. 

Particularly in light of tie-breaking issues, it is of interest to identify subsets of $\Dmc(\A,\Ct)^n$ for which the partial  but resolute aggregators are well defined. Compared with preference aggregation, the so called   {\em restricted domains} are unexplored in the context of judgment aggregation.  The exception to this is the work of Dietrich and List \cite{Domains} and the earlier \cite{List2003}.  Dietrich and List identify four restricted domains for which the issue-by-issue majority gives a complete and consistent judgment set:   single-plateaued profiles, single-canyoned profiles, unidimensionally aligned profiles and unidimensionally ordered profiles. 

The single-plateaued profiles and single-canyoned profiles are identified by the existence of at least one total order $\geq$, called {\em structuring order},  over the agenda $\A$. Informally, single-plateauedness requires that every individual's judgment set contains an "interval" (a plateau) of judgments with respect to $\geq$. For example, if $\ai_i \geq \ai_{j}  \geq \ai_{k}$ is part of the  structuring order, and $\Js$ is such that $\{\ai_i, \ai_k\} \subseteq \Js$,  then it has to be the case that $\ai_j \in \Js$ otherwise the judgment set will not have a plateau and a profile that includes such a  judgment set will not be single-plateaued.  Single-canyonedness on the other hand, requires that the rejected judgments by each agent form such an interval (a canyon). So if  if $\ai_i \geq \ai_{j}  \geq \ai_{k}$ and $\{\ai_i, \ai_k\} \not\subseteq \Js$, then the judgment $\ai_j$ must also be rejected, \ie~$\neg\ai_j \in \Js$.  Every single-canyoned profile is also a single-plateaued profile. 

It is not so simple to identify the agendas and constraints for which all of the profiles in $\Dmc(\A,\Ct)$ are single-plateaued although there clearly is a relation between the logical relations among the agenda issues and the existence of the structuring order. 

 The unidimensionally ordered profiles are identified by the existence of at least one total order $\succsim$, also called a structuring order,  over the set of agents in $\Pf$. A profile is  unidimensionally ordered if for all agents $\Js_i \in  \Pf$ it holds that $\Js_{i-1} \succsim \Js_{i} \succsim \Js_{i+1}$ and for each judgment, the individuals accepting it are
all adjacent to each other.  Identifying if a profile is  the unidimensionally aligned is very simple: there exists a total order over the judgment sets in  $\Pf$ such that   $\Js_1 \succsim \Js_2 \succsim \cdots \succsim \Js_n$  and for all $n>i>1$,  $\Js_i$ is between judgment sets $\Js_{i-1}$ and $\Js_{i+1}$. All unidimensionally aligned profiles are also unidimensionally ordered profiles.

\newpage

\section{Properties of aggregators }  
\label{ch:name}  
Having introduced the various specific and classes of judgment aggregators, we  present and discuss the properties of judgment aggregators that can be encountered in the literature under various contexts. Already in Chapter~\ref{ch:mathtest} we encountered two such properties: universal domain and majority-preservation. The study of aggregator properties serves both an academic and a practical purpose. Desirable and undesirable combinations of aggregator properties help us characterise the existence of aggregators. Researchers in philosophy, political science and related fields, may identify good and bad properties of aggregators, typically expressed as relations that should or should not hold between the aggregated profile and the resulting collective judgment sets. Once mathematically formalised, researchers can identify whether, and how many,  functions that satisfy these properties exist. If it is determined that no functions exists, \ie~an impossibility result is reached, the researchers can refocus their efforts.   Practically, knowing which aggregators satisfy which  properties helps us match specific aggregators to specific  (types of) judgment aggregation  problems. 

The study of aggregator properties in judgment aggregation is somewhat paradoxical. Since the first decade of research in judgment aggregation had been marked by a pursuit for characterisation results, most properties are defined for resolute aggregators only. On the other hand, most aggregators are not resolute.  This has now begun to change and we have some properties studied for irresolute aggregators. We present both types of properties here. In the last section we given an overview of which properties are known to hold for which aggregators, and point out to the gaps in the literature. 

We should mention that, at present there exists no characterisation results on any of the irresolute aggregators. An exception is the $\RMWA$ aggregator for whose characterisation Nehring and Pivato have obtained some results \cite{NehringPivato16}.  

\subsection{Resoluteness-independent properties}

Some properties are very intuitive and simple and can be defined regardless of whether the aggregator is resolute or irresolute. We begin introducing aggregator properties by introducing this class of properties  whose desirability is very easy to defend.  

\paragraph{Dictatorship.} The first desirable aggregator property is that of {\em non-} \linebreak {\em dictatorship}. This is a very simple and intuitive requirement  that the aggregator does not always select as collective the judgment set representing one particular agent regardless of what are the  judgment sets of the other agents in the profile. 

  Consider profiles $\Pf=(\Js_{\ast}, \Js_1, \ldots, \Js_{n-1})$ and  $\Pf'=(\Js_{\ast}, \Js'_1, \ldots, \Js'_{n-1})$. The agent $\Js_{\ast}$ is a dictator, and $F$ a dictatorial aggregator if $F(\Pf) = F(\Pf)=\Js_{\ast}$ regardless of what the other judgment sets $\Js_i \in \Pf$ and $\Js'_i \in \Pf$

\paragraph{Responsivness} Related but not the same to non-dictatorship is the desirable property of {\em responsiveness}  \cite{DietrichList2005,GrossiP:2014}. Non-imposition is the requirement that any one rational judgment set has a "chance" of being selected as a collective judgment set. More precisely there exists at least one profile $\Pf \in \Dmc(\A,\Ct)^{n}$ for every $\Js \in \Dmc(\A,\Ct)$ such that $F(\Pf) = \{ \Js\}$. 

\paragraph{Unanimity.} Another simple, intuitive, and related desirable property,  is that of {\em unanimity}. Assume that every single agent in the profile has chosen the exact same judgment set. It is hard to argue that this judgment set, and no other, must be selected as the collective. This is the property of unanimity. If $\Pf  \in \Dmc(\A,\Ct)^{n} $ is such that $\Pf = (\Js_1, \ldots, \Js_n)$  and for every $1 \leq i \leq n$,  $\Js_i = \Js$ for some $\Js  \in \Dmc(\A,\Ct)$, then necessarily $\F(\Pf) = \{ \Js\}$. It is very easy to observe that $F$ satisfies unanimity then it will necessarily satisfy non-imposition. We need to also note that, for the class of distance-based aggregators, the satisfying of unanimity is due to the function $d$ satisfying the identity of the indiscernible property and the function $\eta$ satisfying the property (iii).

\paragraph{Anonymity.} The last property in this class that we will consider is {\em anonymity}. An aggregator is anonymous if the choice of collective judgment sets does not depend on the order in which the agents are listed in the profile.  Let $\Pf$ and $\Pf'$ be two profiles that differ only on the order in which the agents in them are listed. For an anonymous $F$, it must hold that $F(\Pf) = F(\Pf')$. The desirability of anonymity is not quite so without reproach as unanimity. Assume that the agents are experts, with some experts being better at making judgments than others. If we want that  the better experts have more influence on what the collective judgments are, then $F$ should not be anonymous. Non-anonymous aggregators are rarely discussed and designed \cite{AAMAS-1-12}.

\subsection{Properties of resolute aggregators}
A resolute aggregator is a function that matches a profile of judgment sets to a single rational collective judgment set. Early on it was shown, in works such as \cite{Dietrich07},  that some basic sets of properties cannot be simultaneously satisfied by one aggregator. Some of these conditions we take as so indispensable that we have here imbedded them in the definition of an aggregator: that the collective judgment set(s) must be both complete and a consistent agenda subset. Others we will now present and discuss.   It is not our goal in this Section to  give a detailed overview of impossibility results, the interested reader can consult \cite{GrossiP:2014} as a   very comprehensive source of information on this topic. Our goal is rather to introduce the reader to the existence of these properties for purposes of completeness and better understanding of the more general  and,  from the view-point of multi agent systems, more practical  properties of irresolute aggregators. 


\subsubsection{Unanimity Preservation, Systematicity}

The first set of resolute property we present are featured in the first works in judgment aggregation showing, together with rationality, anonymity and non-dictatorship,  the impossibility of  existence for an aggregator to satisfy all properties. We begin with {\em unanimity preservation}, which is often simply called unanimity. 
 Intuitively what the unanimity tries protect are the issue-unanimities, namely, when agents all agree on the judgment of one issue, than that judgment must be in the collective judgment set, regardless of what the judgments on the other issues are. At first glance, this property is desirable, however, arguments can be found that question this desirability. We give an illustrative example. 
 
 \begin{example}\label{ex:unp} Consider the judgment aggregation problem with an agenda with pre-agenda $\A^{+}=\{ p, q, r, s, t\}$, constraints $\Ct = \{ t\leftrightarrow (p \vee q \vee r \vee s)\}$, and the profile $\Pf$ given in Table~\ref{tab:unp}. The profile   $\Pf$ has unanimity both on judgment $\neg p$ and on judgment $t$. For a (resolute) function that satisfies the unanimity preservation, it must be the case that $\{ \neg p, t\} \subseteq F(\Pf)$.   However, the first issue is not the same as the last issue, in terms of how logically"entrenched" the judgments on each issue are with respect to the judgments on the rest of the issues in the agenda. 
  
 \begin{table}[h!]
 \centering
 \begin{tabular}{r|ccccc}\\
 Voters   & $p$ & $q$ & $r$ & $s$ & $t$\\\hline
\multicolumn{6}{c}{      $\Ct = \{ t\leftrightarrow (p \vee q \vee r \vee s)\}$}\\\hline
 $\Js_1$ & -      &   +   &-      &-       & + \\
  $\Js_2$ & -      &   -   &+      &-       & + \\
   $\Js_3$ & -      &   -   &-      &+      & + \\ 
 \end{tabular}
 \caption{Profiles illustrating the unanimity preservation} \label{tab:unp}
 \end{table}
 
 Were the first four issues in the agenda premises, and the last issue a conclusion, the  premise based procedure would yield $\textrm{PBP}(\Pf) = \{\neg p, \neg q, \neg r, \neg s, \neg t\}$, observing the unanimity preservation for $\neg p$, but violating it for $t$. Since the unanimity preservation requires that unanimity is preserved on every issue, $\textrm{PBP}$ fails to satisfy this property. 
 \end{example}

\paragraph{Unanimity preservation} A resolute aggregator $F$ satisfies the unanimity preservation if and only if, for every $\ai \in \A$ and every $\Pf \in \Dmc(\A,\Ct)^n$, it holds:  if $\ai \in \Js_i$ for every $\Js_i \in \Pf$, then $\ai \in \F(\Pf)$. The judgment $\ai$ is called a {\em unanimously supported judgment}.

The next property we consider is actually comprised of two properties: neutrality and independence of irrelevant alternatives. 

The independence of irrelevant alternatives is satisfied when the collective judgment on each issue is obtained by aggregating (only) the individual judgments on that issue and no other judgments have a weigh in on that judgment. Formally we can express this requirement as follows. 

\paragraph{Independence of Irrelevant Alternatives (IIA).}  A resolute aggregator $\F$ satisfies IIA if and only if, for any two profiles $\Pf_1, \Pf_2 \in \Dmc(\A,\Ct) $, and any issue $\{ \ai, \neg \ai\} \in \A$  such that $\Pf_1^{\downarrow \{ \ai, \neg \ai\} } = \Pf_2^{\downarrow \{ \ai, \neg \ai\} }$ it holds:   $\ai \in \F(\Pf_1)$ if and only if $\ai \in \F(\Pf_2)$. 

The IIA property is very strong, and certainly of questionable desirability, since it requires that aggregation of judgments on an issue, logically related to judgments on other agenda issues, should not depend on those other judgments. This is expressed in the requirement that when two profiles have exactly the same agents giving exactly the same judgments on one issue, \ie~ $\Pf_1^{\downarrow \{ \ai, \neg \ai\} } = \Pf_2^{\downarrow \{ \ai, \neg \ai\} }$, should select exactly the same collective judgments for this issue, regardless of how much or little the profiles differ on the rest of the judgments. 

Because IIA is intuitively such a strong requirement, a weakening was proposed in \cite{Mongin2008} called {\em Independence of Irrelevant Propositional Alternatives (IIPA)}. Instead of requiring that the aggregator satisfies the independence on aggregation on every agenda issue, the IIPA requires that independence only on those issues that are propositional variables. The  IIPA property is only defined for agendas closed under propositional variables in the logic framework with $\Ct = \{\top\}$. We will omit the formal definition of this property. 

The second part of systematicity is the property of neutrality. While anonymity is concerned with the insensitivity of the aggregator to the order of of agents in the profile, neutrality is concerned with the insensitivity of the aggregator to the order, or rather "name", "type", "nature" of the issues in the agenda.  Neutrality was considered as simple and weak requirement in the early work in judgment aggregation, so it was not formalised. We find two formalisations, or versions,  of neutrality in \cite{Grandi2013} called {\em Issue-Neutrality} and {\em Domain-Neutrality} defined for aggregators in the binary framework. 
Intuitively the neutrality properties are  concerned with profiles in which two issues have the same "pattern" of judgments.  The issue-neutrality is concerned with the "positive pattern", when all agents in the profile either accept or reject both issues. The domain-neutrality is concerned with the "negative pattern", when all agents in the profile reject the second issue when they accept the first and vice-versa. In the case of the "positive pattern" a issue-neutral aggregator would require that the same "pattern"is observed in the collective judgment set, while a domain-neutral aggregator would require the same in the case of the "negative pattern". We give an example in the logic framework.  

\begin{example}\label{ex:neut} Consider the agenda with pre-agenda $\A^{+} = \{ p \wedge r, p\wedge s, q, p\wedge q, t\}$ with $\Ct = \{\top\}$, so again the same one of Example~\ref{ex:scw}. Consider the profiles $\Pf_1$ (on the top side) and $\Pf_2$ (on the bottom side), given in Table~\ref{tab:neut}. 

 \begin{table}[h!]
 \centering
 \begin{tabular}{r|ccccc}\\
Agents   & $p\wedge r$ & $p\wedge s$ & $q$ & $p\wedge q$ & $t$\\\hline
 \multicolumn{6}{c}{      $\Ct = \{\top\}$}\\\hline
 $\Js_1$ & -      &   +   &+     &+       & - \\
  $\Js_2$ & +      &   -   &-      &-       & + \\
   $\Js_3$ & +      &   -   &+    &+      & + \\ 
 \end{tabular}
   \begin{tabular}{r|ccccc}\\
Agents   & $p\wedge r$ & $p\wedge s$ & $q$ & $p\wedge q$ & $t$\\\hline
 \multicolumn{6}{c}{      $\Ct = \{\top\}$}\\\hline
 $\Js_1$ & -      &   +   &+     &+       & + \\
  $\Js_2$ & +      &   -   &-      &-       & - \\
   $\Js_3$ & +      &   -   &+    &+      & - \\ 
  \end{tabular}
 \caption{Profiles $\Pf_1$ and $\Pf_2$ used to illustrating the issue- and domain-neutrality principles.} \label{tab:neut}
 \end{table}
 
 In the profile $\Pf_1$ (on the top side)  we observe a "positive pattern":   for every agent $\Js_i \in \Pf_1$ we have that $p \wedge r \in \Js_i$ if and only if $t \in \Js_i$.   In the profile $\Pf_2$ (on the lbottom side)  we observe a "negative pattern":   for every agent $\Js_i \in \Pf_1$ we have that $p \wedge r \in \Js_i$ if and only if $\neg t \in \Js_i$. 
\end{example}

We give the formal definitions of the issue- and domain-neutrality properties in the logic framework. Let us state the following three conditions for for any two issues $\{\ai, \neg \ai \} \subset \A$ and  $\{\psi, \neg \psi \} \subset \A$ and every profile $\Pf \in \Dmc(\A, \Ct)^n$: 
\begin{itemize}
\item[A.]  $\ai \in \Js_i$ if and only if $\psi \in \Js_i$.
\item[B.] $\ai \in \F(\Pf)$ if and only if $\psi \in \F(\Pf)$.
\item[C.]  $\ai \in \Js_i$ if and only if  $\neg\psi \in \Js_i$.
\item[D.]  $\ai \in \F(\Pf)$ if and only if $\neg \psi \in \F(\Pf)$.
\end{itemize} 

\paragraph{Issue-Neutrality} An aggregator $\F$ satisfies issue-neutrality if and only if   conditions $A$ and $B$  hold.  \paragraph{Domain-Neutrality} An aggregator $\F$ satisfies domain-neutrality if and only if   conditions $C$ and $D$  hold. 

Already from Example~\ref{ex:neut} we can observe that when the neutrality is not quite as "weak" property as it may appear. Namely in Example~\ref{ex:neut} we can observe the "patterns" in the profile can occur between two issues that are not logically related. These "patterns" can also occur between issues that are very strongly logically related, such as judgments $\ai$ and $\psi$ for example for which  the constraint $\ai \leftrightarrow \psi$ holds. The intuition behind neutrality that the "name" of the issue does not matter only what judgments were cast for it, is weakened in the presence of logical relations. 

The version of neutrality  considered to build systematicity is the issue-neutrality \cite{GrossiP:2014}. This version can also be considered a "weaker" requirement because it does not "insist" that particular collective judgments are selective as collective, just that they have to be the same. Systematicity has very early been cast-off the sphere of interest as a very strong requirement for aggregators. 

 \paragraph{Systematicity} An aggregator satisfies systematicity if and only if it satisfies independence of irrelevant alternatives and issue-neutrality. 
 
\begin{figure}[h!]
    \begin{center}
        \includegraphics[width=\textwidth]{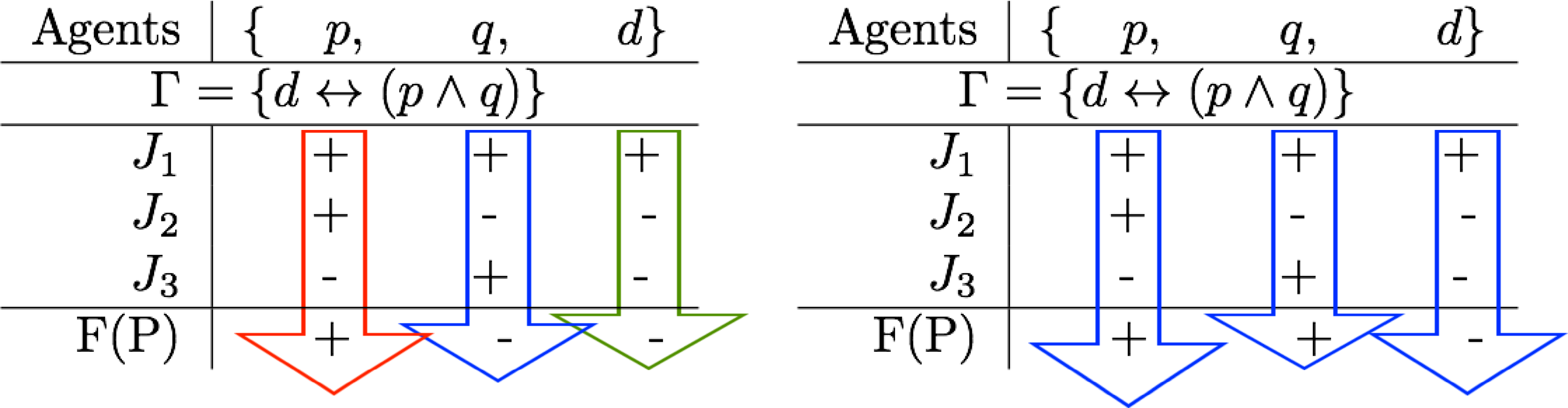} 
    \end{center}
\caption{Illustrating the difference between the properties of indipendence of irrelevant alternatives (left-hand side) and systematicity (right-hand side).}\label{fig:IIA}
\end{figure}

 Intuitively, while independence of irrelevant alternatives stipulates that the judgments on each issue have to be aggregated separately from the others (illustration Figure~\ref{fig:IIA} left-hand side table), systematically requires that this issue-based aggregation is done in the same way  (illustration Figure~\ref{fig:IIA} right-hand side table).
 
It was shown that the systematicity, unanimity preservation, rationality, and non-dictatorship yield an impossibility result, see in \cite{Dietrich07} . Impossibility results regarding the combination of unanimity preservation, rationality, independence of irrelevant alternatives and non-dictatorship can be found in \cite{Dietrich07,Dokow2010a}.

\subsubsection{Monotonicity}\label{sec:resmon}
The next set of properties we consider are {\em monotonicity} properties. There are more than one version of this property defined in the literature. Intuitively the  monotonicity property states that if a judgment is assigned as a collective judgment   for one profile  it should also be a assigned as collective for any profile in which there is even more support for that judgment. Sometimes the intuition is given as the paradigm:  an increase in support for an issue should not damage that issues chances of being collectively supported. Depending on whether the increase in support is for one judgment or a whole judgment set, different notions of monotonicity can be defined. 

Monotonicity over whole judgment sets is defined in \cite{Rationality2008}. We state the definition here and call it {\em set monotonicity} to differentiate it from the other monotonicity properties. 

\paragraph{Set monotonicity.} A resolute aggregator $\F$ is set monotonic if and only if for every two profiles $\Pf, \Pf' \in \Dmc(\A, \Ct)^n$, with \linebreak $\Pf = (\Js_1, \ldots,\Js_i, \ldots,  \Js_n)$ and $\Pf' = (\Js_1, \ldots,\Js, \ldots,  \Js_n)$, if $\F(\Pf) = \Js$, then $\F(\Pf') = \Js$. 

Set monotonicity guarantees that if an agent  $\Js_i$ replaces her judgment set with the collective judgment set $\Js$, and no other change is made in the profile, then $\Js$ continues to be the collective judgment set. In contrast to strengthening the support for a whole judgment set,  we can consider the case when support is strengthen for a particular judgment. This gives us an another type of monotonicity, defined in  \cite{DietrichList2005}, which we here call {\em issue monotonicity}. 
If one agent in the profile changes only one of her judgments from $\neg \ai$ to $\ai$, then if $\ai$ was in the collective judgment set for the old profile, it should remain in the collective judgment set in the new profile. 

We first define the concept of {\em  $\ai$-strengthening of a profile}. A profile $\Pf \in \Dmc(\A, \Ct)^n$ is an $\ai$-strengthening of profile $\Pf' \in \Dmc(\A, \Ct)^n$, $\ai \in \A$,  if \linebreak $\Pf = (\Js_1, \ldots, \Js_i, \ldots, \Js_n)$ and $\Pf' = (\Js_1, \ldots, \Js'_i, \ldots, \Js_n)$  with $\Js_i \setminus \Js'_i = \ai$. 

Intuitively, a profile is  $\ai$-strengthening of another profile if it differs from it on exactly one judgment in exactly one judgment set. Example~\ref{ex:strengthening} gives two such profiles.

\begin{example}\label{ex:strengthening} Consider the judgment aggregation problem of the "doctrinal paradox", Example~\ref{ex:doctrinal}. The profile  $\Pf$ (left-hand side on Table~\ref{tab:strengthening}) is the  "doctrinal paradox" profile. The profile $\Pf'$ (reft-hand side on Table~\ref{tab:strengthening}) is an $\neg p$-strengthening of $\Pf$ because it differs from $\Pf$ only on one judgment set, that is $\Js_3$ and $\Js'_3$ and $q\in \Js_3$, while $\neg q \in \Js'_3$.
\begin{table}[h!]
\centering
\begin{minipage}{0.45\textwidth}
\begin{tabular}{r|ccc}
Agents & $p$ & $q$ & $d$ \\ \hline
 \multicolumn{4}{c}{$\Ct=\{ (p\wedge q) \leftrightarrow d\}$}\\\hline
 $\Js_1$ & + & + & + \\
 $\Js_2$ & + & - & - \\
  $\Js_3$ & - & + & - \\
\end{tabular}
\end{minipage}
\begin{minipage}{0.45\textwidth}
\begin{tabular}{r|ccc}
Agents & $p$ & $q$ & $d$ \\ \hline
 \multicolumn{4}{c}{$\Ct=\{ (p\wedge q) \leftrightarrow d\}$}\\\hline
 $\Js_1$ & + & + & + \\
 $\Js_2$ & + & - & - \\
  $\Js'_3$ & - & - & - \\
\end{tabular}
\end{minipage}
\caption{Profile $\Pf'$ on the right-hand side is a $\neg p$-strengthening of a profile $\Pf$ on the left-hand side. }\label{tab:strengthening}
\end{table}
\end{example} 

Clearly, not all profiles will have their $\ai$-strengthening for every issue  $\ai $ in the agenda, and some might not even have an  $\ai$-strengthening for any agenda issue $\ai$. The monotonicity properties defined around the idea of $\ai$-strengthening of a profile apply their requirements only on profiles that do have an  $\ai$-strengthening profile.

\paragraph{Issue monotonicity.} A resolute aggregator $\F$ is issue monotonic if for every profile $\Pf'$ and its  $\ai$-strengthening profile $\Pf$, for any $\ai \in \A$, if $\ai \in \F(\Pf')$, then $\ai \in \F(\Pf)$. 

A weakening of issue monotonicity was also proposed in \cite{DietrichList2005}. Intuitively, while issue monotonicity requires that the strengthening of support of one judgment {\em never} "pushes out" that judgment from the collective judgment set, weak monotonicity requires that the same judgment should  {\em sometimes}  not be "pushed out". We give the formal definition. The difference in the text of the definitions is slight. 

\paragraph{Weak issue monotonicity.} A resolute aggregator $\F$ is issue monotonic if for every profile $\Pf'$ and its  $\ai$-strengthening profile $\Pf$, for any $\ai \in \A$, it is {\em not the case} that if $\ai \in \F(\Pf')$, then $\neg \ai \in \F(\Pf)$. 

Grandi and Endriss \cite{Grandi2013} define the issue monotonicity   for the binary framework and also define another version of monotonicity, which we call {\em neutral monotonicity} to distinguish it.  We give the definition here for the logic framework. Intuitively neutral monotonicity stipulates that if for some judgments $\ai, \ai' \in \A$, if the support for $\ai'$ in the profile is larger than the support for $\ai$ and $\ai$ is in the collective judgment set, then so should $\ai'$ be in it.   

\paragraph{Neutral monotonicity.} A resolute aggregator $\F$ is neutral monotonic if and only if for every $\Pf \in \Dmc(\A, \Ct)^n$  and any two $\ai, \psi \in \A$, if \linebreak $N(\ai, \Pf) < N(\psi, \Pf)$ and $\ai \in \F(\Pf)$, then $\psi \in \F(\Pf)$. 

Monotonicity properties are desirable because, each in their own way,  ensure  that the aggregator is responsive to the individual judgment sets. Another source of desirability of (weak) issue monotonicity, in particular, is that  it is necessary to ensure  that an aggregator si {\em non-manipulable}\cite{DietrichList2005}.   We discuss briefly the non-manipulability condition. Grossi and Pigozzi devote a section in \cite{GrossiP:2014} to this topic (Chapter 5).

\subsubsection{Manipulability}

Assume that an agent  has an incentive that a certain judgment, or judgment set, is selected among, or as,  the collective judgment sets. An agent {\em manipulates}, when she does not include that certain judgment, but its opposite in her judgment set, or when she does not "declare" her most "desirable" judgment set. An aggregator is non-manipulable when no agent can get a more desirable collective judgment (set)  by manipulating.   
Dietrich and List show in  \cite{DietrichList2005} that an aggregator is non-manipulable when it satisfies independence of irrelevant alternatives and at least weak monotonicity.  They also show that for a path-connected agenda an aggregator does not exist that satisfies universal domain, collective rationality, responsiveness, non-dictatorship and non-manipulability.

Non-manipulability is a desirable conditions, as it ensures the "fairness" of the aggregator. Studies of manipulability is where game theory and judgment aggregation intersect. In voting, or preference aggregation problems, the "incentives" are built into the aggregation problem - the agent is represented with a preference order over the set of alternatives. With judgment aggregation problems this is not the case. Dietrich and List suggest that a distance function can be used in lieu of incentives, every agent's judgment sets is his most preferred one and the closer a rational judgment set is to that most desirable judgment set, the happier the agent is with it \cite{DietrichList2005}. 

At this point we would like to give some pointers to the literature in computational social choice that studies  various complexity problems encountered under  manipulability under different contexts.  

In \cite{EndrissGP12} the Hamming distance is used to model the incentives for manipulation and the complexity of manipulability for the premise based procedure is considered. 

 In \cite{Baumeister2015} issues of manipulability and bribery are considered, also for  Hamming distance incentives, for the uniform quota rules and the premise based procedure. Intuitively, bribery is the problem of finding the minimal number of agents to change their judgment set in some minimal fashion so that a particular judgment (set) is selected (called {\em constructive manipulation}) or a particular particular judgment (set) is not selected (called {\em destructive manipulation}).

In  \cite{BotanEtAlAAMAS2016} the question of group manipulation is studied, again under Hamming distance constructed preferences, for (issue) neutral aggregators. Intuitively, group manipulation is the manipulation version in which the agents do not manipulate on their own, but together with other agents. 

In contrast to the work we so far mentioned, the context of interest in \cite{AlonFMT13} is when the "manipulator" is not one of the agents but the "designer" of the agenda, the {\em agenda setter}, and {\em bundling attack}.  Intuitively bundling attacks are manipulations  by changing the agenda, attempts to get a particular judgment, or judgment set set, be in, or out of,  the collective judgment set, by combining several issues into one.  In the judgment aggregation problem the agenda is given as-is. To study these types of manipulation problems, we must model also the phase in which the agenda is set.  Alon {\em et al. } consider the special case when the agenda is simple, \ie~ no issue is  logically dependent on any of the other issues. 

Lastly we mention the \cite{Dietrich2016} article that also studies the manipulability problem for the agenda setter from the social choice theoretic aspect, showing new impossibility results in this context.  Dietrich studies the question of {\em agenda sensitivity}, whether the collective judgment sets can be controlled by changing the agenda, intuitively explained as in some logically equivalent way. Examples of logically equivalent ways to change the agenda are when  some of the agenda  issues are replaced by some logically equivalent formulas, or when the issues are  replaced with a conjunction of issues.

As even this brief overview of studies on manipulability intimates, these problems are many and widely unexplored in computational judgment aggregation, but are of interest in contexts of competitive self-interested agents, but not in contexts in which the agents are for example experts that merely contribute their opinions but are individually unaffected by the collective judgment set.  


\subsection{Properties of irresolute aggregators}

In this section we consider properties of irresolute aggregators.   The purpose of defining these properties is to be able to differentiate between the aggregators and match them with particular problem contexts. Theres exists one result in which  irresolute aggregator properties are used to give a characterisation of an aggregator, and we give an overview of that result here as well. 

\subsubsection{Generalising properties}
To define the when  an irresolute aggregator property  generalises a resolute aggregator property we first observe that for every  irresolute aggregator $F$ there exist a subset of  profiles, possibly empty,  for which $F$ is resolute. 

Recall that an aggregator $\F$ that satisfies universal domain is  defined\footnote{Recall that $ \mathcal{P}^{\ast}(\Dmc(\A,\Ct))$ denotes the power set of $\Dmc(\A,\Ct)$ excluding the empty set.}  as $F: \Dmc(\A, \Ct)^n \rightarrow \mathcal{P}^{\ast}(\Dmc(\A,\Ct))$, there exists a set $S_{\F} \subseteq  \Dmc(\A, \Ct)^n $, such that for every $\Pf \in S$, $\F(\Pf)$ is a singleton sit. We call this set $S_{\F}$  {\em  the resolute domain restriction for $\F$}.   

An irresolute aggregator property $\textrm{IProp}_{\F}$ generalises a resolute aggregator property $\textrm{RProp}_{\F}$ if an only if the following relation hods: if $\F$ satisfies $\textrm{RProp}_{\F}$ for every profile in the resolute domain restriction $S_{\F}$, then  $\F$ satisfies $\textrm{IProp}_{\F}$ for every profile in   $S_{\F}$. 
%
%

\subsubsection{Unanimity preservation}\label{sec:una}

We begin with the property of unanimity preservation. 
Typically,  there is always more than one way to generalise a resolute aggregator property into an irresolute one. The two most intuitive generalisations of the resolute unanimity preservation property are  to require that the unanimously supported judgment  is in {\em all} of the collective judgment sets, which we call the {\em strong unanimity principle} or to require that this judgment is in {\em at least one} of the collective judgment sets, which we call the {\em weak unanimity principle}.  We give an example from \cite{LangPSTV15}.

\begin{example} Consider the pre-agenda $\A^+=\{ p, p\rightarrow (q \vee r), q, r, p \rightarrow (s \vee t), s,t,p \rightarrow (u \vee v), u,v\}$ with constraitns $\Ct=\{\top\}$. Consider the profile  $\Pf$ given in Table~\ref{tab:unap}. The collective judgment sets of $\RMCSA(\Pf)$ and $\RRA(\Pf)$ are indicated bellow the profile in Table~\ref{tab:unap} as well. 

\begin{table}[h!]
\centering
\begin{tabular}{r|cccccccccc}
{\small Agents}& $p$ & $p \rightarrow (q \vee r)$ & $q$ & $r$ & $p \rightarrow (s \vee t)$ & $s$ & $t$ & $p \rightarrow (u \vee v)$ & $u$ & $v$ \\\hline
 \multicolumn{11}{c}{      $\Ct = \{\top\}$}\\\hline
$\Js_1$&+ & + & + & - & + & + & - & + & + & -\\
$\Js_2 $&+ & + & - & + & + & - & + & + & - & +\\
$\Js_3$&+ & -   & - & -  & -  & - & -   & -  & - & - \\ \hline
m(\Pf) & + & + & - & -  & + & - & - & +  & - & - \\ 
\hline\hline
$\RMCSA$& - & + & - & -  & + & - & - & +  & - & - \\ \hline
$\RRA$& + & + &+ & +  & + & + & + & +  & + & + \\  
               & + & + &+ & -  & + & + & - & +  & + & - \\ 
               & + & + &+ & -  & + & + & - & +  & - & +\\ 
               & + & + &+ & -  & + & - & + & +  & + & - \\
               & + & + &+ & -  & + & - & + & +  & - & + \\             
               & + & + &- & +  & + & + & - & +  & + & - \\ 
               & + & + &- & +  & + & + & - & +  & - & +\\ 
               & + & + &- & +  & + & - & + & +  & + & - \\
               & + & + &- & + & + & - & + & +  & - & + \\
               & + & - &- & -  & - & - & - & -  & - & - \\    
\end{tabular} \caption{A profile showing how  various aggregators "behave" with respect to the unanimity principle.} \label{tab:unap}
\end{table}
There is one unanimously supported judgment in $\Pf$, that is the judgment $p$. Observe that this judgment is in any of   judgment set in $\RMCSA(\Pf)$, which for this profile is just one set,  but it is included in every judgment set in $\RRA(\Pf)$.  The collective judgment sets $\RMSA(\Pf)$ for this particular profile, are such that $\RMSA(\Pf) = \RMCSA(\Pf) \cup \RRA(\Pf)$, so the unanimously supported judgment is in at least one collective judgment set among the judgment sets $\RMSA(\Pf)$.
\end{example}
We give the formal definitions of weak and strong unanimity principle that differ only on how many of the collective judgment sets for a profile must contain the unanimously supported judgment, if such a judgment exists in the profile. 

\paragraph{Weak Unanimity Principle.}An aggregator $\F$ satisfies the weak unanimity principle when for every $\Pf \in \Dmc(\A,\Ct)$ and any $\ai \in \A$: if $\ai \in \Js_i$ for all $\Js_i \in \Pf$, then there exists at least one $\Js \in \F(\Pf)$ such that $\ai \in \Js$.
 
\paragraph{Strong Unanimity Principle} An aggregator $\F$ satisfies the weak unanimity principle when for every $\Pf \in \Dmc(\A,\Ct)$ and any $\ai \in \A$: if $\ai \in \Js_i$ for all $\Js_i \in \Pf$, then for every $\Js \in \F(\Pf)$ such that $\ai \in \Js$.

It is obvious that if an aggregator $\F$ satisfies the strong unanimity principle, it will also necessarily satisfy the weak unanimity principle. The reverse does not holds. If $\F$ satisfies either the weak or the strong unanimity principle, then it will also satisfy the resolute unanimity principle for every  profile from the   resolute domain restriction \linebreak $S_{\F} \subseteq \Dmc(\A,\Ct)$. 

The strong and weak unanimity principles were defined in \cite{ADT09,TARK11}. In Figure~\ref{tab:overview}, we give the known results regarding which aggregators satisfy or fail to satisfy the unanimity principles.  These results were given in \cite{LangPSTV15}, where proofs for each result are given.   Figure~\ref{tab:overview} contains the overview of which aggregators satisfy which of the rest of the irresolute aggregator properties we introduce. 

\subsubsection{Monotonicity} 

Monotonicity as a property for irresolute aggregators was first considered in \cite{LangPSTV15} that generalises the resolute issue monotonicity defined in \cite{DietrichList2005}. Intuitively, irresolute monotonicity "guarantees" that if a judgment is in all collective judgment sets for a profile, then it should remain to be so when the support for that judgment increases in the profile {\em ceteris paribus}.  The intuition behind the desirability of the monotonicity property does not change - no judgment's selection as collective should be hurt by increased support. 

We already gave four definitions  of resolute monotonicity and each of them can be generalised into an irresolute aggregator property. The one generalised in  \cite{LangPSTV15}  is the issue monotonicity and it is but one way in which to   issue monotonicity could be generalised. The requirements of this generalisation are as follows.  For a pair of profiles in which one is the $\ai$-strengthening of the other, if $\ai$ is in all collective judgments for the profile, it should be in all the collective judgment sets of that profile's $\ai$-strengthening. 
 
The  {\em ceteris paribus} notion is captured by considering profiles in which one is a $\ai$-strengthening of the other, as defined in Section~\ref{sec:resmon}. We give the formal definition. 

An aggregator $\F$ satisfies (irresolute) monotonicity if and only of for every profile $\Pf \in \Dmc(\A,\Ct)^n$ and every profile $\Pf'$ that is an $\ai$-strengthening of $\Pf$ for some $\ai \in \A$ it holds: if $\ai \in \Js$ for every $\Js \in \F(\Pf)$, then  $\ai \in \Js'$ for every $\Js' \in \F(\Pf')$.
 
It can be directly observed that if an aggregator satisfies irresolute monotonicity, then it will satisfy issue monotonicity on the   resolute domain restriction for that aggregator. 
Of course, one can define a monotonicity property  requiring that for the pair of profiles, if $\ai$ is in at least one of the collective judgment sets for the first profile, it should also be in at least one of the collective judgment sets for that profiles $\ai$-strengthening, but to really capture monotonicity under such a definition perhaps it would better to require that the number of collective judgment sets containing $\ai$  does not decrease. Both of these versions of monotonicity would be weaker, \ie~implied by the monotonicity property defined in  \cite{LangPSTV15}.

\subsubsection{Independence properties}

The independence of irrelevant alternatives (IIA) has been criticised as being somewhat unnatural for judgment aggregation \cite{BR06} because it effectively requires that the aggregations of judgments are   issue by issue and do  not inter-influence each-other, but  since the issues in the agenda are logically related, this effectively also means disregarding the logic relations among the issues. The weakening of the IIA requirement that respects the logic relations among issues was defined only recently in \cite{AAAI} and it is called {\em agenda separability}. 

\paragraph{Agenda Separability}
 Recall that among the agenda properties we presented in   Section~\ref{sec:agprop}, we considered independent partitions and syntactical  independent partitions of an agenda, which were two ways to express that the agenda can be split into two groups of issues that are logically independent.   
Intuitively under agenda separability we would  require that judgments on  one group of issues do not influence the collective judgments selection on another group of issues, but only when the first group is not logically related to the second and not indiscriminately as was the case under IIA.  Another way to look at the intuition behind agenda separability  is  to consider two agendas $\A_1$ and $\A_2$ who can be considered independent partitions of their union. Aggregating the profiles of such two agendas separately, or "lumped together" as one profile for $\A_1 \cup \A_2$, should not matter. Observe that, if $\A$ has two independed partitions $\A_1$ and $\A_2$, then necessarily $\Ct$ can also be partitioned into two, possibly empty, sets of formulas $\Ct_1$ and $\Ct_2$ that do not share variables.

\begin{example} Consider the  following judgment aggregation problem, given in \cite{CostantiniEtAlAAAI2016}. Party goers need to choose one drink and one snack.    The options are between beer ($b$) and champagne ($c$) for drinks, and potato crisps ($p$) and caviar ($ k$)  for snacks. The  agenda and constraints\footnote{The operator $\veebar$ is the logic operator exclusive or, defined as $\ai_1 \veebar \ai_2 \equiv (\ai_1 \wedge \neg \ai_2) \vee (\neg \ai_1 \wedge \ai_2)$, namely either $\ai_1$ is true or $\ai_2$ is true but they cannot both be true nor they can both be false at the same time.} are \linebreak $\A  = \{ b, \neg b, c, \neg c, p, \neg p, k, \neg k\}$ and  $\Ct=\{  b \veebar c,    k \veebar   p\}$  respectively. The profile as given in Table~\ref{tab:coctail}. 
\begin{table}[h!]
\centering
\begin{tabular}{r|cccc}
Agents& $b$ & $c$ & $p$ & $k$\\\hline
\multicolumn{5}{c}{      $\Ct = \{b \veebar c,    k \veebar   p\}$}\\\hline
$\Js_1-\Js_{11}$      & + & - & + & - \\ 
$\Js_{12}-\Js_{21}$  & - & + & - & + \\ 
$\Js_{22} - \Js_{23}$&+ & - & - &+\\\hline
m(\Pf) & + &-&-&+
\end{tabular}
\caption{An example of an agenda that has independent partitions.}\label{tab:coctail}
\end{table}

This agenda $\A$ can be partitioned into two independent partitions  $\A_1=\{  b, \neg b, c, \neg c\}$ and $\A_2=\{ p, \neg p, k, \neg k\}$; intuitively, these are the drink part and the snack part. We have that $\Dmc(\A_1, \Ct) = \{ \{ b, \neg c\}, \{ \neg b, c\}\}$ and $\Dmc(\A_2, \Ct) = \{ \{ p, \neg k\}, \{ \neg p, k\}\}$. The union of any set from $\Dmc(\A_1, \Ct)$ with any set from $\Dmc(\A_2, \Ct)$ is consistent.
\end{example}

We give the formal definition.  An aggregator $\F$ satisfies agenda separability if and only if for all independent partitions $\A_1$ and $\A_2$ of an agenda $\A$ , and all profiles $\Pf \in \Dmc(\A,\Ct)^n$, we have  
\begin{equation}\F(\Pf) =  \{ \Js^1 \cup \Js^2  ~|~ 	\Js^1 \in \F(\Pf^{\SA_1}) \mbox{ and } \Js^2 \in \F(\Pf^{\SA_2}) \}.\end{equation}

We give an example of an aggregator, that does not satisfy  agenda separability. 

\begin{example}\label{ex:asep} Consider the judgment aggregation problem and aggregator $\RY$ from Example~\ref{ex:y}. The profile is given in Table~\ref{tab:y1}. For this profile and the aggregator $\RY$ we obtain that
\[\RY(\Pf) = \textrm{ext}(\{ q, \neg (p \wedge q) \})  =\left\{
\begin{array}{rrrrr}
 \{\neg (p\wedge r), & \neg (p\wedge s), & q,  &   \neg (p\wedge q), & t\},       \\
 \{\neg (p\wedge r), & \neg (p\wedge s), & q,  &   \neg (p\wedge q), & \neg t\}
\end{array}
\right\}.
\]
The agenda from this problem has two independent partitions $\A_1 = \{ p\wedge r, \neg(p\wedge r), p\wedge s, \neg (p\wedge s), q, \neg q, p \wedge q, \neg (p \wedge q)\}$ and $\A_2 = \{t, \neg t\}$. Table~\ref{tav:sepy} gives the profiles $\Pf^{\downarrow \A_1}$ (top side) and $\Pf^{\downarrow \A_2}$ (bottom side) and respective aggregations with the $\RY$ aggregator. Since the profile $\Pf^{\downarrow \A_2}$  is majority-consistent, we cannot obtain the set  $ \{\neg (p\wedge r),   \neg (p\wedge s),   q,      \neg (p\wedge q),   \neg t\}$ from a union of  a $\Js^1 \in \RY(\Pf^{\downarrow \A_1})$ and a $\Js^2 \in \RY(\Pf^{\downarrow \A_2})$. 

\begin{table}[h!]
\centering
\begin{minipage}{0.7\textwidth}
\begin{tabular}{r|cccc}
 Agents       &\{ $\;\;\; p\wedge r $,& $\;\;\;  p \wedge s$, &  $\;\;\; q$, &   $\;\;p\wedge q$\} \\\hline
  \multicolumn{5}{c}{      $\Ct = \{\top\}$}\\\hline
  $\Js_1 -\Js_6$                 &  $\;\;\;\;$+           &  $\;\;\;$+             &  $\;\;\;$+   &  $\;$ +            \\
  $\Js_7 -\Js_{10}$            & $\;\;\;\;$+           & $\;\;\;$+             & $\;\;\;$ -   &  $\;$ -             \\
 $\Js_{11} -\Js_{17}$        & $\;\;\;\;$-           & $\;\;\;$-             &   $\;\;\;$+   &  $\;$ -               \\\hline
$\RY(\Pf^{\downarrow \A_1})$  & $\;\;\;$   & $\;\;\;\;$  & $\;\;\;\;$+ & $\;\;\;$-     \\  
\end{tabular}
\end{minipage}
 
\begin{minipage}{0.5\textwidth}
\begin{tabular}{r|c}
Agents & $t$   \\ \hline
 \multicolumn{2}{c}{$\Ct = \{\top\}$}\\\hline
   $\Js_1 -\Js_6$  & +   \\
  $\Js_7 -\Js_{10}$   & +   \\
 $\Js_{11} -\Js_{17}$ & -   \\\hline
$\RY(\Pf^{\downarrow \A_2})$&+     \\  
\end{tabular}
\end{minipage}
\caption{Profiles $\Pf^{\downarrow \A_1}$  and $\Pf^{\downarrow \A_2}$ exemplifying that $\RY$ does not satisfy agenda separability.  }\label{tav:sepy}
\end{table}
\end{example} 

Agenda separability is an irresolute aggregator property and it is not a generalisation of IIA in the same sense as the monotonicity and unanimity properties we already considered.  In  \cite{AAAI} it was shown that if an aggregator satisfies IIA for every profile on the   resolute domain restriction, then it will also satisfy agenda separability.

Agenda separability ensures a mild protections from manipulability agains an agenda setter: if an aggregator satisfies it, an agenda setter cannot manipulate the results of the aggregation by adding "unrelated" issues to the agenda. As we see in the  Example~\ref{ex:asep}, under the $\RY$ aggregator, an agenda setter that does not want a decision made on the truth-state of on issue, although a clear majority is of the opinion that the issue should be answered "yes", can accomplish this by appending this issue to an existing agenda.

\paragraph{Overlapping agenda separability}
A stronger independence property is considered in \cite{AAAI}, called {\em overlapping agenda separability} satisfied only (as far as we know) by two aggregators: $\RMSA$ and $\RRA$.  

Some agendas cannot be partitioned into independent partitions, but their level of mutual logical entrenchment is "localised". The agenda property of having {\em independent overlapping decompositions} was defined in \cite{AAAI} to capture this concept. Intuitively, an agenda $\A$ has independent overlapping decompositions $\A_1$ and $\A_2$ such that, if  $\A_1 \cap\A_2$ is removed from both $\A_1$ and $\A_2$, the issues in the resulting sub-agendas are no longer logically dependent on each other. We give the formal definition from \cite{AAAI}.

Let $\A$ be an agenda  
and let $\A = \A_1 \cup \A_2$ (but not necessarily $\A_1 \cap \A_2 = \emptyset$). We say that $\{\A_1, \A_2\}$ is an \emph{independent overlapping decomposition} (IOD) of $\A$ if and only if
for every $\Js^1 \in \Dmc(\A_1,\Ct)$, for every $\Js^2 \in \Dmc(\A_2,\Ct)$
\begin{equation}
\mbox{ if ~~} \Js^1 \cap \A_2 = \Js^2 \cap \A_1 \mbox{~~ then ~~} \Js^1 \cup \Js^2 \in \Dmc(\A,\Ct).
\end{equation}  

We give an example of a judgment aggregation problem with an agenda that has an independent overlapping decomposition. The example is also from \cite{AAAI}.

\begin{example} Consider the pre-agenda $\A ^+ = \{ p, p\rightarrow q, p \rightarrow r, q, r, s, s\rightarrow q, s \rightarrow r\}$ and $\Ct=\{\top\}$. This agenda $\A$ has an independent overlapping decomposition. The pre-agendas of the decompositions are:
\[ \A ^+_1 = \{ p, p\rightarrow q, p \rightarrow r, q, r \},\]
\[ \A ^+_2 = \{q, r, s, s\rightarrow q, s \rightarrow r\}.\]

Observe that, whenever a judgment set from $\Dmc(\A_1, \Ct)$ has the same judgments on issues $q$ and $r$ as a judgment set from $\Dmc(\A_2, \Ct)$, their union will be a consistent and complete judgment set from $\Dmc(\A, \Ct)$. 
\end{example}
 
 An aggregator $\F$ satisfies overlapping agenda separability when the aggregating the two profile $\Pf^{\downarrow \A_1}$ and $\Pf^{\downarrow \A_2}$ separately or together gives the same results, as long as the collective judgments on the issues from $\A_1 \cap \A_2$ are the same in every judgment set in $\F(\Pf^{\downarrow \A_1})$ and in $\F(\Pf^{\downarrow \A_2})$. We give the formal definition from \cite{AAAI}.

An aggregator $\F$ satisfies \emph{overlapping agenda separability} (OAS) if and only if for every agenda $\A$  
and every independent overlapping decomposition $\{\A_1, \A_2\}$ of $\A$, for every profile $\Pf$ over $\A$ it holds that:
\begin{equation}
\begin{array}{l}
\mbox{ if for every } \Js^1 \in \R(\rest{\Pf}{\A_1}), \mbox{ for every } \Js^2 \in \R(\rest{\Pf}{\A_2}):\Js^1 \cap \A_2 = \Js^2 \cap \A_1 \\
 \mbox{ then } \F(\Pf) =  \{ \Js^1 \cup \Js^2  ~|~ 	\Js^1 \in \F(\Pf^{\downarrow \A_1}) \mbox{ and } \Js^2 \in \F(\Pf^{\downarrow \A_2})\}.
\end{array}
\end{equation}

The desirability of the overlapping agenda separability property (OAS) is not quite as straightforward as with agenda separability. OAS is related to the idea of {\em forgetting} in knowledge representation and reasoning, see for example \cite{ZHANG2009}. Aggregators that satisfy OAS have a certain robustness. Say a judgment aggregation problem is aggregated and afterwards it was shown that not all issues in the agenda need collective judgments assigned to them. If these now irrelevant issues are an  independent overlapping decomposition of the agenda, then the aggregation results on the issues for which we still need collective judgments, are still valid. They would be the same if the partial profile only on these issues is aggregated separately. 
 
\subsubsection{Reinforcement and Homogeneity}
The last two irresolute aggregator properties we consider might was well have been defined for resolute aggregators.  These properties are of relevance in preference aggregation and voting theory studies, where they are used to attain characterisation results for aggregators, however, in judgment aggregation they were not  considered until  \cite{LangPSTV15} and \cite{CostantiniEtAlAAAI2016}.
An intuitive way to look at reinforcement and homogeneity is as   properties concerned with "horizontally partitioning the profile", whereas  agenda separability and unanimity  can be seen  as properties concerned with   "vertically partitioning the profile".  

\paragraph{Reinforcement.} If when aggregating two profiles  on the same agenda and constraints, some of the collective judgment for the first profile are the same as some of the collective judgment sets of the  second profile, then if we combine the two profiles, these "shared" judgment sets is what we should obtain as collective. An aggregator that ensures this property is satisfied is said to satisfy {\em reinforcement}. 

Recall the definition of $\Pf_1+\Pf_2$ operator from Section~\ref{sec:op}:
The sum of profiles  $\Pf_1$ and $\Pf_2$ where $\Pf_1 \in \Dmc(\A,\Ct)^{n_1}$, $\Pf_1=(\Js_1, \ldots, \Js_{n_1})$  and $ \Pf_2 \in \Dmc(\A,\Ct)^{n_2}$, $\Pf_2=(\Js'_1, \ldots, \Js'_{n_2})$  is a profile $\Pf = \Pf_1 + \Pf_2$, such that   $\Pf \in \Dmc(\A,\Ct)^n$, where $n = n_1 + n_2$ and $\Pf= (\Js_1, \ldots, \Js_{n_1}, \Js'_1, \ldots, \Js'_{n_2})$.

Formally, an aggregator $\F$ satisfies reinforcement if for any two profiles  $\Pf_1 \in \Dmc(\A,\Ct)^{n_1}$ and $\Pf_1=(\Js_1, \ldots, \Js_{n_1})$  such that\footnote{The profiles do not share agents but of course some judgment sets may be included in both, as different agents may have selected the same judgment set.} $\Pf_1 \sqcap \Pf_2 = \emptyset$, it holds that $\F(\Pf_1 + \Pf_2) = \F(\Pf_1) \cap \F(\Pf_2)$.

\paragraph{Homogeneity}  The homogeneity property is somewhat a special case of the reinforcement property. Instead of any two profiles being combined, we consider profiles containing the same judgment sets being added several times and how does this affect the aggregation results. An aggregator that satisfies homogeneity should not change the collective judgment sets assigned to a profile regardless of how many times the profile is duplicated. We give the formal definition.

Let us denote with $k\Pf = \underbrace{\Pf + \cdots + \Pf}_{k}$. An aggregator $\F$ satisfies homogeneity if and only if for every $\Pf\in \Dmc(\A,\Ct)^n$ it holds $\F(k\Pf) = \F(\Pf)$. 

It is not difficult to observe that if an aggregator satisfies reinforcement, then it will satisfy homogeneity. The reverse does not hold.

 \subsubsection{Overview of aggregator-property satisfaction}
\begin{figure}
   \includegraphics[width=\textwidth]{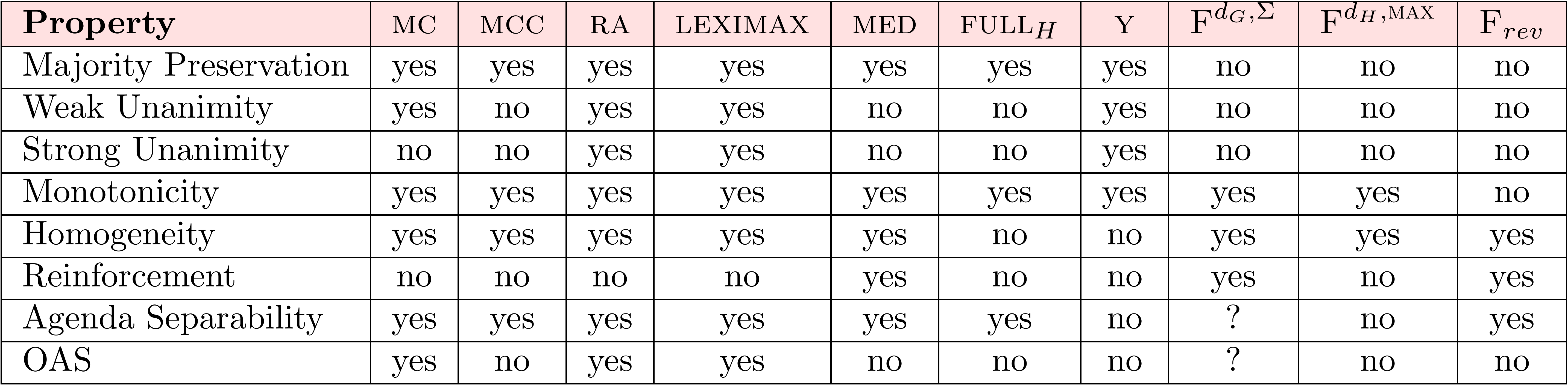}
 \caption{Known results and gaps of aggregators properties satisfaction.}\label{tab:overview}
 \end{figure}

 We now give an overview of what is known in the literature regarding which aggregator satisfies which properties.  The results are given in Figure~\ref{tab:overview}. The  "?" in a cell denotes that no result is as of yet known, "yes" clearly denotes that he aggregator satisfies the property, while "no" denotes that the aggregator does not satisfy the property.
 All aggregators satisfy anonymity. The results are from \cite{AAAI} regarding the agenda separability properties and from  \cite{LangPSTV15} regarding almost all else. The results regarding {\sc leximax} and the agenda separability properties follow from the property of $\textrm{\sc leximax} \subseteq \RRA$, namely that {\sc leximax} refines $\RRA$, and $\RRA$ satisfying both of these separability properties. 
 
 Furthermore, we know from \cite{CostantiniEtAlAAAI2016} that the whole class of binomial rules satisfies reinforcement, and consequently homogeneity, but fails majority-preservation. We do not know anything about the specific aggregators from this class satisfying the rest of the properties. The whole class of distance-based aggregators, as well as the whole class of scoring aggregators clearly fails majority-preservation as well. We also know from \cite{LangPSTV15} that the whole class of scoring aggregators satisfies reinforcement, and hence homogeneity. The class of distance-based aggregators $\F^{d,\max}$ satisfies homogeneity. 
 
 \newpage


\section{Relations with other aggregation methods} 
\label{ch:name2}  
Judgment aggregation is closely related to preference aggregation, and voting, as well as with belief merging. Out goal in this section is clarify the relations in order to make clear for which problems which aggregation theory is best suited. Judgment aggregation generalises preference aggregation. That means that a preference aggregation problem can be represented as a judgment aggregation problem.    The relation with belief merging is not straight forward. Both theories are concerned with aggregating truth values of formulas, however under belief merging there is no agenda. 



\subsection{Preference aggregation}\label{sec:pref}

Preference aggregation problems are the problems of choosing one option by aggregating individual  preferences over a collection of available options. The agents whose preferences are aggregated are also called {\em voters} and the act of casting individual choices is known as {\em voting}. The options are also called alternatives or candidates. The term candidates comes from the best known application of preference aggregation as a method of choosing a representative such as a president of a country. 

Preference aggregation is a much bigger research field, and much older, in comparison to judgment aggregation. Although discussions on how to elect a winner by voting can be traced back to antique Greece, it has been two documents that mark the begining of studying how individual choices should be aggregated in a fair way to elect a representative winner. There two documents are the "Sur les \'{E}lections au Scrutin" by Jean-Charles de Borda\footnote{The original text in french can be accessed here \url{http://gerardgreco.free.fr/IMG/pdf/MA_c_moire-Borda-1781.pdf}} published in 1781 and the 1785 "Essai sur l'Application de l'Analyse a la Probabilite des Decisions Rendues a la Pluralite des Voix"  by Nicolas de Condorcet\footnote{The original text in french can be accessed here \url{http://gallica.bnf.fr/ark:/12148/bpt6k417181}.}  It was Kenneth Arrow who axiomatised the preference aggregation theory and proved the basic impossibility theorems \cite{Arrow:63}. 

We give a basic introduction to preference aggregation theory and show how a preference aggregation problem can be represented as a judgment aggregation problem, as per the standard method \cite{Dietrich07}. We show when a judgment aggregator generalises a preference aggregator and give an overview of the known relations between the methods we represented as per \cite{ADT2013}. 

\subsubsection{Preference aggregation problem} 

A preference aggregation problem is defined by a set of options and a list of voters. The set of options, or candidates is $\Alt=\{o_1,\ldots,o_m\}$. A {\em vote} is a preference order $\succ$ over $\Alt$ that is total, strict and transitive. An order is a total order over $\Alt$  if and only if all the elements of $\Alt$ are ranked, it is   strict  if and only if for every $x,y\in \Alt$ either $x \succ y$ or $y \succ x$. An order is transitive if and only if for  every $x,y,z\in \Alt$ if $x \succ y$ and $y \succ z$, then necessarily $x \succ z$. Let $\Pmc$ be the set of all total, strict and transitive orders for $\Alt$. In general, votes can also be weak orders over $\Pmc$, when strictness is not required, however these are not representable within a judgment aggregation framework with the standard method. 

A voter is represented with an order $\succ_i \in \Pmc$. A profile of votes $\V = (\succ_1, \ldots, \succ_n)$ is a collection of voters $\V \in \Pmc^n$.

A preference aggregation problem is a pair $<\Alt, \V>$.  A solution to a preference aggregation problem is an order $\succ \in \Pmc$ that is representative of $\V$, we call it a {\em collective order}. A voting problem is also a pair $<\Alt, \V>$, but here the solution is one option, called {\em a winner} from $\Alt$ instead of a preference order over $\Alt$. Here, we will be concerned with voting problems.  Clearly, every voting problem can be represented as a preference aggregation problem, with the winner being the undominated alternative in the collective order. To determine that collective order one applies a {\em voting function}, or a vote aggregator, to the profile of votes $\V$. A vote aggregator $R$ assigns a nonempty set of options from $\Alt$ to a profile of votes $R: S \rightarrow \mathcal{P}^{\ast}(\Alt)$, where $S \subseteq \Pmc^n$.  As in judgment aggregation, $R$ is resolute if and only if for every $\V \in S$, $R(\V)$ is a singleton, and $R$ satisfies universal domain if $S = \Pmc^n$. 

We give two examples of voting aggregators: the Condorcet method and the Borda method. We choose to present these two because the are the most well known and studied methods in the social choice literature.

\subsubsection{The Condorcet method and the Borda method}

An example of a vote aggregator that does not satisfy universal domain is the Codorcet method. The Condorcet method considers all pairs of options $x,y \in \Alt$. The winner, called a {\em Condorcet winner}, is the option for which there is a majority that ranks it higher than every other option in $\Alt$. This method is easier to understand with  an example.

\begin{example}\label{ex:condwin} Consider the set of options $\Alt = \{ a,b,c,d\}$ and a five agents profile:
\[
\V_1 = \left(
\begin{array}{c}
 a \succ_1 c \succ_1 d \succ_1 b,\\
 b\succ_2 c \succ_2 d \succ_2a,\\
 d\succ_3 a \succ_3 c \succ_3 b,\\
 a\succ_4 b \succ_4 d \succ_3 c,\\
 d\succ_5 a \succ_5  c \succ_5  b\\
 \end{array}
\right)
\]
If we look at pairwise comparisons we obtain the following numbers. 
\begin{itemize}
\item For option $a$ we obtain that: 
\begin{itemize}
\item 3 agents voted $a \succ b$ and 2 agents voted $b \succ a$.
\item 4 agents voted $a \succ c$ and 1 agent voted $c \succ a$.
\item 2 agents voted $a \succ d$ and 3 agents voted $d \succ a$.
\end{itemize}
\item For option $b$ we obtain that: 
\begin{itemize}
\item 2 agents voted $b \succ a$ and 3 agents voted $a\succ b$.
\item 2 agents voted $b \succ c$ and 3 agents voted $c \succ b$.
\item 2 agents voted $b \succ d$ and 3 agents voted $d \succ b$.
\end{itemize}
\item For option $c$ we obtain that: 
\begin{itemize}
\item 1 agents voted $c \succ a$ and 4 agents voted $a \succ c$.
\item 3 agent  voted $c \succ b$ and 2 agents voted $b \succ c$.
\item 2 agents voted $c \succ d$ and 3 agents voted $d \succ c$.
\end{itemize}
\item For option $d$ we obtain that: 
\begin{itemize}
\item 3 agents voted $d \succ a$ and 2 agents voted $a \succ d$.
\item 3 agents voted $d \succ b$ and 2 agents voted $b \succ d$.
\item 3 agents voted $a \succ d$ and 2 agents voted $c \succ d$.
\end{itemize}
\end{itemize}

We see that for option $d$ and only for option $d$ there is a majority of agents  that ranks it higher than each of the other alternatives $a$, $b$ and $c$. Therefore $d$ is the Condorcet winner. 
\end{example}

The Condorcet method does not satisfy universal domain because there exist profiles that do not have a Condorcet winner. The profile that does not have a Condorcet winner, has a {\em Condorcet cycle}. We illustrate this case with an example. 

\begin{example} Consider a set of options $\Alt={a,b,c} $ and a three agent profile:
\[
\V_2 = \left(
\begin{array}{c}
 a \succ_1 b \succ_1 c,\\
 b\succ_2 c \succ_2 a,\\
 c\succ_3 a \succ_3 b \\
 \end{array}
\right)
\] 

In this profile we have 2 agents preferring $a$ to $b$ and 1 agent   preferring $a$ to $c$. For $b$ we have 2 agents preferring $b$ to $c$ and  1 agent   preferring $b$ to $a$. The alternative $c$ does not have a pairwise majority either: there are two agents who prefer $c$ to $a$, but only one agent that prefers $c$ to $b$. So none of the alternatives is the Condorcet winner. To understand why this phenomena is called a (Condorcet) cycle we need to draw what is known as the {\em profile's majority graph}.

 \begin{figure}[h!]
    \begin{center}
        \includegraphics[width=\textwidth]{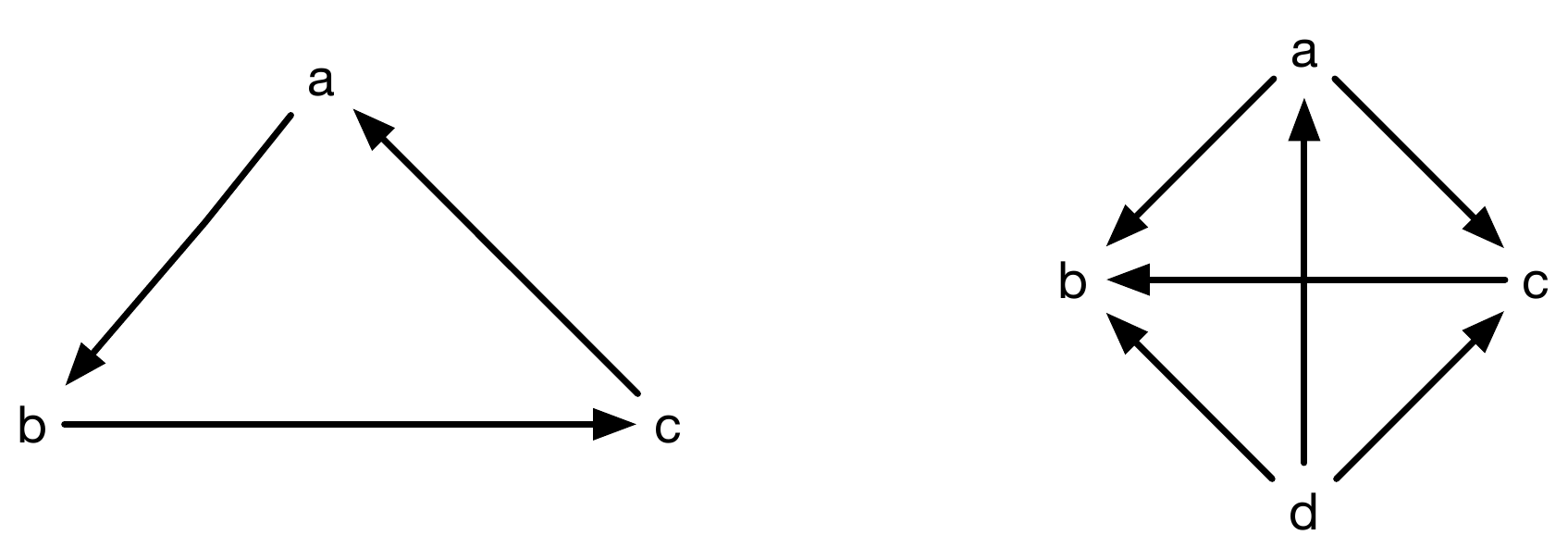} 
    \end{center}
\caption{Majority graph for profile $\V_2$ (left-hand side) and profile $\V_1$ from Example~\ref{ex:condwin} (right-hand side).}\label{fig:majgr}
\end{figure}
\end{example}

The {\em majority graph} is a graph in which the set of nodes is the set of options $\Alt$, and there exists a directed edge from $x \in \Alt$ to $y \in \Alt$ if and only if there is a majority of agents in $\V$ who prefer $x$ to $y$.  So for this $\V_2$ we obtain the majority graph given on the left-hand side in Figure~\ref{fig:majgr}. Observe that each one of the options has both incoming and outgoing edges. In contrast,  consider the majority graph for the profile $\V_1$ from Example~\ref{ex:condwin} given on the right-hand side in Figure~\ref{fig:majgr}. The profile $\V_1$ has a Condorcet winner, the option $d$. Observe how in the majority-graph, option $d$ has no incoming edges. This is a characteristic of the Condorcet winner: an option is a Condorcet winner for a profile $\V$ if and only if that option has no incoming edges in the majority graph of profile $\V$. 

Since the Condorcet winner has to "defeat" every other option in a pairwise comparison, it follows that when a profile has a Condorcet winner, this winner is unique. Therefore the Condorcet winner is a resolute vote aggregator. 
A vote profile that has a Condorcet winner is called {\em Condorcet consistent}. A vote aggregator is Condorcet-preserving if and only if it selects the Condorcet winner and nothing but the Condorcet winner when the profile aggregated has one. 
We can already intimate that there is a connection between Condorcet consistency in voting and majority  consistency in judgment aggregation. Such a relation does exist and we discuss it in Section~\ref{sec:majcon}.

The second voting method we consider, which historically proceeds the Condorcet method,   is the Borda method. The Borda method is an example of a vote aggregator that is irresolute and not Condorcet-preserving. Under the Borda method we assign {\em scores} to each option in $\Alt$ based on its position with respect to the other options in a vote $\succ_i$. Assume that there are $m$ options in $\Alt$, \ie~ $|\Alt| = m$.  The highest ranked option  in $\succ_i$ gets a score of $m-1$, the second highest ranked gets the score of $m-2$ and so on, with the last ranked option getting the score of $0$. The scores of an option from each voter are summed up, and the option with the highest score is the {\em Borda winner}. We illustrate the Borda method with an example.

\begin{example}
Consider the set of options $\Alt$ and the profile $\V_1$ from Example~\ref{ex:condwin}. The Borda scores from each of the votes for each of the options are given in Table~\ref{tab:borda}.
\begin{table}[h!]
\centering
\begin{tabular}{|r|c|c|c|c|}\hline
\rowcolor{red!25}                    & $a$ &$b$ &$c$ &$d$\\\hline
$\succ_1$ &3 &0&2&1\\\hline
$\succ_2$ &0 &3&2&1\\\hline
$\succ_3$ &2 &0&1&3\\\hline
$\succ_4$ &3 &2&0&1\\\hline
$\succ_5$ &2 &0&1&3\\\hline\hline
$\sum$     &10&5&6&9\\\hline
\end{tabular}
\caption{Borda scores for the profile $\V_1$ from Example~\ref{ex:condwin}.}\label{tab:borda}
\end{table}

As we can observe in the last row of the Table~\ref{tab:borda}, the option with the highest Borda score, 10, is the option $a$. Recall that this profile does have a Condorcet winner, option $d$, thus the Borda method is not Condorcet-preserving.
\end{example}
 
We now show how a voting problem can be represented as a judgment aggregation problem.

\subsubsection{Voting problems as judgment aggregation problems}

To gain an intuition of how a voting problem can be represented as a judgment aggregation problem recall how in the Condorcet method we did not work with the total preference orders over the entire set $\Alt$, but compared the options pair by pair, or pairwise. We can represent the set $\Alt$ as a set of ranked pairs of options, instead. So $\Alt = \{ a,b,c\}$ would become $\{ a\succ b, a\succ c, b\succ c\}$. Instead of creating a preference order over $\Alt$, each agent can express their preferences by saying whether she agrees or not that $a \succ b$, that $a\succ c$ and that $b\succ c$. Changing the elicitation and expression of information in this way, does not influence how much information is contained in the set of options and in each vote. Now observe that $a\succ b$ is nothing but a proposition, to which an agent assigns a truth-value. 

We use a special propositional variable $\pp{x}{y}$ to denote "option $x$ is preferred to option $y$", or "option $x$ is ranked higher than to option $y$".  Given a set of options $\Alt=\{ o_1, \ldots o_m\}$ we build the {\em preference pre-agenda }  $\A_{\Alt}^+$ as
\begin{equation}\label{eq:prag}
\A_{\Alt}^+ = \{\pp{o_i}{o_j} \mid o_i,o_j \in \Alt \textrm{ and } i>j \}.
\end{equation}

By requiring that  $i>j$  in (\ref{eq:prag}), we avoid having both $\pp{x}{y}$ and $\pp{y}{x}$ in the pre-agenda. We want this because in the agenda we will have $\neg \pp{x}{y}$ which has the same meaning as  $\pp{y}{x}$.

\begin{example} For $\Alt = \{ a,b,c,d\}$ we obtain   $\A_{\Alt} =\{ \pp{a}{b}, \neg \pp{a}{b},$ \linebreak$\pp{a}{c}, \neg \pp{a}{c},\pp{a}{d}, \neg \pp{a}{d}, \pp{b}{c}, \neg \pp{b}{c},\pp{b}{d}, \neg \pp{b}{d}, \pp{c}{d}, \neg \pp{c}{d}\}$.
\end{example}

A  total, strict and transitive order $\succ$ over $\Alt$ corresponds  to a judgment set $\Js $ when:
\begin{itemize}
\item  $x\succ_i y$ if and only if $\pp{x}{y} \in \Js_i$,
\item $\Js$ is complete for $\A_{\Alt}$, 
\item $\Js$ is consistent for
\begin{equation}  \Ct = \textrm{Tr} = \{ (\pp{x}{y} \wedge \pp{y}{z} \rightarrow) \mid x,y,z \in \Alt\}.
\end{equation}
\end{itemize}
"
The last condition is the transitivity constraint, ensuring that the transitivity of the "translated" preference order is preserved. The second condition ensures that the "translated" preference order is still total and strict. 
Hence, every profile of votes $\V \in \Pmc$ for a set of options $\Alt$ has a corresponding profile of judgments  $\Pf \in \Dmc(\A_{\Alt}, \textrm{Tr} )$. 

Sometimes in voting, the agents are required to give preference orders that are not necessarily complete, nor transitive and strict, but in which  a top preferred option exists. To capture those votes in judgment aggregation, we need a weaker constraint than $\textrm{Tr}$. The $\textrm{W}$ or {\em winner constraint} describes that there exist one option that in undominated, for that is preferred to any other option:
\begin{equation}
\textrm{W} = \{ \underset{o_i \in \Alt}{\bigvee} \underset{i\neq j}{\underset{o_j\in\Alt}{\bigwedge}} \pp{o_i}{o_j}\}.
\end{equation}

We include and example from \cite{ADT2013} to illustrate the difference between the $\textrm{Tr}$ and $\textrm{W}$ constraints. 
\begin{example} Consider $\Alt = \{a,b,c,d,e\}$ and a judgment set for $\A_{\Alt}$ \linebreak
 $\Js=\{  \pp{a}{b}, \pp{a}{c},\pp{b}{c} , \pp{d}{b},\pp{c}{e},\pp{e}{b}\}$.
 
 $\Js$ does not satisfy the $\textrm{Tr}$ constraint because $\pp{b}{c} \wedge \pp{c}{e} \wedge \pp{e}{b} $ violates  $(\pp{b}{c} \wedge \pp{c}{e}) \rightarrow\pp{b}{e} $. But an undominated option does exist, the option $a$, so $\Js$ satisfies the $\textrm{W}$ constraint.
\end{example}

\subsubsection{Majority consistency an Condorcet consistency}\label{sec:majcon}
There is a connection between the Condorcet winners of a vote profile $\V $, when they exist, and the elements of $m(\Pf)$, where $\Pf$ is a judgment profile corresponding to $\V$. To observe this, we go back to the majority graph. 
There is an edge from option $x$ to option $y$ in the majority graph for $\V$ if and only if there is a majority in $\V$ that consider   $x\succ y$. But  there is a majority in $\V$ that ranks $x$ higher than $y$ if and only if there is a majority of agents that selected $\pp{x}{y} $ in their judgment sets in the judgment profile $\Pf$ that corresponds to the vote profile $\V$. But then, by definition $\pp{x}{y} \in m(\Pf)$. Thus we can observe that  is an edge from option $x$ to option $y$ in the majority graph for $\V$ if and only if  $\pp{x}{y} \in m(\Pf)$.

Next observe that  there is a Condorcet winner for $\V\in\Pmc^n$  if and only if the corresponding $\Pf \in \Dmc(\A_{\Alt}, \textrm{Tr})^n$ is a majority consistent profile. If there is no Condorcet winner, the majority graph has no option that has only incoming edges, therefore  at least one cycle must exist in graph. Note that  there is a cycle between three options if and only if transitivity  fails. For example if there is a majority for which $a\succ b$ and there is a majority for which $b \succ c$ and there is a majority for which $c \succ a$, then $(\pp{a}{b}\wedge \pp{b}{c}) \rightarrow \pp{a}{c}$ is violated and the $m(\Pf)$ will not be consistent.

Next we present the definition from \cite{ADT2013} regarding when a vote aggregator is generalised by a judgment aggregator. 

\subsubsection{Vote aggregators and judgment aggregators}
We have shown how to represent a voting problem with a judgment aggregation problem, but how to determiner the winners from the collective judgment sets? 

The Condorcet winner of $\V$ will be one of the elements of the $m(\Pf)$, where $\Pf$ is the corresponding judgment profile to the voting profile $\V$, it will be the $x \in \Alt$  that always occurs on either the left side of the $\pp{o_i}{o_j}$ or on the right side of $\neg\pp{o_i}{o_j}$. This is the {\em undominated option} and it is the winner in a collective judgment set selected by a judgment aggregator. 

We define the winners of a collective judgment set $\Js \in \F(\Pf)$ for $\Pf \in \Dmc(\A_{\Alt}, \Ct)^n$:

\begin{equation}
\textrm{Winner}(\Js) = \{ x \mid  \pp{y}{x}\not\in \Js \textrm{ for any } y \in \Alt, y\neq x\}.
\end{equation}

Now we can define when a judgment aggregator generalises a vote aggregator, as in \cite{ADT2013}: when  a judgment aggregator always assigns judgment sets to a profile $\Pf$ that have the same winners as a vote aggregator would assign to a corresponding vote profile $\V$. Formally $\F$ generalises $R$ if and only if for every $\Pf \in \Dmc(\A_{\Alt},\Ct)^n$, $\Ct \in \{ \textrm{Tr}, \textrm{W}\}$ that corresponds to a vote profile $\V \in \Pmc^n$ it holds
\begin{equation}
R(\V) = \{ x \mid x\in \textrm{Winner}(\Js), \Js \in \F(\Pf)\}.
\end{equation}

\begin{figure}[h!]
   \includegraphics[width=\textwidth]{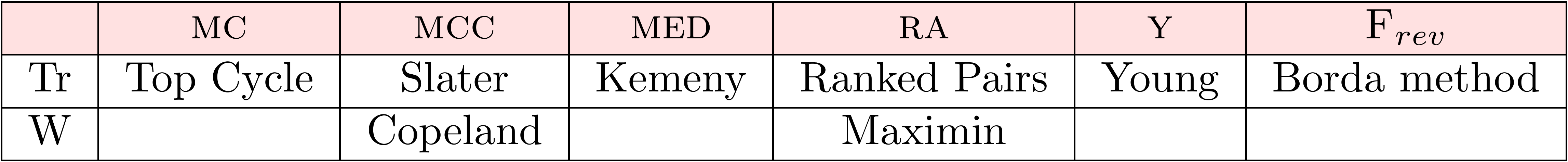}
\caption{Generalisation results between vote aggregators and judgment aggregators from \cite{ADT2013}.}\label{tab:resvot} 
\end{figure}

Figure~\ref{tab:resvot} summarises the relations between voting aggregators and judgment aggregators investigated in  \cite{ADT2013}. The proofs as well as definitions for the vote aggregators can be found in \cite{ADT2013}. Furthermore, it is shown \cite{Dietrich:2013} that $\REVS$ generalises the Borda method under the $\textrm{Tr}$ constraint.

The judgment representation of an aggregation problem is less succinct than  the preference aggregation. It is easy to see that if $|\Alt|=m$, then $|\A_{\Alt}| = (m-1)\cdot m$, \ie~twice (one for the positive one for the negative formula) the number of combinations of 2 elements from $m$ elements without repetition. 

\subsection{Belief Merging}

Belief merging is a knowledge representation discipline concerned with aggregating or merging several sets of formulas into one consistent set \cite{KoniecznyP11}. Judgment aggregation is concerned with aggregating sets of formulas, however unlike belief merging, in judgment aggregation we are interested in a particular set of formulas given by the agenda. We introduce the belief merging framework and highlight the differences. 
 
Recall that $\La_p$ is a set of well formed formulas of propositional logic. A knowledge base $K \subset \La_p$ is a finite set of formulas. A profile $E$ is a non-empty multi-set (a bag) of knowledge bases $E = \{ K_1, \ldots, K_n\}$. The set $\mathbb{E}$ is the set of all possible profiles. In addition to the knowledge bases, an integrity constraint $\mu \in \La_p$ is also defined.

Note how we did not require that a knowledge base is consistent. In belief merging there are no issues as in judgment aggregation, one knowledge base  can have formulas that  do not share formulas with another knowledge base. 

A belief merging problem is a pair $\langle E, \mu\rangle$. Clearly a judgment aggregation problem in the logic framework can be considered a belief merging problem, when the agenda is omitted. We can construct an agenda, when given a profile of knowledge bases, by including in it every formula that occurs in a $K_i$ in $E$, and its negation. However, then $K_i$ may be incomplete judgment sets for this agenda. 

A knowledge base is consistent if and only if $\bigwedge K$ is a consistent formula of propositional logic, denoted $K \nvDash \bot$. Let $\ai_i = \bigwedge K_i$.  Similarly, a profile $E=\{ K_1, \ldots, K_n\}$ is consistent if $\underset{1 \leq i \leq n}{\bigwedge} \ai_i$ is a consistent formula of propositional logic. We write $E \nvDash \bot$.

 The principle object of study within belief merging are operators that aggregated a profile of belief bases $E$ into a belief base $K$ that satisfies the integrity constraint $\mu$. A belief merging operator is defined as $\Delta: \mathbb{E} \times \La_p \rightarrow \mathcal{P}(\La_p)$. Belief merging operators are typically written $\Delta_{\mu}(E)$ instead of $\Delta(E,\mu)$.

In judgment aggregation, and social choice in general, an aggregator is first defined and then characterised by the properties that only it satisfies. In belief merging a minimal set of properties, or {\em postulates}, is defined such that a function is considered a belief merging operator if and only if it satisfies these properties. We give the postulates here. 
 Let  $\sqcup$ be the multi-set union operator: for $E_1 = \{ K_1, \ldots, K_n\}$ and $E_2 = \{K_{n_1}, \ldots, K_n\}$, $E_1 \sqcup E_1 = \{ K_1, \ldots, K_n,K_{n_1}, \ldots, K_n\}$.
\begin{description}
\item \textsc{(IC0)} $ \Delta_{\mu}(E) \models \mu$.\\
This postulate stipulates that the result of the merging must satisfy the integrity constraints.
\item \textsc{(IC1)} If $\mu$ is consistent, then $ \Delta_{\mu}(E)$  is consistent.\\
This postulate stipulates that the result of the merging must be a consistent set of formulas. 
 \item\textsc{(IC2)}  If  $E$ is consistent with $\mu$, then  $ \Delta_{\mu}(E) \equiv E\wedge \mu$.\\
 This postulate stipulates  that if the knowledge bases are consistent with each-other, their conjunction is consistent, and consistent with the integrity constraints, then merging them is just taking the conjunction of the profile with the integrity constraints.
 \item\textsc{(IC3)} If $E_1 \equiv E_2$ and $\mu_1 \equiv \mu_2$, then $\Delta_{\mu_1}(E_1) \equiv \Delta_{\mu_2}(E_2)$.\\
 This postulate describes the desirability of the merging operator to consistent and always aggregate the same knowledge bases under the same constraints in the same manner.
 \item\textsc{(IC4)} If $K_1 \models \mu$ and $K_2\models \mu$, then  $\Delta_{\mu}(\{ K_1,K_2\}) \wedge K_1$ is consistent if and only if   $\Delta_{\mu}(\{ K_1,K_2\}) \wedge K_2$ is consistent.\\
 This postulate guarantees that no knowledge base is given preferential treatment when they are consistent with the integrity constraints. 
\item \textsc{(IC5)} $\Delta_{\mu}(E_1) \wedge \Delta_{\mu}(E_2) \models \Delta_{\mu}(E_1 \sqcup E_2)$.\\
This postulate expresses the requirement that if two groups $E_1$ and $E_2$ agree on some formulas  then these formulas will be chosen if we join the two groups. 
\item  \textsc{(IC6)} If $\Delta_{\mu}(E_1) \wedge  \Delta_{\mu}(E_2) $ is consistent, then $\Delta_{\mu}(E_1 \sqcup E_2) \models \Delta_{\mu}(E_1) \wedge \Delta_{\mu}(E_2) $.\\
This postulate is the second direction of postulate \textsc{(IC5)}. Together these two postulate stipulate that if two groups agree on at least one formula (that it is true), then the result of merging  two groups is precisely the set of formulas on which they al agree. 
\item  \textsc{(IC7)}  $\Delta_{\mu_1}(E) \wedge \mu_2 \models \Delta_{\mu_1 \wedge \mu2}(E)$.\\
This postulate  guarantees that if a   profile merged under constraint $\mu_1$ is consistent with constraint $\mu_2$, then the result of this merge is the same as merging the profile under both constraints. 
\item \textsc{(IC7)} If $\Delta_{\mu_1}(E) \wedge \mu_2$ is consistent, then $\Delta_{\mu_1 \wedge \mu2}(E) \models \Delta_{\mu_1}(E) $. \\
This postulate is the second direction of the previous one. Together they act  as a kind of  reasoning monotonicity property. They guarantee that if a formula is among the results of the merge, this formula is consistent with a second integrity constraint, and then the profile is aggregated by a stronger constraint, the formula will still be a result of the merge.  
\end{description}

Everare {em et al.} \cite{EveraereKM15} show a correspondence between belief merging postulates and properties of judgment aggregator. Thus the postulate \textsc{(IC0)}   is the requirement that the collective  judgment sets are consistent, which is satisfied by construction of the judgment aggregators. Satisfying  \textsc{(IC1)}  corresponds to satisfying the universal domain property. 

Postulate   \textsc{(IC2)} corresponds to a judgment aggregator property defined in \cite{EveraereKM15} and called {\em consensuality}. A judgment profile $\Pf$ is consensual if and only if there exists at least one judgment $\ai \in \A$ that is unanimously supported in $\Pf$. A judgment  aggregator satisfies consensuality if and only if for every consensual profile on some $\ai \in \A$, we have that $\ai \in \Js$ for every $\Js \in \F(\Pf)$. Quite obviously, the consensuality property is the same as the strong unanimity principle we defined in Section~\ref{sec:una}, introduced in \cite{ADT09,TARK11}.

Postulate   \textsc{(IC3)} corresponds to the judgment aggregator property anonymity, while postulates  \textsc{(IC4)} resembles a neutrality property, but it is not quite the same. This postulate requires a new judgment aggregator property to capture it.  Intuitively, we can expect that this property would not be satisfied by many of the judgment aggregators we introduced, in particular those from the class of distance-based aggregators.  

Postulates   \textsc{(IC5)} and  \textsc{(IC6)}  correspond to the reinforcement judgment aggregator property, defined as consistency in \cite{EveraereKM15}. 

Lastly postulates   \textsc{(IC7)} and  \textsc{(IC8)} also do not correspond to any of the judgment aggregator properties. Two new properties are defined in \cite{EveraereKM15} to capture \textsc{(IC7)} and  \textsc{(IC8)}  in judgment aggregation, called {\em Sen's property $\alpha$} and {\em Sen's property $\beta$} respectively. These properties describe what should happen if some issues are added or removed from the agenda. We give the definitions of these properties in our framework. 

\paragraph{Sen's property $\alpha$.} Consider $\Pf \in \Dmc(\A,\Ct)^n$ and consider agenda $\A'$ such that $\A' \subset \A$. An aggregator $\F$ satisfies Sen's property $\alpha$ if and only if  for any $\ai \in \A'$,  if $\ai \in \Js$ for every $\Js \in \F(\Pf)$, then $\ai \in \Js'$ for every $\Js' \in \F(\Pf^{\downarrow \A'})$.

\paragraph{Sen's property $\beta$.} Consider $\Pf \in \Dmc(\A,\Ct)^n$ and consider agenda $\A'$ such that $\A' \subset \A$. An aggregator $\F$ satisfies Sen's property $\beta$ if and only if for any  $\ai_1, \ai_2 \in \A'$ such that $\{\ai_1,\ai_2\} \subseteq  \Js$ for every $\Js \in  \F(\Pf^{\downarrow \A'})$ it holds that $\ai_1 \in \Js$  if and only if $\ai_2 \in \Js$ for every $\Js \in \F(\Pf)$.

Although reminiscent of the agenda separability properties, the Sen's properties have not been previously considered in judgment aggregation and they are obviously more demanding then agenda separability properties. This impression is also echoed in \cite{EveraereKM15} where it is argued that the Sen's properties are undesirable since they do not  take into consider logical relations between the issues in the agenda. 

  \newpage

\section{Aggregation problems in MAS }  

\label{ch:mas}  

In this section we give an overview of different types of collective decision-making problems in multi agent systems that can be modelled with judgment aggregation, and those that probably should not. We discern between problems based on where the judgment sets are coming from and what type of issues are in the agenda. We associate different types of problems with different aggregator properties. Judgment aggregation has been used to decide collective goals based on collectively supported beliefs \cite{COIN10,Synthese12new}, as a method to aggregate arguments \cite{AwadBTR14,BoothAR14}, it has been considered as a method to aggregate graphs \cite{EndrissG14a}, and as a tool for determining collective annotations  \cite{EndrissFernandezACL2013}. Here we give a more general categorisation of problems that can be handled by judgment aggregation.  What is perhaps best illustrated in this section is that research in judgment aggregation has so far apparently opened more problems than it has solved, including the  problem of designing efficient judgment aggregators remaining open.

\subsection{When to use  judgment aggregation?}

 Judgment aggregation is a good tool to model complex collective decision-making problems. We use the word "complex" to denote problems in which multiple interdependent decisions have to be made at the same time on issues that are of different epistemic types, \eg\linebreak preferences, estimates, goals, beliefs, etc. 
It is best suited for decision problems in which we need a  decision on the truth state of multiple issues.

 If we know precisely for which issues we need a truth state decision, and perhaps what constraint those truth-states should satisfy,  then these are the issues that will form the agenda and the constraints respectively. If we are interested in what is true for several different sources of information, without having questions on which we are looking for answers, then it is better to use  belief merging. 

We showed that preferences can be aggregated as judgments, however we also showed that representing the preferences as orders over options if far more efficient than representing the aggregation problem with judgments and transitivity constraints. Therefore, for choosing one option from a set of options, based on input from various agents, it is best to use voting or preference aggregation directly. An exception is the case when the options are not independent and choosing one option can imply choosing or excluding other options. We discuss this special case in the next section.


\subsection{Types of agenda issues}
The agenda in judgment aggregation is defined as a set of propositional logic formulas. This a very abstract way of representing issues. On one hand this abstraction allows decision problems involving many different kinds of issues to be modelled in the judgment aggregation framework. In contrast, preference aggregation models  only preferences between options. An issue $\ai$ can model  "option $a$ is better than option $b$, which expresses preference, or  $\ai$ can model "the value of the British pound in January 2017 will be lower than the value of the British pound in January 2016", expressing a value judgment. 

Although this information on the epistemic nature of issues is lost in the agenda representation, it must be recaptured by the properties of the judgment aggregator. Thus, for example, when aggregating preferences one agent's preference should influence the selection of collective judgment sets as much as every other agent's preferences, thus requiring that the aggregator is anonymous. In contrast, some agents may be better at giving value judgments than others, and the aggregator should give these agents more influence on deciding the collective judgment sets. 

We consider various types of agenda issues that can be encountered and suggest when appropriate aggregator properties that should be satisfies when aggregating judgments on these issues. 

\subsubsection{Subjective judgments} 

Votes are best aggregated by vote aggregators or preference aggregators, but that approach holds best when we need to choose one from a set of options. There are natural problems in which we need to chose a combination of options, with certain combinations being feasible and others not, due to resource limitations for example. This problem is known as {\em combinatorial voting} \cite[Chapter 9]{handbook}. We illustrate it with an example. 

\begin{example} We need to select a meal to be served at a banquette. The meal consists of a starter, main course and desert. the caterer provides a list of options for staters $\Alt_S$, mains  $\Alt_M$ and desert  $\Alt_D$ and all participants make their preference, but they may only choose one of each. We have a combinatorial set of options in this problem $\Alt_S \times \Alt_M \times \Alt_D$, called a {\em combinatorial domain}. 
Assume  $\mathcal{O}_S= \{ a,b,c\}$, $\Alt_M = \{x,y\}$, $\Alt_D=\{ \alpha, \beta, \gamma\}$,  then the agents need to select one from  options  $ \Alt= \{ (a,x,\alpha), (a,x,\beta), (a,x, \gamma), $ \linebreak $(a,y,\alpha), (a,y,\beta), (a,y, \gamma),  (b,x,\alpha), (b,x,\beta), (b,x, \gamma), (b,y,\alpha), (b,y,\beta),$ \linebreak $ (b,y, \gamma), (c,x,\alpha), (c,x,\beta), (c,x, \gamma), (c,y,\alpha), (c,y,\beta), (c,y, \gamma)\}$.  

We can model this problem as a judgment aggregation problem.  We can form an agenda such that   $\A^+ =\{p_{o}\mid p\in \Alt_S \cup \Alt_M \cup \Alt_D\}$, so  for each option an agent can give a judgment $p_{o}$ to express choosing that option and  $\neg p_{o}$ to express not choosing it.  We use the constraints $\Ct =\{ (a \veebar b \veebar c) \wedge (x \veebar y) \wedge (\alpha \veebar \beta \veebar \gamma)\}$ to ensure that only one of each courses is selected.

We can easily express further limitations on what combination can be selected by adding constraints. 
\end{example}

Grandi considers judgment aggregation for combinatorial voting \cite{GrandiIJCAI2011}. \linebreak
Combinatorial voting problems can be found in many resource allocation scenarios, where bundles of resources need to be allocated to a group of agents. 

We should point out that the reader should not confuse combinatorial voting with the only apparently similar  {\em committee voting}. In committee voting there is only one set of options, but we want to select $k$ winners from them not only one. An example of a committee voting problem is when we need to select  representatives for 160 seats in parliament and there are 3000 candidates that are running for a seat. More on committee voting, also known as multiwinner elections can be found in \cite{Elkind:2014}.

The combinatorial voting  agenda  whose pre-agenda is defined for a combinatorial domain $D_1 \times \cdots \times D_k$ as $\A^+ =\{p_{o}\mid p_{o}\in (D_1 \cup \cdots \cup D_k) \}$ contains only issues that express preferences, although these issues are not of form $\pp{o_i}{o_j}$. 

Epistemically, preferences are question whose true answer is {\em subjective}, meaning that there exists no one true answer for everyone, but every agent is an authority on themselves.  For example, one person may prefer combining beer and potato chips, while another may prefer beer and caviar. Even if there is a consensus that beer and caviar are not a match, it is not wrong to make that preference, choice or have that desire. Therefore,  aggregating subjective judgments should be done in such a way that as many of the preferences as possible are respected. Aggregators that satisfy unanimity principles and majority-preservation are adequate for these kinds of problems, as also argued in \cite{Rabinowicz16}. 

\subsubsection{Value judgments}
In contrast to subjective judgments, value judgments do have an objective true-state. For example proposition $\ai$ denoting "the weight of the ox is 453kg" is a value judgment. Two agents may have different opinions regarding  whether the weight of the ox is or is not 453kg, however only one of them can be right and this can be established objectively by measuring the ox.   

When dealing with intelligent autonomous agents, agendas with issues that require value judgments can be expected to occur perhaps more often than agendas with issues that require subjective judgments when a group of agents needs to estimate the truth-state of various parameters and propositions in order to make decisions about what to do, for example. The type of collective judgment sets that we are looking for when aggregating value judgments are different then when aggregating subjective judgments. Here it is not so relevant how many of the individual judgments coincide with the collective judgments, hence unanimity preservation and majority preservation are not  essential. We would ideally want selected as collective, those judgments that have the highest probability of being the same with the objective truth. 

The property of judgment aggregators to select the collective judgments with highest probability of being the objective truth is called {truth-tracking}. The truth-tracking properties of judgment aggregators is a very \linebreak under-explored area, with the exception of some work on the premise-based, conclusion-based agenda's and the median aggregator\cite{Hartmann2012,PigozziH07,PigozziH07a}. It was shown in \cite{AAMAS-2-12}, through experiments with robots,  that when the premises are propositional variables, the premised based procedure is the better than the conclusion-based procedure and the median aggregator. Further empirical and probabilistic analysis is needed to identify the truth-tracking properties of the remaining judgment aggregators.   

\subsubsection{Equivalence relations}
A special type of an agenda issue  is the {\em equivalence relation} which is the question "Is    $x=y$?". Some equivalence relations require subjective judgments, for example "Is ordering in indian food the same  as going out for thai food?", other's require   value judgments, \eg~ "Are these two pictures both pictures of the Eiffel tower? ". The aggregation of judgments on equivalences is an important problem in artificial intelligence (AI) in general since a lot of classification of equivalent objects is at the core of many AI problems\footnote{ Classification tasks usually would involve value judgments.}. 

The aggregation of equivalence relations was considered in \cite{Fishburn1986}. An equivalence relation is a binary predicate that is symmetric, reflexive and transitive. Namely $P$ is an equivalence relation if for all $x,y,z$ it holds that $P(x,x)$ is true (reflexivity), if $P(x,y) \leftrightarrow P(y,x)$ holds (symmetry), and  $(P(x,y) \wedge P(y,z)) \rightarrow P(x,z)$ holds (transitivity).  Given  a set of items  $I$ we can construct an agenda $\A_{I}$ as follows

\begin{equation}
\A_{I} = \{ p_{x=y}, \neg  p_{x=y} \mid x,y \in I, x\neq y\}
\end{equation}
A judgment $p_{x=y}$ denotes that $x$ is equal to $y$, while $\neg  p_{x=y} $ denotes that $x$ is not equal to $y$.
To enforce symmetry and transitivity we construct the following set of constraints

\begin{equation}
\Ct = \{ (p_{x=y} \wedge p_{y=z} )\rightarrow p_{x=z} \mid x,y,z \in I\} \cup \{ (p_{x=y} \leftrightarrow  \mid x,y \in I\}. 
\end{equation}
With which kind of aggregator judgments on equivalence relations are best aggregated would depend on whether they are value or subjective judgments.

\subsubsection{Controlled judgments}
We briefly would like to make notice of a particular type of issues that can occur in collective decision making problems among artificial agents. Some issues are such that one agent, or one group of agents, can force their truth-value. For example, I am the only person who can determine whether the judgment for the issue "I can lift the book on the desk." is true or falls. Other agents may have guesses and means of estimating whether this issue is true or false, but I am the only one that is an authority on its truth-state.  If a group of agents are making a collective plan to accomplish a goal, and only one of the agents is able to execute action $a$, then this agent is the only one that can give a judgment on the issue "action $a$ is feasible". 

If an agenda contains only issues that admit controlled judgments, then there is no collective decision making problem. If however, an agenda contains both  controlled judgments and either value judgments or subjective judgments. Aggregating the judgments in such an agenda issue by issue does not guarantee a consistent collective judgment set. This is the so called 
Liberal, or Sen's paradox in judgment aggregation \cite{Dietrich2008}. We illustrate it with an example, with the same structure as  to the one in 
\cite{Dietrich2008}.

\begin{example} Consider the following issues:
\begin{description}
\item $\ai_1$: Is action $a$ feasible? 
\item $\ai_1 \rightarrow\ai_2$: Is it the case that if action $a$ is feasible, then plan $p$ is feasible? 
\item $\ai_2$: Is plan $p$ feasible? 

Consider three agents with agent one being the authority on  $\ai_1$ and agent two, the planner being the authority on  $\ai_2$. We have the agenda $\A = \{ \ai_1, \neg \ai_1, \ai_1 \rightarrow\ai_2, \neg(\ai_1 \rightarrow\ai_2), \ai_2, \neg \ai_2\}$ and $\Ct= \{\top\}$.  The agents give incomplete judgment sets. A profile is given in Table~\ref{tab:sen}.

\begin{table}
\centering
\begin{tabular}{r|ccc}
Voters & $\ai_1$ & $\ai_1 \rightarrow \ai_2$ & $\ai_2$\\\hline
\multicolumn{4}{c}{      $\Ct = \{\top\}$}\\\hline
$\Js_1$ & + & & - \\ 
$\Js_2$ & & + & -\\
$\Js_3$ & &   &+\\ \hline
$m(\Pf)$  & +&+   &-
\end{tabular}
\caption{An example of the Liberal paradox in judgment aggregation.} \label{tab:sen}
\end{table}

\end{description}
\end{example}
Note that the Liberal paradox occurs also in judgment aggregation problems where the judgments are of different kinds but one expert or group of expert only is consulted in the profile and decides the collective judgment for an issue. 

 It is not at all clear how agendas with controlled judgments and in general agendas with  issues that admit different types of judgment  should be handled in judgment aggregation. These agendas occur for example when determining which goal a group to pursue based on what beliefs the group upholds about the world, a problem modelled as a judgment aggregation problem in \cite{Synthese12new}, where it is recommended that a two-step procedure is used. 
 Different types of issues need to be treated differently  as argued in \cite{Slavkovik14}, however non-neutral aggregators, with the exception of the binomial rules of \cite{CostantiniEtAlAAAI2016} have not yet been developed.

\subsection{Types of information sources}
In addition to what types of judgments the issues in the agenda admit, we must also consider how the judgment sets are formed. The intuitive collective decision making problem is the problem of a group of agent reaching an agreement of which judgments should be assigned to which issues. In this example, the judgment sets of the profile regardless of the type of the judgments,  are formed by each decision maker, in response to the issues in the agenda. The agents can be instructed to construct full judgment sets and to  the constraints associated with the agent and further more, they can consider the collective judgment sets after they are formed and set a tie-breaking method. 

However, judgment aggregation can also be used to combine information from several information sources, like knowledge bases, user reviews, databases etc.  While the collective judgment sets are used by many in the first case, here they are used by the agenda setter. Although merging information is the domain of belief merging, if we are only interested in only the boolean answers of specific questions, then we should use judgment aggregation. One has to be aware however, that, as shown in \cite{EveraereKM15} aggregating the full information bases may not give the same outcome as aggregating judgments on particular questions. It is however not reasonable to expect that every source will be such that judgments to all issues can be extracted from it. Furthermore, it cannot be expected that every source will abide to the agenda constraints. We illustrate this case of judgment set sources with an example. 

\begin{example}\label{ex:review} Assume we are trying to decide whether to visit Sicily in July. There are multiple reviews and travel guides online we can consult. A rational agent would base her decision on reasons. So what is a good reason to "visit Sicily", an issue we mark $d$? 
Here is an example of reasons:
\begin{description}
\item $p_1$ : I have the desire to travel.
\item $p_2$ : There are good beaches in Sicily.
\item $p_3$ : Accommodation is affordable.
\item $p_4$ : There are interesting sites to visit. 
\item $p_5$: Hotels are cheaper than $50$ euros per night .
\end{description} 

Clearly  issue $p_1$ admits  a controlled judgment - only I can speak of my desires. Issues $p_2$, $p_3$ and $p_4$ are subjective judgments, while issue $p_5$ is a value judgment since it can objectively be determined what the prices of accommodation are. 
The constraints would be $\Ct = \{ \big( p_1 \wedge (p_2 \vee p_4) \wedge p_3 \big) \leftrightarrow d, p_5 \rightarrow p_3\}$ and the pre-agenda is   $\A^+ = \{ d, p_1,p_2,p_3,p_4,p_5\}$. 

However, not all information sources would abide by our agenda constraints. Some people may consider hotels cheaper than $80$ euros per night affordable, for others, Scilly is good place to visit (judgment $d$) because of completely different reasons, like good food, although they do not think that there are good beaches or affordable accommodation there. 
\end{example}

We see from Example~\ref{ex:review} that, since the agenda setter is  the user of the aggregated information, what matters is that her conditions are satisfied, not that other people agree that this is a good way to make a decision. Thus although we aggregate subjective judgments  in the case of information sources, it is not paramount to use majority-preserving aggregators, or aggregators that satisfy the unanimity principle(s). 

 To handle aggregation problems as in   Example~\ref{ex:review}  we need aggregators that are of type $\F: \mathcal{P}(\A)^n \rightarrow \mathcal{P}^{\ast}(\Dmc(\A,\Ct)) $, so aggregators that assign a set of non-empty judgment sets consistent with the agenda constraint $\Ct$ to a  profile of possibly incomplete and inconsistent (with $\Ct$) sets of judgments. The $\RMSA$, $\RMCSA$, $\RMWA$, $\RY$ and $\RRA$ aggregators can be easily extended to become this type of aggregators. 


\newpage


\renewcommand{\refname}{\spacedlowsmallcaps{References}} 

\bibliographystyle{unsrt}

\bibliography{Bibliography.bib} 


\end{document}